\documentclass[review]{elsarticle}
\def\thetitle{Conformance Checking of\\Mixed-paradigm Process Models}

\def\thekeywords{Process mining \sep Conformance checking \sep Declarative process models \sep Imperative process models \sep Mixed-paradigm process models \sep Hybrid process models}
\usepackage[english]{babel}
\usepackage{floatrow}
\usepackage{soul}
\usepackage{subcaption}
\usepackage{url}
\usepackage{colortbl}
\usepackage[table]{xcolor}
\usepackage{amssymb}
\usepackage{stmaryrd}
\usepackage{amsmath}
\usepackage{mathrsfs}
\usepackage{amssymb}
\usepackage{amsthm}
\usepackage{xfrac}
\usepackage[textsize=scriptsize,backgroundcolor=yellow!40]{todonotes}
\usepackage[nonumberlist,acronym,sanitize=none]{glossaries}
\glsdisablehyper
\usepackage{comment}
\usepackage[pdftex, colorlinks=true, hyperfootnotes=true, hyperindex=true,
            plainpages=false, pagebackref=false, pdfpagelabels=true, pdfstartview=FitH,
            linkcolor=blue, citecolor=blue, urlcolor=blue,
            bookmarks, bookmarksopen, bookmarksdepth=3]{hyperref}
\usepackage[capitalise,nameinlink]{cleveref}
\crefname{algocf}{Alg.}{Algs.}
\Crefname{algocf}{Algorithm}{Algorithms}
\usepackage{tkz-base}
\usetikzlibrary{decorations.pathmorphing,trees,snakes,arrows,shapes,automata,petri}
\usepackage{paralist}
\usepackage{multicol}
\usepackage{booktabs}
\usepackage{enumitem}
\usepackage{multirow}
\usepackage{rotating}
\usepackage{etaremune}
\usepackage[ruled,linesnumbered,algo2e,noend]{algorithm2e}
\SetAlFnt{\tiny\ttfamily}
\usepackage{marginnote}
\usepackage{mathtools}
\usepackage{mathabx}
\usepackage{adjustbox}
\usepackage{ifthen}
\usepackage[normalem]{ulem}
\usepackage[mathlines,switch]{lineno}
\usepackage{esvect}
\usepackage{pgfplots}
\usepackage{makecell}
\usepackage{diagbox}
\usepackage[scientific-notation=false,group-separator={,}]{siunitx}
\usepackage{microtype}
\usepackage{epstopdf}
\usepackage{marvosym}
\RequirePackage{amsthm}
\theoremstyle{definition}
\newtheoremstyle{mystyle}
{6pt}
{3pt}
{}
{}
{\bfseries}
{.}
{.5em}
{\thmname{#1}\thmnumber{ #2}\thmnote{ (#3)}}
\theoremstyle{mystyle}
\newtheorem{definition}{Definition}

\RequirePackage{cleveref}
\crefname{definition}{Def.}{Defs.}
\Crefname{definition}{Definition}{Definitions}
\crefname{proposition}{Prop.}{Props.}
\Crefname{proposition}{Proposition}{Propositions}
\crefname{theorem}{Th.}{Ths.}
\Crefname{theorem}{Theorem}{Theorem}
\crefname{corollary}{Cor.}{Cors.}
\Crefname{corollary}{Corollary}{Corollaries}
\crefname{lemma}{Lemma}{Lemmata}
\Crefname{lemma}{Lemma}{Lemmata}
\crefname{example}{Example}{Examples}
\Crefname{example}{Example}{Examples}
\RequirePackage{xparse}

\NewDocumentCommand{\DrawAbsence}{ +O{$\paramx$} }{%
\begin{tikzpicture}[baseline=(current bounding box.center)]\node[DECLARE.task,DECLARE.existcon=$0..m$]{#1};\end{tikzpicture}%
}

\NewDocumentCommand{\DrawAltprec}{ O{$\paramx$} O{$\paramy$} }{%
\begin{tikzpicture}[baseline=(current bounding box.center),node distance=3.5*\DTZU]
 \node[DECLARE.task,] (1) {#1};
 \node[DECLARE.task,right=of 1] (2) {#2};
   
 \path (1) edge [DECLARE.alt.prec] node {} (2);
\end{tikzpicture}%
}

\NewDocumentCommand{\DrawAltresp}{ O{$\paramx$} O{$\paramy$} }{%
\begin{tikzpicture}[baseline=(current bounding box.center),node distance=3.5*\DTZU]
 \node[DECLARE.task,] (1) {#1};
 \node[DECLARE.task,right=of 1] (2) {#2};
   
 \path (1) edge [DECLARE.alt.resp] node {} (2);
\end{tikzpicture}%
}

\NewDocumentCommand{\DrawAltsucc}{ O{$\paramx$} O{$\paramy$} }{%
\begin{tikzpicture}[baseline=(current bounding box.center),node distance=3.5*\DTZU]
 \node[DECLARE.task,] (1) {#1};
 \node[DECLARE.task,right=of 1] (2) {#2};
   
 \path (1) edge [DECLARE.alt.succ] node {} (2);
\end{tikzpicture}%
}

\NewDocumentCommand{\DrawAtmostone}{ O{$\paramx$} }{%
\begin{tikzpicture}[baseline=(current bounding box.center)]\node[DECLARE.task,DECLARE.existcon=$0..1$]{#1};\end{tikzpicture}%
}

\NewDocumentCommand{\DrawChnprec}{ O{$\paramx$} O{$\paramy$} }{%
\begin{tikzpicture}[baseline=(current bounding box.center),node distance=3.5*\DTZU]
 \node[DECLARE.task,] (1) {#1};
 \node[DECLARE.task,right=of 1] (2) {#2};
   
 \path (1) edge [DECLARE.chn.prec] node {} (2);
\end{tikzpicture}%
}

\NewDocumentCommand{\DrawChnresp}{ O{$\paramx$} O{$\paramy$} }{%
\begin{tikzpicture}[baseline=(current bounding box.center),node distance=3.5*\DTZU]
 \node[DECLARE.task,] (1) {#1};
 \node[DECLARE.task,right=of 1] (2) {#2};
   
 \path (1) edge [DECLARE.chn.resp] node {} (2);
\end{tikzpicture}%
}

\NewDocumentCommand{\DrawChnsucc}{ O{$\paramx$} O{$\paramy$} }{%
\begin{tikzpicture}[baseline=(current bounding box.center),node distance=3.5*\DTZU]
 \node[DECLARE.task,] (1) {#1};
 \node[DECLARE.task,right=of 1] (2) {#2};
   
 \path (1) edge [DECLARE.chn.succ] node {} (2);
\end{tikzpicture}%
}

\NewDocumentCommand{\DrawCoex}{ O{$\paramx$} O{$\paramy$} }{%
\begin{tikzpicture}[baseline=(current bounding box.center),node distance=3.5*\DTZU]
 \node[DECLARE.task,] (1) {#1};
 \node[DECLARE.task,right=of 1] (2) {#2};
   
 \path (1) edge [DECLARE.coex] node {} (2);
\end{tikzpicture}%
}

\NewDocumentCommand{\DrawEnd}{ O{$\paramx$} }{%
\begin{tikzpicture}[baseline=(current bounding box.center)]\node[DECLARE.task,DECLARE.existcon=End]{#1};\end{tikzpicture}%
}

\NewDocumentCommand{\DrawExistence}{ O{$\paramx$} O{$\paramy$} }{%
\begin{tikzpicture}[baseline=(current bounding box.center)]\node[DECLARE.task,DECLARE.existcon=$n..\ast$]{#1};\end{tikzpicture}%
}

\NewDocumentCommand{\DrawInit}{ O{$\paramx$} }{%
\begin{tikzpicture}[baseline=(current bounding box.center)]\node[DECLARE.task,DECLARE.existcon=Init]{#1};\end{tikzpicture}%
}

\NewDocumentCommand{\DrawNotchnsucc}{ O{$\paramx$} O{$\paramy$} }{%
\begin{tikzpicture}[baseline=(current bounding box.center),node distance=3.5*\DTZU]
 \node[DECLARE.task,] (1) {#1};
 \node[DECLARE.task,right=of 1] (2) {#2};
   
 \path (1) edge [DECLARE.chn.succ,DECLARE.neg] node {} (2);
\end{tikzpicture}%
}

\NewDocumentCommand{\DrawNotcoex}{ O{$\paramx$} O{$\paramy$} }{%
\begin{tikzpicture}[baseline=(current bounding box.center),node distance=3.5*\DTZU]
 \node[DECLARE.task,] (1) {#1};
 \node[DECLARE.task,right=of 1] (2) {#2};
   
 \path (1) edge [DECLARE.coex,DECLARE.neg] node {} (2);
\end{tikzpicture}%
}

\NewDocumentCommand{\DrawNotsucc}{ O{$\paramx$} O{$\paramy$} }{%
\begin{tikzpicture}[baseline=(current bounding box.center),node distance=3.5*\DTZU]
 \node[DECLARE.task,] (1) {#1};
 \node[DECLARE.task,right=of 1] (2) {#2};
   
 \path (1) edge [DECLARE.succ,DECLARE.neg] node {} (2);
\end{tikzpicture}%
}

\NewDocumentCommand{\DrawParticipation}{ O{$\paramx$} }{%
\begin{tikzpicture}[baseline=(current bounding box.center)]\node[DECLARE.task,DECLARE.existcon=$1..\ast$]{#1};\end{tikzpicture}%
}

\NewDocumentCommand{\DrawPrec}{ O{$\paramx$} O{$\paramy$} }{%
\begin{tikzpicture}[baseline=(current bounding box.center),node distance=3.5*\DTZU]
 \node[DECLARE.task,] (1) {#1};
 \node[DECLARE.task,right=of 1] (2) {#2};
   
 \path (1) edge [DECLARE.prec] node {} (2);
\end{tikzpicture}%
}

\NewDocumentCommand{\DrawResex}{ O{$\paramx$} O{$\paramy$} }{%
\begin{tikzpicture}[baseline=(current bounding box.center),node distance=3.5*\DTZU]
 \node[DECLARE.task,] (1) {#1};
 \node[DECLARE.task,right=of 1] (2) {#2};
   
 \path (1) edge [DECLARE.resex] node {} (2);
\end{tikzpicture}%
}

\NewDocumentCommand{\DrawResp}{ O{$\paramx$} O{$\paramy$} }{%
\begin{tikzpicture}[baseline=(current bounding box.center),node distance=3.5*\DTZU]
 \node[DECLARE.task,] (1) {#1};
 \node[DECLARE.task,right=of 1] (2) {#2};
   
 \path (1) edge [DECLARE.resp] node {} (2);
\end{tikzpicture}%
}

\NewDocumentCommand{\DrawSucc}{ O{$\paramx$} O{$\paramy$} }{%
\begin{tikzpicture}[baseline=(current bounding box.center),node distance=3.5*\DTZU]
 \node[DECLARE.task,] (1) {#1};
 \node[DECLARE.task,right=of 1] (2) {#2};
   
 \path (1) edge [DECLARE.succ] node {} (2);
\end{tikzpicture}%
}
 \modulolinenumbers[5]
\newcolumntype{d}{>{\columncolor{gray!10}}c}
\newcolumntype{m}{>{\columncolor{gray!10}}l}
\newcommand{\grayrow}{\rowcolor{gray!20}} %
\newenvironment{iiilist}%
{\begin{inparaenum}[\itshape(i)\upshape]}%
{\end{inparaenum}}

\RequirePackage{xparse}
\NewDocumentEnvironment{AuthNote}{+o+o}{%
	\IfValueT{#2}{\marginnote{\scriptsize{#2}}}%
	\begin{scriptsize}
		\colorbox{gray}%
		{\color{white} Note\IfValueT{#1}{ (#1)}:}%
		\quad%
		\color{brown}
}{%
	\normalcolor
	\end{scriptsize}
}

\RequirePackage{pifont}

\RequirePackage{lipsum}
\newcommand{\LipsumGray}[1][]{{\color{gray}\ifthenelse{\equal{#1}{}}{\lipsum}{\lipsum[#1]}}}

\RequirePackage{siunitx}
\newcolumntype{D}[1]{S[
	table-omit-exponent,
	round-mode=places,
	round-integer-to-decimal,
	round-precision={#1}]} 
\tikzset{
	transition/.style={
		rectangle,thick,draw=black!75,minimum size=2em,text width=3em,align=center
	},
	silenttransition/.style={
		rectangle,thick,fill=black,minimum height=2em
	},
	finalmarking/.style={
		fill=gray
	}
}

\usetikzlibrary{arrows.meta,decorations.markings}
\def\DTZU {2ex}
\tikzset{
	DECLARE.task/.style={
		rounded corners=1ex,
		minimum width=3em,
		minimum height=1ex,
		align=center,
		font=\sffamily,
		inner sep=5,
		draw
	},
	DECLARE.existcon/.style={
		label={[style=draw,yshift=-1\pgflinewidth]above:{\tiny \textit{#1}}}
	},
	DECLARE.neg/.style={
		semithick,
		postaction={decorate,decoration={markings,
				mark=at position .5 with {\arrow[xshift=0.15*\DTZU]{Bar[width=1.5*\DTZU]}},
				mark=at position .5 with {\arrow[xshift=-0.15*\DTZU]{Bar[width=1.5*\DTZU]}}
		}}
	},
	DECLARE.resex/.style={
		semithick,
		{Circle[length=1*\DTZU,width=1*\DTZU]}-,
		shorten <=-0.5*\DTZU
	},
	DECLARE.resp/.style={
		semithick,rounded corners,
		{Circle[length=1*\DTZU,width=1*\DTZU]}-{Triangle[length=1*\DTZU,width=1*\DTZU]},
		shorten <=-0.5*\DTZU
	},
	DECLARE.alt.resp/.style={
		semithick,rounded corners,
		{Circle[length=1*\DTZU,width=1*\DTZU]}-{Triangle[length=1*\DTZU,width=1*\DTZU]},
		shorten <=-0.5*\DTZU,
		double distance=0.25*\DTZU
	},
	DECLARE.chn.resp/.style={
		semithick,rounded corners,
		preaction={draw, shorten <= -0.5*\DTZU, double distance=0.25*\DTZU, {Circle[length=1*\DTZU,width=1*\DTZU]}-{Triangle[length=1*\DTZU,width=1*\DTZU]}},
		{-{Triangle[length=1*\DTZU,width=1*\DTZU]}}
	},
	DECLARE.prec/.style={
		semithick,rounded corners,
		{-{Triangle[length=1*\DTZU,width=1*\DTZU]Circle[length=1*\DTZU,width=1*\DTZU]}},
		shorten >=-0.5*\DTZU
	},
	DECLARE.alt.prec/.style={
		semithick,rounded corners,
		postaction={draw, shorten >= 0.5*\DTZU, double distance=0.25*\DTZU, {-{Triangle[length=1*\DTZU,width=1*\DTZU]}}},
		{-{Circle[length=1*\DTZU,width=1*\DTZU]}},
		shorten >= -0.5*\DTZU
	},
	DECLARE.chn.prec/.style={
		semithick,rounded corners,
		preaction={draw, shorten >= 0.5*\DTZU, double distance=0.25*\DTZU, {-{Triangle[length=1*\DTZU,width=1*\DTZU]}}},
		{-{Circle[length=1*\DTZU,width=1*\DTZU]}},
		shorten >= -0.5*\DTZU
	},
	DECLARE.succ/.style={
		semithick,rounded corners,
		{Circle[length=1*\DTZU,width=1*\DTZU]}-{Triangle[length=1*\DTZU,width=1*\DTZU]Circle[length=1*\DTZU,width=1*\DTZU]},
		shorten <=-0.5*\DTZU,
		shorten >=-0.5*\DTZU
	},
	DECLARE.coex/.style={
		semithick,rounded corners,
		{Circle[length=1*\DTZU,width=1*\DTZU]}-{Circle[length=1*\DTZU,width=1*\DTZU]},
		shorten <=-0.5*\DTZU,
		shorten >=-0.5*\DTZU
	},
	DECLARE.alt.succ/.style={
		semithick,rounded corners,
		postaction={draw, shorten >= 0.5*\DTZU, shorten <= -0.5*\DTZU, double distance=0.25*\DTZU, {Circle[length=1*\DTZU,width=1*\DTZU]}-{Triangle[length=1*\DTZU,width=1*\DTZU]}},
		{-{Circle[length=1*\DTZU,width=1*\DTZU]}},
		shorten >= -0.5*\DTZU
	},
	DECLARE.chn.succ/.style={
		semithick,rounded corners,
		preaction={draw, shorten >= 0.5*\DTZU, shorten <= -0.5*\DTZU, double distance=0.25*\DTZU, {Circle[length=1*\DTZU,width=1*\DTZU]}-{Triangle[length=1*\DTZU,width=1*\DTZU]}},
		{-{Circle[length=1*\DTZU,width=1*\DTZU]}},
		shorten >= -0.5*\DTZU
	}
}

\def\Pn {\ensuremath{\mathcal{P}\!\mathcal{N}}}

\def\WfN {\ensuremath{\mathcal{W}\!f\!\mathcal{N}}}
\newcommand{\Fires}[3]{\ensuremath{#1\xrightarrow[\Pn]{#2} #3}}
\def\Dp {\ensuremath{\mathcal{D}\!\mathcal{P}}}
\def\Mp {\ensuremath{\mathcal{M}\!\mathcal{P}}}
\RequirePackage{xparse}
\NewDocumentCommand{\MultiS}{m o}{%
  \IfValueTF{#2}{%
    \ensuremath{\mathfrak{M}#1\left|#2\right|}%
  }{
    \ensuremath{\mathfrak{M}#1}%
  }%
}
\def\LTL {\textsc{LTL}}

\def\LTLf {\ensuremath{\LTL_f}}

\newacronym{po}{PO}{Partial Order}
\newacronym{tl}{TL}{Temporal Logic}  
\newacronym{ltl}{\LTL}{Linear Temporal Logic}
\newacronym{ldl}{LDL}{Linear Dynamic Logic}
\newacronym{ldlf}{LDL$_f$}{Linear Dynamic Logic over Finite Traces}
\newacronym{fol}{FOL}{First Order Logic}
\newacronym{ltlf}{\LTLf}{Linear Temporal Logic on Finite Traces}
\newacronym{ltlp}{LTLp}{Linear-time Temporal Logic with Past}
\def\ltlpf {\ensuremath{\textrm{LTLp}_f}}
\newacronym{ltlpf}{\ltlpf}{Linear-time Temporal Logic with Past on Finite Traces}
\newacronym{mso}{MSO}{Monadic Second Order Logic}
\newacronym{rex}{RE}{regular expression}
\def\MultiSetFunctor {\ensuremath{\mathbb{M}}}
\newglossaryentry{multiset}{
	name={multi-set},description={a collection possibly containing multiple units of the same element},
	symbol={\MultiSetFunctor}}

\def\PowerSetFunctor {\ensuremath{\mathbb{P}}}
\newglossaryentry{powerset}{
	name={power-set},description={the collection of sets generated by all combinations without repetition of elements in a set},
	symbol={\PowerSetFunctor}}

\def\Autom {\ensuremath{\mathscr{A}}}
\def\Au {\Autom}
\newacronym[symbol=\Autom,longplural={finite state automata}]{fsa}{FSA}{finite state automaton}
\newacronym[symbol=\Autom,longplural={deterministic finite-state automata}]{dfa}{DFSA}{deterministic finite-state automaton}
\newacronym[symbol=\Autom,longplural={nondeterministic finite-state automata}]{nfa}{NFSA}{nondeterministic finite-state automaton}
\def\AutomInitState {\ensuremath{s_0}}

\newglossaryentry{fsainit}{name={initial state},description={initial state of the automaton},
	symbol=\AutomInitState}

\def\LanguageFunctor {\ensuremath{\mathscr{L}}}

\newcommand{\LanguageFunc}[1] {\ensuremath{\LanguageFunctor\!\left(#1\right)}}

\newcommand{\RegExp}[1] {\texttt{#1}}

\newcommand{\Step}[3]{\ensuremath{#1\xrightarrow[\Au]{#2} #3}}
\newcommand{\StepAu}[4]{\ensuremath{#1\xrightarrow[#2]{#3} #4}}

\newglossaryentry{satis}{name={satisfaction},description={evaluation as true of a formula on a structure}}
\newglossaryentry{verif}{name={verification},description={evaluation of a formula on a structure}}
\newcommand{\DclrSty}[1] {\textsc{#1}}
\def\Declare {\DclrSty{Declare}}
\newglossaryentry{declare}{%
	name={\Declare},description={a declarative process modelling language and notation}}
\def\DeclaModel {\ensuremath{\mathcal{M}}}
\newglossaryentry{declamodel}{%
	name={declarative \glsentrytext{promod}},description={\glsentrydesc{promod}, expressed by means of constraints},
	symbol={\DeclaModel}
}

\newglossaryentry{mindeclamodel}{%
	name={discovered \glsentrytext{declamodel}},description={\glsentrydesc{declamodel}, discovered from an \glsentrytext{evtlog}},
	symbol={\DeclaModel}
}
\newglossaryentry{minerful}{%
	name={\DclrSty{miner}ful},description={A declarative process discovery algorithm}}
\newacronym{mf}{Mf}{\gls{minerful}}
\newglossaryentry{minerfulVac}{%
	name={MINERful Vacuity Checker},description={\glsentrytext{minerful}} algorithm with semantical vacuity detection}
\newacronym{mfv}{Mf-Vchk}{\gls{minerfulVac}}
\newacronym{dmm}{DMM}{Declare Maps Miner}
\newglossaryentry{decmapmin}{%
	name={Declare Maps Miner},description={the declarative process discovery algorithm \glsentrytext{decmapmin}}}
\newacronym{dmm2}{DM2}{Declare Miner 2}
\newglossaryentry{decmapmin2}{%
	name={Declare Miner 2},description={improvement of \glsentrytext{decmapmin} algorithm}}
\newglossaryentry{janus}{%
	name={Janus},description={the declarative process discovery algorithm \glsentrytext{janus}}}
\def\Subsum {\ensuremath{\sqsubseteq}}

\newglossaryentry{subsum}{%
	name={subsumption},description={is subsumed by},%
	symbol={\Subsum}}

\newglossaryentry{relaxop}{%
	name={relaxation},description={relaxation operator, climbing the \glsentrytext{subsum} hierarchy}}

\newglossaryentry{actv}{%
	name={activation},description={the activation of a constraint}}
\newglossaryentry{activator}{name={activator},description={the event that signals the occurrence of the activation in the trace}}

\newglossaryentry{target}{%
	name={target},description={target}}

\def\Cns {\ensuremath{C}}
\newglossaryentry{con}{%
	name={constraint},description={a temporal business rule},
	symbol={\Cns}
}

\newglossaryentry{welldef}{%
	name={well-defined},description={of \glsentrytext{con}s for which a finite non-empty trace exists that complies with them}
}
\newglossaryentry{cnspar}{%
	name={parameter},description={a parameter of a \glsentrytext{con}},
}
\newglossaryentry{cnsarity}{%
	name={arity},description={number of parameters of a \glsentrytext{con}},
}
\newglossaryentry{exi}{
	name={existence},
	description={constrains single tasks}
}
\newglossaryentry{exicon}{
	name={\glsentrytext{exi} \glsentrytext{con}},
	description={constrains single tasks}
}
\newglossaryentry{posicon}{
	name={position \glsentrytext{con}},
	description={constrains the position of tasks}
}
\newglossaryentry{cardicon}{
	name={cardinality \glsentrytext{con}},
	description={limits the number of tasks}
}
\newglossaryentry{rela}{
	name={relation},
	description={constraint on pairs of tasks}
}
\newglossaryentry{relacon}{
	name={\glsentrytext{rela} \glsentrytext{con}},
	description={constraint on pairs of tasks}
}
\newglossaryentry{unirelacon}{
	name={unidirectional \glsentrytext{relacon}},
	description={constraint on pairs of tasks, out of which one is the activation, as the other is the target}
}
\newglossaryentry{unifwrelacon}{
	name={\glsentrytext{fw}-\glsentrytext{unirelacon}},
	description={constraint on pairs of tasks, having the first parameter as the activation, and the second one as the target}
}
\def\FwCns {\ensuremath{\mathit{fw}}}
\newglossaryentry{fw}{
	name={forward},
	description={forward constraint},
	symbol={\FwCns}
}

\newglossaryentry{unibwrelacon}{
	name={\glsentrytext{bw}-\glsentrytext{unirelacon}},
	description={constraint on pairs of tasks, having the second parameter as the activation, and the first one as the target}
}
\def\BwCns {\ensuremath{\mathit{bw}}}
\newglossaryentry{bw}{
	name={backward},
	description={backward constraint},
	symbol={\BwCns}
}

\newglossaryentry{corelacon}{
	name={coupling \glsentrytext{con}},
	description={constraint based on pairs of relation constraints}
}
\newglossaryentry{nega}{
	name={negative},
	description={of a constraint, that negates a coupling relation constraint}
}
\newglossaryentry{negacon}{
	name={\glsentrytext{nega} \glsentrytext{con}},
	description={constraint negating a coupling relation constraint}
}
\def\CnsTmp {\ensuremath{\mathcal{C}}}
\newglossaryentry{cnstemp}{%
	name={template},description={the template of a \glsentrydesc{con}},
	symbol={\CnsTmp}}
\def\CnsTmpPrm {\ensuremath{\CnsTemp'}}
\def\CnsTmpSec {\ensuremath{\CnsTemp''}}
\newcommand{\CnsTmpFunc}[2] {\ensuremath{\CnsTmp(#1\ifthenelse{\equal{#2}{}}{}{,#2})}}
\newcommand{\CnsTmpFuncPrm}[2] {\ensuremath{\CnsTmpPrm(#1\ifthenelse{\equal{#2}{}}{}{,#2})}}
\newcommand{\CnsTmpFuncSec}[2] {\ensuremath{\CnsTmpSec(#1\ifthenelse{\equal{#2}{}}{}{,#2})}}

\newglossaryentry{cnstype}{%
	name={type},description={the type of a \glsentrydesc{cnstemp}}}
\def\CnsRep {\ensuremath{\mathfrak{C}}}
\newglossaryentry{cnsrep}{name={repertoire},description={the repertoire of \glsentrytext{declare} \glsentrytext{temp}s},
	symbol={\CnsRep}}
\newglossaryentry{cnsuniv}{name={\glsentrytext{con}s universe},description={the set of \glsentrytext{declare} \glsentrytext{temp}s over the process alphabet reflected in the log}}
\def\CnsInstRelation {\ensuremath{\Gamma}}

\newglossaryentry{cnsinst}{%
	name={\glsentrytext{cnstemp} instantiation relation},description={the assignment relation instantiating \glsentrytext{cnstemp}s into \glsentrytext{con}s, namely assigning \glsentrytext{task}s to \glsentrytext{cnspar}s.},
	symbol={\CnsInstRelation}}

\newcommand{\CnsInterpFun} {\ensuremath{\mathscr{I}}}
\newglossaryentry{cnsinterp}{
	name={interpretation function},description={function interpreting a \glsentrytext{declamodel}},
	symbol={\CnsInterpFun}}

\def\RelaConTemp {\ensuremath{\mathcal{R}}}
\newglossaryentry{relacontemp}{%
	name={relation template},description={the template of a relation \glsentrydesc{con}},
	symbol={\RelaConTemp}}

\def\ExiConTemp {\ensuremath{\mathcal{E}}}
\newglossaryentry{exicontemp}{%
	name={existence template},description={the template of an existence \glsentrydesc{con}},
		symbol={\ExiConTemp}}

\def\Supp {\ensuremath{\sigma}}

\newglossaryentry{support}{%
	name={support},description={the support of a \glsentrydesc{con}},
	symbol={\Supp}}

\def\Conf {\ensuremath{\kappa}}
\newglossaryentry{conf}{%
	name={confidence},description={the confidence level of a \glsentrydesc{con}},
	symbol={\Conf}}

\def\IntF {\ensuremath{\iota}}
\newglossaryentry{intf}{%
	name={interest factor},description={the interest factor of a \glsentrydesc{con}},
	symbol={\IntF}}

\def\CnsEvalFunctor {\ensuremath{\eta}}
\newglossaryentry{cnseval}{
	name={evaluation},description={evaluation of a \glsentrytext{con} or a \glsentrytext{declamodel} over a \glsentrytext{evttrace} or an \glsentrytext{evtlog}},
	symbol={\CnsEvalFunctor}}

\def\ExiTxt {Existence}

\def\AbseTxt {Absence}
\def\UniqTxt {AtMostOne}
\def\ExacTxt {Exactly}
\def\ChoiceTxt {Choice}
\def\ExChoiceTxt {ExclusiveChoice}

\def\InitTxt {Init}
\def\EndTxt {End}
\def\ResExTxt {RespondedExistence}

\def\RespTxt {Response}
\def\AltRespTxt {AlternateResponse}

\def\ChaRespTxt {ChainResponse}
\def\PrecTxt {Precedence}
\def\AltPrecTxt {AlternatePrecedence}
\def\AltPrecTxtShort {Alt.Precedence}
\def\ChaPrecTxt {ChainPrecedence}
\def\CoExiTxt {CoExistence}
\def\SuccTxt {Succession}

\def\NotCoExiTxt {NotCoExistence}
\def\NotSuccTxt {NotSuccession}
\def\NotChaSuccTxt {NotChainSuccession}

\def\ExacTmp {\ensuremath{\textsc{\ExacTxt}}}

\def\ExChoiceTmp {\ensuremath{\textsc{\ExChoiceTxt}}}

\def\PrecTmp {\ensuremath{\DclrSty{\PrecTxt}}}

\def\NotSuccTmp {\ensuremath{\DclrSty{\NotSuccTxt}}}
\def\NotChaSuccTmp {\ensuremath{\DclrSty{\NotChaSuccTxt}}}
\newcommand{\Exi}[2] {\ensuremath{\DclrSty{\ExiTxt}(#1,#2)}}

\newcommand{\Abse}[2] {\ensuremath{\DclrSty{\AbseTxt}(#1,#2)}}
\newcommand{\Uniq}[1] {\ensuremath{\DclrSty{\UniqTxt}(#1)}}
\newcommand{\Exac}[2] {\ensuremath{\textsc{\ExacTxt}(#1,#2)}}
\newcommand{\Choice}[1] {\ensuremath{\textsc{\ChoiceTxt}(#1)}}
\newcommand{\ExChoice}[1] {\ensuremath{\textsc{\ExChoiceTxt}(#1)}}

\newcommand{\Ini}[1] {\ensuremath{\DclrSty{\InitTxt}(#1)}}
\newcommand{\End}[1] {\ensuremath{\DclrSty{\EndTxt}(#1)}}
\newcommand{\ResEx}[2] {\ensuremath{\DclrSty{\ResExTxt}(#1,#2)}}

\newcommand{\Resp}[2] {\ensuremath{\DclrSty{\RespTxt}(#1,#2)}}
\newcommand{\AltRes}[2] {\ensuremath{\DclrSty{\AltRespTxt}(#1,#2)}}

\newcommand{\ChaResp}[2] {\ensuremath{\DclrSty{\ChaRespTxt}(#1,#2)}}

\newcommand{\Prec}[2] {\ensuremath{{\DclrSty{\PrecTxt}}(#1,#2)}}
\newcommand{\AltPrec}[2] {\ensuremath{\DclrSty{\AltPrecTxt}(#1,#2)}}
\newcommand{\AltPrecShort}[2] {\ensuremath{\DclrSty{\AltPrecTxtShort}(#1,#2)}}
\newcommand{\ChaPrec}[2] {\ensuremath{\DclrSty{\ChaPrecTxt}(#1,#2)}}
\newcommand{\CoExi}[2] {\ensuremath{\DclrSty{\CoExiTxt}(#1,#2)}}
\newcommand{\Succ}[2] {\ensuremath{\DclrSty{\SuccTxt}(#1,#2)}}

\newcommand{\NotCoExi}[2] {\ensuremath{\DclrSty{\NotCoExiTxt}(#1,#2)}}
\newcommand{\NotSucc}[2] {\ensuremath{\DclrSty{\NotSuccTxt}(#1,#2)}}
\newcommand{\NotChaSucc}[2] {\ensuremath{\DclrSty{\NotChaSuccTxt}(#1,#2)}}

\newglossaryentry{fulfilment}{name={fulfilment},description={satisfaction of a constraint on a trace in which the activation occurs}}
\newacronym[\glslongpluralkey={Business Processes}]{bp}{BP}{Business Process}
\newacronym{bpi}{BPI}{Business Process Intelligence}
\newacronym{bpm}{BPM}{Business Process Management}
\newacronym{bpms}{BPMS}{Business Process Management System}
\newacronym{bpmn}{BPMN}{Business Process Model and Notation}
\newacronym{cpn}{CPN}{colored Petri net}
\newacronym{kpi}{KPI}{Key Performance Indicator}
\newacronym{ocbc}{OCBC}{Object-centric Behavioral Constraints}
\newacronym{soa}{SOA}{Service-Oriented Architecture}
\newacronym{pn}{PN}{Petri net}
\newacronym{wf}{WF}{workflow}
\newacronym{wfms}{WfMS}{Workflow Management System}
\newacronym{wfn}{WfN}{Workflow net}
\newacronym{xes}{XES}{eXtensible Event Stream}
\newacronym{yawl}{YAWL}{Yet Another Workflow Language}
\newglossaryentry{task}{%
	name={task},description={the non-divisible, elementary activity}}
\def\paramx {\ensuremath{x}}
\def\paramy {\ensuremath{y}}

\def\letterx {\ensuremath{a}}
\def\lettery {\ensuremath{b}}

\newcommand{\Task}[1] {\ensuremath{\scalebox{0.85}{\normalfont\textsf{#1}}}}

\def\taskc {\Task{c}}

\def\taski {\Task{i}}

\def\taskm {\Task{m}}
\def\taskn {\Task{n}}
\def\tasko {\Task{o}}
\def\taskp {\Task{p}}

\def\taskr {\Task{r}}
\def\tasks {\Task{s}}

\def\taskw {\Task{w}}
\def\taskx {\Task{x}}

\def\taskeu {\Task{\EUR}}
\newglossaryentry{promod}{%
	name={process model},description={the model of a process}
}
\def\LogAlph {\ensuremath{\Sigma}}
\newglossaryentry{logalph}{
	name={log alphabet},description={the process alphabet, as reflected in a log},%
	symbol={\LogAlph}}
\def\Evt {\ensuremath{e}}
\newglossaryentry{evt}{
	name={event},description={a record of an instantaneous fact during the process enactment},%
	symbol={\Evt}}
\def\Trc { \ensuremath{\varrho} }

\newglossaryentry{trace}{
	name={trace},description={a sequence of \glsplural{evt}},%
	symbol={\Trc}}
\def\EvtLog {\ensuremath{L}}

\newglossaryentry{evtlog}{
	name={event log},description={a collection of \glstext{evttrace}s},%
	symbol={\EvtLog}}
\journal{Information Systems}
\begin{document}

\begin{frontmatter}

\title{\thetitle
}
\author[address1]{Boudewijn~F.~van~Dongen}
\ead{b.f.v.dongen@tue.nl}

\author[address2]{Johannes~De~Smedt\corref{mycorrespondingauthor}}
\cortext[mycorrespondingauthor]{Corresponding author}
\ead{johannes.desmedt@kuleuven.be}

\author[address4]{Claudio~Di~Ciccio}
\ead{diciccio@di.uniroma1.it}

\author[address5]{Jan~Mendling}
\ead{jan.mendling@wu.ac.at}

\address[address1]{Eindhoven University of Technology, The Netherlands}
\address[address2]{KU Leuven, Belgium}
\address[address4]{Sapienza University of Rome, Italy}
\address[address5]{Vienna University of Economics and Business, Austria}

\begin{abstract}
Mixed-paradigm process models integrate strengths of procedural and declarative representations like Petri nets and {\Declare}. They are specifically interesting for process mining because they allow capturing complex behaviour in a compact way. A key research challenge for the proliferation of mixed-paradigm models for process mining is the lack of corresponding conformance checking techniques. In this paper, we address this problem by devising the first approach that works with intertwined state spaces of mixed-paradigm models. More specifically, our approach uses an alignment-based replay to explore the state space and compute trace fitness in a procedural way. In every state, the declarative constraints are separately updated, such that violations disable the corresponding activities. Our technique provides for an efficient replay towards an optimal alignment by respecting all orthogonal {\Declare} constraints. We have implemented our technique in ProM and demonstrate its performance in an evaluation with real-world event logs.
 \end{abstract}

\begin{keyword}
\thekeywords
\end{keyword}

\end{frontmatter}
%
%
%
%
%
%
\section{Introduction}
\label{sec:intro}
%
%
%

Alternative representations of models and computer programs have been investigated since the 1980s. Much of this research on computer programs is driven by a distinction between declarative (\emph{which rules} have to be considered) and procedural (\emph{which sequence of steps} have to be taken)~\cite{DBLP:journals/ijmms/GilmoreG84}. Empirical research on the mutual strengths and weaknesses of declarative and procedural programs has found that none of them is generally more effective, but that there are programming tasks that benefit more from one or the other~\cite{DBLP:journals/ijmms/GilmoreG84}. This finding has been further developed into the cognitive fit theory~\cite{DBLP:journals/isr/VesseyG91}: in essence, it states that cognitive effectiveness depends on the fit between representational paradigm and task. Much of these findings has been replicated for declarative and procedural process models~\cite{fahland2009declarative,Fahland2010477}. A separate conclusion from this research is the fact that some behavior can be compactly represented in a declarative way while it is complex as a procedure, and vice versa. Following the principle of minimum description length~\cite{barron1998minimum} inspired the idea to represent parts of the behavior in a declarative and parts in a procedural way.

The idea of mixing the declarative and procedural paradigms in single process models has been further developed into different formalizations of so-called mixed-paradigm process models \cite{westergaard2013mixing,de2015fusion,de2015mixed}, sometimes also referred to as hybrid models \cite{maggi2014dhybrid,DBLP:conf/otm/SlaatsSMR16}.
These models are proposed not only for modeling but also as a target language for process mining algorithms. Those algorithms have in common that the respective process models are composed of fragments constructed with procedural and declarative process mining techniques. The formalization of such models builds either on modular subprocesses with independent state spaces~\cite{maggi2014dhybrid} or intertwined state spaces \cite{westergaard2013cpn,de2015mixed}, in which the execution of activities changes the state in both model fragments simultaneously.

The formal definition of the internal mechanism of mixed-paradigm process models is a challenging task. The mentioned state-space representations have provided a foundation for mining and model checking of mixed-paradigm models~\cite{maggi2011monitoring,de2015fusion,desmedt2016modelchecking}.
The formalization of conformance checking remains an open yet critical issue for the further proliferation of mixed-paradigm approaches.
Due to the specifics of their state space, previous procedural and declarative approaches like~\cite{rozinat2008conformance,van2012replaying,de2014alignment} cannot be readily applied for mixed-paradigm conformance checking (MPCC).
Still, Conformance Checking for Mixed-Paradigm processes models is required for the following reasons. 
First, MPCC is required to compare conformance of mixed-paradigm models and classical declarative or procedural models. Second, MPCC is required for direct discovery algorithms. Third, MPCC allows for the assessment of the relative contribution of both paradigms to a mixed-paradigm model.

This paper presents the first approach addressing this research gap. We build on a procedural technique for the alignment-based replay over Petri nets~\cite{adriansyah2013memory}, which uses integer linear programming (ILP)-based state space traversals to find optimal alignments between logs and model.
The Petri net controls the state space exploration, while in every state the declarative constraints are separately updated, such that violations disable the corresponding activities. 
This technique provides an efficient replay for optimal alignments while respecting all orthogonal {\Declare} constraints to compute trace fitness.
Furthermore, it also allows for the replay of any declarative constraint-based process model defined in finite state machines, 
thus improving on \cite{de2014alignment} and ensuring compatibility with \cite{DBLP:conf/caise/DongenCCT17}.

The paper is structured as follows. \Cref{sec:background} presents the general motivation of using mixed-paradigm process models. \Cref{sec:preliminaries} describes formal preliminaries. \Cref{sec:approach} defines our conformance checking technique for mixed-paradigm models. \Cref{sec:evaluation} presents our evaluation using various real-world event logs. \Cref{sec:conclusion} concludes the paper.
\section{Background}
\label{sec:background}
%
%
%
This section discusses the background of our work. First, we explain the potential benefits of combining different modeling paradigms. Then, we discuss proposals of mixed-paradigm process models and present our running example. Finally, we identify challenges for the development of a conformance checking approach based on mixed-paradigm process models.

\subsection{Strengths and Weaknesses of Different Modeling Paradigms}
Computer programs and business process models have in common that they describe how problems can be solved. In essence, we can distinguish two paradigms used to define programs and process models: the procedural and the declarative. A \emph{procedural} representation builds on an explicit notion of state: we can identify at which position a program or process model is standing and what the next steps are that are available to proceed with. A \emph{declarative} representation hides an implicit notion of state: we cannot observe at which position a program or process model is standing, but when a next step is chosen, we can judge if it is permissible. 

Various languages of each paradigm have been defined and analyzed in prior research. Programming languages that are procedural include COBOL, Pascal and Java, while declarative programming languages include Prolog and SQL~\cite{van2009programming}. The debate over which of the two paradigms is superior was largely settled by cognitive experiments by Gilmore and Green who demonstrated that one paradigm was supportive for programming tasks where the other one was weak, and vice versa~\cite{DBLP:journals/ijmms/GilmoreG84}. Similar ideas found their way into research on business process models, which has been largely dominated by procedural languages such as Petri Nets \cite{murata1989petri}, Event-Driven Process Chains (EPCs) \cite{van1999formalization}, Yet Another Workflow Language (YAWL)~\cite{Aalst.Hofstede/InfSyst2005:YAWL} and Business Process Model and Notation (BPMN)~\cite{white2004introduction}. Languages like {\Declare}~\cite{pesic2008constraint}, EM-BrA$^2$CE \cite{goedertier2007bra2ce}, Dynamic Condition Response (DCR) Graphs~\cite{hildebrandt2011declarative}, Declarative Process Intermediate Language (DPIL)~\cite{Schonig2015125}, and Guard State Milestone (GSM)~\cite{hull2011introducing} were introduced exactly for the reason that it might be difficult to describe flexible behavior with procedural models, but potentially easy with declarative ones. Experiments comparing Declare with Petri nets largely confirmed the earlier findings of Gilmore and Green mutual strengths and weaknesses for process models~\cite{fahland2009declarative,Fahland2010477}.

The benefits of an effective representation are highly important for process mining and automatic process discovery. Mining algorithms generally struggle to strike a balance between different quality criteria including fitness, recall, simplicity and generalization~\cite{DBLP:books/sp/Aalst16}. Certain behavior observed in an event log might be easier to represent though a procedural model; other parts would be best specified through a declarative model. Arguably, a discovered process model might show the best balance between the quality criteria if the more flexible behavior is captured in a declarative way and the less adaptable behavior in a procedural way. Following up on this idea in other areas of mixing paradigms~\cite{DBLP:conf/coopis/Carlsen98,adams2006worklets,denti2005multi}, various proposals for mixed-paradigm process models have been made~\cite{westergaard2013mixing,de2015mixed,debois2015hybrid}, which we discuss in turn.

Different degrees of mixing can be distinguished for mixed-paradigm process models. First, \emph{modular mixing} approaches provide a mix by the help of sub-processes.
Following that approach, atomic sub-processes can be modeled in either paradigm without influencing each others state space \cite{pesic2007declare,slaats2016semantics}. Arguably, the modular approach does not truly integrate both paradigms in a single model.
Second, process mining using \emph{flexible mixing} approaches initially discover both procedural \cite{van2004workflow,leemans2013discovering} and declarative models \cite{lamma2007inducing,maggi2012efficient,DiCiccio2015} in order to then obtain a mixed-paradigm model. This approach is incorporated in Fusion Miner/FusionMINERful \cite{de2015fusion,desmedt2016modelchecking} and extracts mixed-paradigm models with intertwined state spaces, which we discuss next.

\subsection{Discovery and Representation of Mixed-Paradigm Process Models}
\label{sec:running:example}
\begin{figure}
	\centering
	\includegraphics[width=\textwidth]{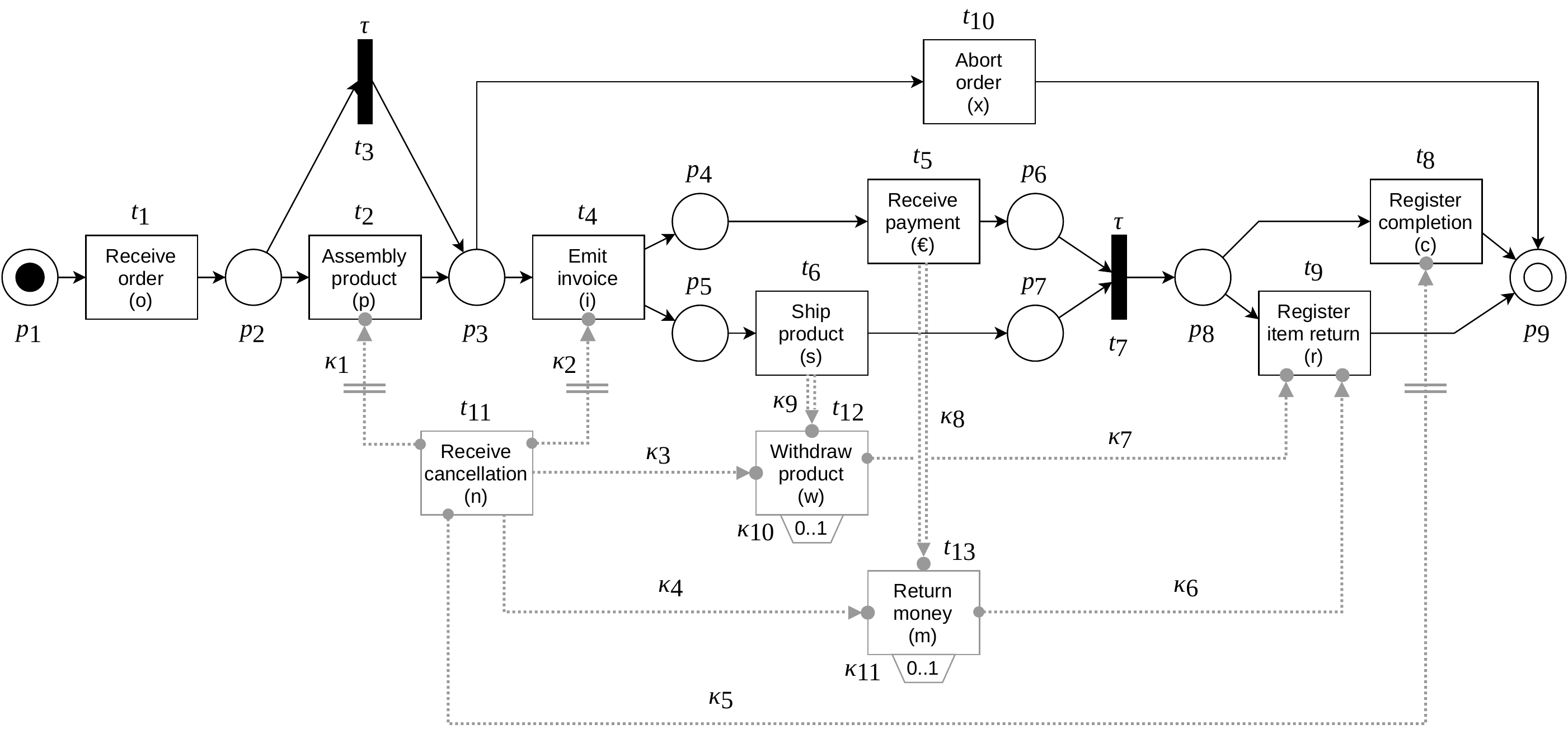}
	\caption{A mixed-paradigm process model.}
	\label{fig:running:example}
\end{figure}
%
%
%
The representation of mixed-paradigm models combines Petri nets with {\Declare} constraints. \Cref{fig:running:example} illustrates a mixed-paradigm process model that is inspired by the process described in \cite[Ch.~4]{Dumas.etal/2018:FundamentalsofBPM}.
The model represents a simplified version of an order-to-cash process. For the sake of readability, we depict the procedural fragment with solid black lines. The declarative constraints are drawn as dotted grey graphical elements. The procedural fragment, represented here as a Workflow Net~\cite{Aalst/JCSC1998:WorkflowNets}, describes the sequential flow of activities. The declarative constraints specify 
the effects of the receipt of a request for the cancellation of the order, which can occur at any time after the start of the instance.
The presence of activities that can take place at any stage of the process tend to clutter the procedural model. They typically require adding several re-routing elements to capture the diverse paths, also including reset and inhibitor arcs in Petri nets~\cite{DBLP:journals/jcss/VerbeekWAH10}, or cancellation events in YAWL~\cite{DBLP:journals/infsof/WynnVAHE09} and BPMN~\cite{chinosi2012}.
While those constructs are suitable to represent the run-time removal of otherwise possible actions, they are not as expressive as declarative constraints, which are naturally suitable to compactly describe such circumstances.
In the following, we describe the mixed-paradigm model focusing first on the procedural fragment and then on the declarative constraints.

A run of the sole procedural part representing the sunny-day scenario begins with the receipt of a new order, followed by the assemblage of the product, the emission of the invoice, the concurrent shipment and payment of the product, and the registering of the completion. Alternative runs of the procedural part exclude the assemblage of the product, lead to the abortion of the order prior to the emission of the invoice, or register that the item was returned.

\begin{sloppypar}
The declarative fragment of the process handles the cases in which a cancellation is requested and better clarify the conditions under which the aforementioned alternative runs may take place. The effect and the required compensation actions change according to the stage at which the cancellation occurs. If the cancellation request comes before the emission of the invoice, it inhibits the execution of the latter activity, thus forcing the run to proceed with the abortion of the order (the upper branch in the model): this is indicated by the
\NotSucc{\Task{Receive cancellation}}{\Task{Emit invoice}}
{\Declare} constraint, graphically depicted as
\begin{scriptsize}
\begin{tikzpicture}[baseline]
\node[DECLARE.task,anchor=base] (1) {Receive cancellation};
\node[DECLARE.task,right=of 1] (2) {Emit invoice};

\path (1) edge [DECLARE.succ,DECLARE.neg,densely dotted,gray] node {} (2);
\end{tikzpicture}%
\end{scriptsize}
in \cref{fig:running:example}.
Likewise, if \Task{Receive cancellation} occurs before \Task{Assembly product} or \Task{Register completion}, it disables those tasks (see the {\NotSuccTmp} constraints in the figure).
If the cancellation request comes after the product has been shipped and the payment received, the product has to be withdrawn and the payment returned. We represent this statement through a set of constraints:
\begin{inparaenum}[(1)]
	\item
	\Prec{\Task{Receive cancellation}}{\Task{Withdraw product}},
	graphically depicted as
	\begin{scriptsize}
		\begin{tikzpicture}[baseline]
		\node[DECLARE.task,anchor=base] (1) {Receive cancellation};
		\node[DECLARE.task,right=of 1] (2) {Withdraw product};
		
		\path (1) edge [DECLARE.prec,densely dotted,gray] node {} (2);
		\end{tikzpicture}%
	\end{scriptsize}%
	,
	imposes that the \Task{Withdraw product} activity can occur only after \Task{Receive cancellation};
	\item
	\AltPrec{\Task{Ship product}}{\Task{Withdraw product}},
	graphically depicted as
	\begin{scriptsize}
		\begin{tikzpicture}[baseline]
		\node[DECLARE.task,anchor=base] (1) {Ship product};
		\node[DECLARE.task,right=of 1] (2) {Withdraw product};
		
		\path (1) edge [DECLARE.alt.prec,densely dotted,gray] node {} (2);			\end{tikzpicture}%
	\end{scriptsize}%
	,
	states that the \Task{Withdraw product} activity can occur only after \Task{Ship product}, and that \Task{Withdraw product} cannot recur before another product is shipped (hence the \textit{alternation});
	\item
	We enforce that for every process instance only one item withdrawal is allowed by the \emph{\Uniq{\Task{Withdraw product}}} constraint (denoted with a \Task{0\ldots1} annotation beneath the activity box);
	by the same line of reasoning, we impose similar constraints on the execution of \Task{Return money}, i.e.:
	\item
	\AltPrec{\Task{Receive cancellation}}{\Task{Return money}};
	\item
	\Prec{\Task{Receive payment}}{\Task{Return money}};
	\item
	\Uniq{\Task{Return money}};
	\item
	finally, the pair of constraints
	\Succ{\Task{Withdraw product}}{\Task{Register item return}}
	and
	\item
	\Succ{\Task{Return money}}{\Task{Register item return}}, depicted as
	\begin{scriptsize}
		\begin{tikzpicture}[baseline]
		\node[DECLARE.task,anchor=base] (1) {Withdraw product};
		\node[DECLARE.task,right=of 1] (2) {Register item return};
		\node[DECLARE.task,right=of 2] (3) {Return money};
		
		\path (1) edge [DECLARE.succ,densely dotted,gray] node {} (2);
		\path (3) edge [DECLARE.succ,densely dotted,gray] node {} (2);
		\end{tikzpicture}%
	\end{scriptsize},
	together impose that if and only if the product was withdrawn and the money returned, then the item return is subsequently registered.
\end{inparaenum}
Notice that \NotSucc{\Task{Receive cancellation}}{\Task{Register completion}} disables the \Task{Register completion} task, thus making it mandatory to execute either \Task{Abort order} or \Task{Register item return}, depending on the status of the process instance when the cancellation request is received.
\end{sloppypar}
 
\subsection{Challenges of Mixed-Paradigm Conformance Checking}
Techniques for the discovery of mixed-paradigm models have so far relied on theoretic or small examples to illustrate the capabilities of mixed-paradigm models to capture an event log in a different way compared with single-paradigm models.
Therefore, it is currently not possible to quantify to what extent mixed-paradigm models are better capable of internalizing strict procedures mixed with loosely-occurring behavior 
in terms of model fitness.
Introducing an adequate conformance checking approach, preferably founded on the same principles as other conformance techniques to ensure comparability, is hence a strong motivation for proposing an alignment-based solution for mixed-paradigm models.
Furthermore, this also helps to fine-tune mixed-paradigm models.
The two main mixed-paradigm discovery algorithms \cite{maggi2014dhybrid,de2015fusion} both employ parameters to determine the ratio between either paradigm -- a requirement that often cannot be gauged straightforwardly.
Hence, mixed-paradigm conformance checking can help in finding this ratio in terms of identifying fitting models (in combination with model checking \cite{desmedt2016modelchecking}) as well as optimize the parameters.
Finally, alignment-based conformance checking can aid in understanding the contribution of each paradigm to the model.
By generating potential alignment issues, problems in model conformance can be pinpointed to either paradigm which can indicate that the ratio of each paradigm needs to be shifted.
In general, this will be linked with either procedural parts or declarative constraints that are too restrictive to allow for replay.

Conformance checking refers to techniques that determine to which degree an event log and a process model are consistent in terms of their behavior. This consistency can be measured by the help of some criteria. For procedural approaches, measures such as fitness, precision and generalization have been defined~ \cite{adriansyah2011conformance,vandenbroucke2013determining,DBLP:journals/is/WeidlichPDMW11} while declarative approaches mainly rely on support and confidence~\cite{DBLP:conf/bpm/CecconiCGM18}.

The challenge of integrating conformance checking for both paradigms root deeper than the definition of quality measures and is related to the execution semantics of intertwined state spaces.
The initial approach that was presented to obtain an intertwined state space was based on conjoining Petri nets and \gls{declare} constraints on the fly \cite{westergaard2013mixing}, which is tractable for models with a small amount of constraints.
Alternatively, conversions to the same execution language can be obtained. However, the conversion turns out to be often intractable \cite{prescherdeclarative}, or resorts to approaches that are too language-specific \cite{desmedt2015ri,DeGiacomo201584}. 
The latter has caused no previous work to apply existing techniques, either procedural or declarative, to mixed-paradigm models.

In this work, we update constraints in every step of the procedural model's state space, which is traversed by using the alignment-based conformance checking technique of \cite{DBLP:conf/caise/DongenCCT17}. That technique applies heuristics in the search space exploration resulting in an execution time that is linear with the trace size in many scenarios. 
\\
Next, we present the formal notions behind mixed-paradigm models and their fundamental constituents, namely Finite State Automata, Workflow nets and {\Declare}.
\section{Preliminaries}
\label{sec:preliminaries}
%
%
%

Mixed-paradigm models, as defined in \cite{westergaard2013mixing,de2015mixed}, are composed of both procedural and declarative process modelling fragments. These fragments can be defined over the same set of activities $A$ to obtain intertwined state spaces.
We instantiate mixed-paradigm models using Petri nets and \gls{declare} for the following reasons.
First, Petri nets are widely used for modelling and formal verification of business processes as much as for process mining.
In this work, we focus on a subset of Petri nets called Workflow nets \cite{van2004workflow}, which exhibit structural properties that are helpful for formal analysis. 
Second, \gls{declare} is the most prominent declarative process modelling approach and extensively used for process mining with declarative constructs.

In the following, the respective process languages are formalized and illustrated. We focus in particular on \acrfullpl{fsa} (\cref{sec:fsa}), \acrfullpl{wfn} (\cref{sec:wfn}), and declarative process models (\cref{sec:declarative}). 
The reader who is knowledgeable about those concepts may choose to skip this section.
\subsection{Finite State Automata}
\label{sec:fsa}
%
%
%
%
%
%
%
%
In order to introduce the execution semantics of declarative constraints which underpin mixed-paradigm models, we first introduce the concept of finite state automata, which serve as state-based representations of {\Declare} constraints. 
These machines allow to express the behavior of each constraint separately and, if needed, can be conjoined into a global automaton incorporating the full behavior of a constraint model.
\begin{definition}[\Acrfull{fsa}]\label{def:fsa}
A (deterministic) \acrfull{fsa} is a finite-state labeled transition system
$\Au = {(T,S,\delta,s_0,S_\textrm{F})}$, where:
\begin{inparadesc}
	\item[$T$]
	is a finite set of symbols; we shall refer to every such symbol $t \in T$ as \emph{transition};
	\item[$S$]
	is a finite non-empty set of states;
	\item[$\delta: S \times T \to S$] is the \emph{transition function}, i.e., a partial function that, given a starting state and a transition, returns the target state;
	\item[$s_0$] is the initial state;
	\item[$S_\textrm{F}\subseteq S$] is the set of final (accepting) states \cite{Chomsky.Miller/InformationandControl1958:FiniteStateLanguages}.
\end{inparadesc}
\end{definition}
Without loss of generality, we assume that $\delta$ is left-total and surjective on $S \setminus \{s_0\}$, that is, the transition function is defined for every pair of states and transitions, and every state is on a path from the initial one -- with the possible exception of the initial state itself. 

\begin{figure}[tb]%
	\centering%
	\begin{subfigure}{0.2\textwidth}%
		\centering%
		\resizebox{1.0\textwidth}{!}{%
\begin{tikzpicture}[->, >=stealth', shorten >=1pt, auto, bend angle=45, initial text = {}]
  \tikzstyle{every state}=[minimum size=1em]

  \node[state, initial, accepting]        (0) {$s_0$};
  \node[state, right=of 0, accepting]     (1) {$s_1$};
  \node[state, below=of 0,fill=gray] (2) {$s_2$};

  \path
  (0) edge [loop above] node [] {$t \in T \setminus\! \left\lbrace t_1,t_2 \right\rbrace$} (0)
  (0) edge [          ] node [] {$t_2$}                                                            (2)
  (2) edge [loop right] node [] {$t \in T$}                                                      (2)
  (0) edge [          ] node [below] {$t_1$}                                                       (1)
  (1) edge [loop below] node         {$t \in T $}                                                (1)
  ;  
\end{tikzpicture} %
		}%
		\caption{}
		\label{fig:constraint:automata:1}%
	\end{subfigure}%
	\hfill
	\begin{subfigure}{0.2\textwidth}%
		\centering%
		\resizebox{1.0\textwidth}{!}{%
\begin{tikzpicture}[->, >=stealth', shorten >=1pt, auto, bend angle=45, initial text = {}]
  \tikzstyle{every state}=[minimum size=1em]

  \node[state, initial, accepting]        (0) {$s_0$};
  \node[state, right=of 0, accepting]     (1) {$s_1$};
  \node[state, below=of 0,fill=gray] (2) {$s_2$};

  \path
  (0) edge [loop above] node []      {$t \in T \setminus\! \left\lbrace t_1,t_2 \right\rbrace$} (0)
  (0) edge [          ] node [below] {$t_1$}                                                            (1)
  (0) edge [          ] node [] {$t_2$}                                                                 (2)
  (2) edge [loop right] node [] {$t \in T$}                                                      (2)
  (1) edge [bend right] node [above] {$t_2$}                                                            (0)
  (1) edge [loop below] node         {$t \in T \setminus\! \left\lbrace t_2 \right\rbrace$}        (1)
  ;  
\end{tikzpicture} %
		}%
		\caption{}
		\label{fig:constraint:automata:2}%
	\end{subfigure}%
	\hfill%
	\begin{subfigure}{0.29\textwidth}%
		\centering%
		\resizebox{1.0\textwidth}{!}{%
\begin{tikzpicture}[->, >=stealth', shorten >=1pt, auto, bend angle=45, initial text = {}]
  \tikzstyle{every state}=[minimum size=1em]

  \node[state, initial, accepting]        (0) {$s_0$};
  \node[state, right=of 0]                (1) {$s_1$};
  \node[state, right=of 1, accepting]     (2) {$s_2$};
  \node[state, below=of 0,fill=gray] (3) {$s_3$};

  \path
  (0) edge [loop above] node []      {$t \in T \setminus\! \left\lbrace t_1,t_2 \right\rbrace$} (0)
  (0) edge [          ] node [above] {$t_1$}                                                            (1)
  (0) edge [          ] node [] {$t_2$}                                                                 (3)
  (3) edge [loop right] node [] {$t \in T$}                                                           (3)
  (1) edge [loop below] node         {$t \in T \setminus\! \left\lbrace t_2 \right\rbrace$}        (1)
  (1) edge [          ] node [below] {$t_2$}                                                            (2)
  (2) edge [bend right] node [above] {$t_1$}                                                            (1)
  (2) edge [loop above] node         {$t \in T \setminus\! \left\lbrace t_1 \right\rbrace$}        (2)
  ;  
\end{tikzpicture} %
		}%
		\caption{}%
		\label{fig:constraint:automata:3}%
	\end{subfigure}%
	\hfill
	\begin{subfigure}{0.2\textwidth}%
		\centering%
		\resizebox{1.0\textwidth}{!}{%
\begin{tikzpicture}[->, >=stealth', shorten >=1pt, auto, bend angle=45, initial text = {}]
  \tikzstyle{every state}=[minimum size=1em]

  \node[state, initial, accepting]        (0) {$s_0$};
  \node[state, right=of 0, accepting]     (1) {$s_1$};
  \node[state, below=of 0,fill=gray] (2) {$s_2$};

  \path
  (0) edge [loop above] node [] {$t \in T \setminus\! \left\lbrace t_1 \right\rbrace$}       (0)
  (0) edge [          ] node [below] {$t_1$}                                                      (1)
  (1) edge [loop below] node         {$t \in T \setminus\! \left\lbrace t_2 \right\rbrace $} (1)
  (1) edge [          ] node [above] {$t_2$}                                                      (2)
  (2) edge [loop right] node [] {$t \in T$}                                                     (2)
  ;  
\end{tikzpicture} %
		}%
		\caption{}%
		\label{fig:constraint:automata:4}%
	\end{subfigure}%
	\caption{Example \glspl{fsa}.}
	\label{fig:constraint:automata:fsa}
\end{figure}
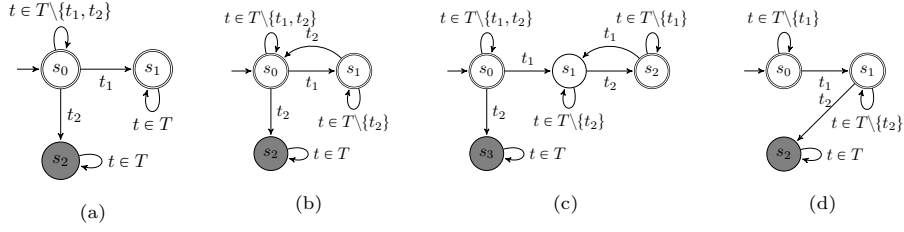

\Cref{fig:constraint:automata:fsa} depicts four \glspl{fsa}. States are represented as circles and transitions as arrows. Accepting states are decorated with a double line. The initial state is indicated with a single, unlabeled incoming arc.
For instance, \cref{fig:constraint:automata:3} is such that
$T \supseteq \{ t_1, t_2 \}$,
$S = \{ s_0, s_1, s_2, s_3 \}$,
$S_\textrm{F} = \{ s_0, s_2 \}$,
$\delta( s_0, t_1 ) = s_1$ and $\delta( s_1, t_2 ) = s_2$ among others.

\begin{definition}[Run of an \gls{fsa}]\label{def:au:run}
Let $\Au = {(T,S,\delta,s_0,S_\textrm{F})}$ be an \gls{fsa} as per \cref{def:fsa}.
A \emph{computation} $\pi$ of $\Au$ is a finite walk on states of $\Au$ starting from the initial state ($s_0$) through the consequent transitions, i.e., a sequence
${\pi = \left\langle \pi_1, \ldots, \pi_n \right\rangle}$
of length $n \in \mathbb{N}$
of tuples
${\pi_i = ( s_{i-1}, t_{i}, s_{i} ) \in \delta}$ with $1 \leqslant i \leqslant n$~\cite{Hopcroft.etal/2006:IntroductiontoAutomataTheoryLanguagesandComputation}.
We shall name the elements of $\pi$ as \emph{steps} and denote them as $\Step{s_{i-1}}{t_i}{s_i}$.
For the sake of conciseness, we shall denote a computation $\pi$ from $s_0$ to $s$ via finite string $t = t_1 \ldots t_n \in T^*$ as $\Step{s_0}{t}{s} = \Step{s_0}{t_1}{\Step{\ldots}{t_n}{s}}$.
A computation $\Step{s_0}{t}{s_\textrm{F}}$ terminating in a final state $s_\textrm{F} \in S_\textrm{F}$ is a \emph{run}.
\end{definition}
%
In \cref{fig:constraint:automata:3}, e.g.,
$ \pi_1 = \Step{s_0}{t_1}{\Step{s_1}{t_2}{\Step{s_2}{t_1}{s_1}}} $,
$ \pi_2 = \Step{s_0}{t_2}{\Step{s_3}{t_2}{s_3}} $,
and
$ \pi_3 = \Step{s_0}{t_1}{\Step{s_1}{t_2}{\Step{s_2}{t_2}{s_2}}} $
are computations,
but only $\pi_3$ is a run because $s_2 \in S_\textrm{F}$ whereas $s_1, s_3 \notin S_\textrm{F}$.
Notice that, in \cref{fig:constraint:automata:1,fig:constraint:automata:2,fig:constraint:automata:3,fig:constraint:automata:4}, we additionally highlight with a grey background colour those states that cannot be in a step of a run, that is, from which accepting states cannot be reached.

\begin{definition}[Language of an \gls{fsa}]\label{def:fsa:lang}
Let $\Au = {(T,S,\delta,s_0,S_\textrm{F})}$ be an \gls{fsa} as per \cref{def:fsa} where $s_\textrm{F} \in S_\textrm{F}$ and $t = t_1 \ldots t_n \in T^*$ be a finite string.
$t$ is \emph{accepted} by $\Au$ if a run
$\pi = \Step{s_0}{t}{s_\textrm{F}}$
exists.
The set of strings accepted by $\Au$ is the \emph{language} of $\Au$, $\LanguageFunc{\Au} \subseteq T^*$.
\end{definition}
For the automaton in \cref{fig:constraint:automata:3}, the language contains the string $\sigma_1=\langle t_1,t_2,t_2\rangle$ as a run exists over this sequence of transitions ($\pi_3$), whereas $\sigma_2=\langle t_2,t_2\rangle$ is not part of the language. 

In the following sections, we will use the \gls{fsa}s to express the behavior of declarative constraints, and use the same set of symbols (transitions) $T$ in the mixed-paradigm models to obtain a coherent alphabet between the declarative and procedural part of the mixed-paradigm models.

\subsection{Workflow nets}
\label{sec:wfn}
%
%
%
The procedural part of the mixed-paradigm models will be constructed out of Workflow nets.
In the following, we recall the formal definition of Petri nets and, thereupon, define Workflow nets.
\begin{figure}
	\centering
	\includegraphics[width=\textwidth]{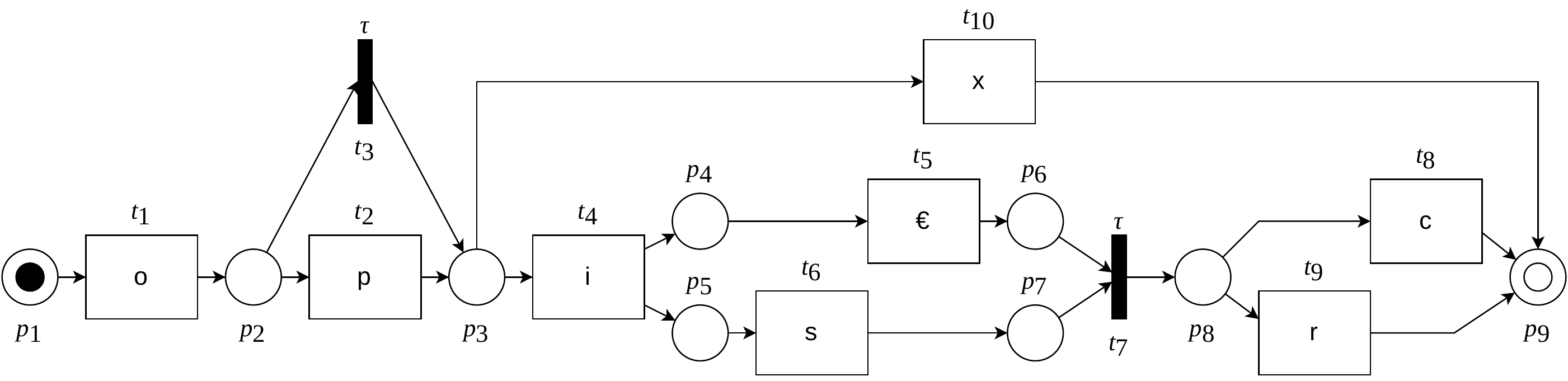}
	\caption[A Workflow net]{A Workflow net. The initial place is indicated by the token of the initial marking. The output place is decorated with an empty disc. Notice that the model in the picture corresponds to the procedural fragment of the model in \cref{fig:running:example}.}
	\label{fig:petrinet}
\end{figure}
\begin{definition}[Petri net]\label{def:pn}
	A \emph{place/transition (P/T) net}~\cite{DeselReisig/ACPN1998:PlaceTransitionPetriNets} is a triple $\mathcal{N} = ( P, T, F )$ such that
	\begin{iiilist}
		\item $P$ and $T$ are disjoint sets of \emph{places} and \emph{transitions}, respectively, and
		\item $F \subseteq (P \times T)\,\cup\,(T \times P)$ is the \emph{flow relation}.
	\end{iiilist}
	Let $\Sigma$ be an alphabet, and  $\tau \notin \Sigma$ be a special symbol denoting a \emph{silent} action.
	A labelled P/T net, hereinafter referred to as \emph{Petri net} for short, is a tuple
	$\Pn = ( \mathcal{N}, A,\, \ell ) = ( P, T, F, A, \ell )$
	where
	\begin{iiilist}
		\item $\mathcal{N}$ is a P/T net,
		\item $A = \Sigma \cup \{ \tau \}$, and
		\item $\ell$ is a mapping $\ell : T \to A$, henceforth also referred to as \emph{labeling} function.
	\end{iiilist}
\end{definition}
Graphically, places are represented as circles and transitions as boxes.
The flow relation depicts arcs connecting places to transitions and vice-versa.
In the model of \cref{fig:petrinet}, e.g.,
$P=\{p_1,\ldots,p_9\}$,
$T=\{t_1,\ldots,t_9\}$, and
$F \supseteq \{
\left( p_1,t_1 \right),
\left( p_2,t_2 \right),\\
\left( p_2,t_3 \right),
\left( t_4,p_4 \right),
\left( t_4,p_5 \right),
\left( t_7,p_9 \right),
\left( t_8,p_9 \right),
\left( t_9,p_9 \right)
\}$.
We shall name the elements of $P \cup T$ as \emph{nodes} when we do not need to distinguish places from transitions.
For every node $x \in P \cup T$, the preset is defined as 
$ \bullet x = \{y\,|\,(y,x)\in F\} $
and the postset as
$ x \bullet = \{y\,|\,(x,y)\in F\} $.
In \cref{fig:petrinet}, e.g.,
$p_1 \bullet = \{ t_1 \}$,
$t_4 \bullet = \{ p_4, p_5 \} $, and
$\bullet p_9 = \{ t_7, t_8, t_9 \}$.
%
%
Labels are typically assigned to transitions to denote tasks
~\cite{Aalst/JCSC1998:ApplicationofPetriNetsToWorkflowManagement}.
Transitions associated with $\ell$ to $\tau$ are named \emph{silent}, because of their indistinguishable and unobservable nature.
They are graphically depicted as narrow black rectangles.
In \cref{fig:petrinet}, e.g.,
$\ell(t_7)= \Task{c}$,
$\ell(t_8) = \Task{r}$, and
$\ell(t_3)=\ell(t_7) = \tau$ as $t_3$ and $t_7$ are silent transitions.
Taking inspiration by the notation of Mealy machines, we shall henceforth use the notation $t/a$ to denote the transition and its label at once whenever needed for the sake of readability (e.g., $t_7/\taskc$, $t_8/\Task{r}$, $t_3/\tau$, $t_7/\tau$).

The state of Petri nets is defined by the distribution of tokens over places, graphically depicted as solid black discs such as the one drawn in \cref{fig:petrinet} inside place $p_1$.
The state is thus represented by a \emph{marking}, namely a function $\mu: P\to\mathbb{N} $ 
that maps places to a number of tokens~\cite{Reisig/1985:PetriNetsAnIntroduction}.
A transition $t\in T$ is said to be ($\mu$-)enabled iff $\forall p\in\bullet t,\, \mu(p)>0$.
In the example of \cref{fig:petrinet}, $t_1$ is enabled, whereas $t_2,\ldots,t_9$ are not.
A $\mu$-enabled transition may yield a \emph{follower marking} $\mu'$ of $\mu$ such that for every $p \in P$:
\[
\mu'(p) =
\begin{cases}
\mu'(p) = \mu(p) - 1    & \text{if } p \in {\bullet}t \\
\mu'(p) = \mu(p) + 1    & \text{if } p \in t{\bullet} \\
\mu'(p) = \mu(p)        & \text{if } p \notin {\bullet}t \cup t{\bullet} \\
\end{cases}
\]
We say transition $t$ \emph{fires} from $\mu$ to $\mu'$ and write $\Fires{\mu}{t}{\mu'}$.
In the example of \cref{fig:petrinet}, the firing of $t_1$ leads to
$ \mu'= \{ \left( p_2, 1 \right) \} \cup \{ \left( p_i, 0 \right) | i \in \{1,3,4,\ldots,9\} \} $.
For the sake of conciseness, we shall use the multi-set notation for markings, indicating the number of tokens assigned by $\mu$ to $p$ as a multiplicity of $p$ if $\mu(p) > 0$: considering the previous example, we shall write $\mu' = \{p_2^1\}$.

A (finite) sequence of transitions
$\varrho = \langle t_1, {\ldots}, t_n \rangle \textrm{ where } t_i \in T \textrm{ for } 1 \leqslant i \leqslant n, n \in \mathbb{N}$
is called a (finite) \emph{firing sequence} \emph{enabled} at marking $\mu$
if there are markings $\mu_1, \ldots, \mu_n$
such that
$ \Fires{\Fires{\Fires{\mu}{t_1}{\mu_1}}{\cdots}{\cdots}}{t_n}{\mu_n} $~\cite{DeselReisig/ACPN1998:PlaceTransitionPetriNets}.
We shall also adopt
$ \Fires{\mu}{\varrho}{\mu_n} $
as a short-hand notation.
In the example of \cref{fig:petrinet},
$\langle t_2, t_4 \rangle$
is a firing sequence enabled at marking
$ \mu' $ defined above.
Indeed,
$ \Fires{\Fires{\mu'}{t_2}{\mu''}}{t_4}{\mu'''} $
where
$ \mu''= \{ p_3^1 \} $
and
$ \mu'''= \{ p_4^1, p_5^1 \} $.

%

To describe the expected run of a process modeled with a Petri net, we introduce the notions of \emph{initial marking} $\mu_0$ and \emph{final marking} $\mu_\textrm{F}$, recalling the notion of initial and final states of \glspl{fsa} seen in \cref{sec:fsa}.
This leads us to the notion of Workflow net, formally defined as follows. 
\begin{definition}[Workflow net]\label{def:wfn}
	A Workflow net $\WfN$~\cite{Hee.etal/ToPNOC2013:BusinessProcessModelingUsingPetriNets} is a Petri net such that
	\begin{iiilist}
		\item there exists one and only one input place $p_\textrm{i} \in P$ such that $\bullet p_\textrm{i} = \emptyset$,
		\item there exists one and only one output place $p_\textrm{o} \in P$ such that $p_\textrm{o} \bullet = \emptyset$,
		\item every node $x \in P \cup T$ is on a walk from $p_\textrm{i}$ to $p_\textrm{o}$,
		\item the initial marking $\mu_0$ is such that $\mu_0(p_\textrm{i}) = 1$ and $\mu_0(p) = 0$ for every $p \in P \setminus \{ p_\textrm{i} \}$,
		\item the final marking set is a singleton $M_\textrm{F} = \{ \mu_{\textrm{F}} \} $ where $\mu_{\textrm{F}}(p_\textrm{o}) = 1$ and $\mu_{\textrm{F}}(p') = 0$ for every $p' \in P \setminus \{ p_\textrm{o} \}$.
	\end{iiilist}
\end{definition}
The model depicted in \cref{fig:petrinet} is a Workflow net having $p_1$ as the input place and $p_9$ as the output place. Graphically, we depict tokens on the net according to the initial marking and draw the final marking with an empty disc.

A firing sequence of a Workflow net is 
enabled by the initial marking.
A firing sequence $\varrho$ is \emph{full} \cite{Aalst.etal/WIRev2012:ConformanceCheckingAndMetrics} when leading from the initial marking to the final marking of the Workflow net, i.e., 
$ \Fires{\mu_0}{\varrho}{\mu_F}$ with $\mu_F \in M_\textrm{F}$.
Some full firing sequences of the Workflow net in~\cref{fig:petrinet} are
$ \varrho_1 = \langle t_1, t_2, t_4, t_5, t_6, t_7, t_8 \rangle $, 
$ \varrho_2 = \langle t_1, t_2, t_4, t_6, t_5, t_7, t_9 \rangle $, 
and
$ \varrho_3 = \langle t_1, t_3, t_{10} \rangle $.
For brevity, we shall name full firing sequences as \emph{runs}, recalling the concept of run for \glspl{fsa}.
An \emph{observable} run $\rho$ is the sub-sequence of $\varrho$ obtained by applying an order-preserving endomorphism that removes silent transitions.
In \cref{fig:petrinet}, the observable runs stemming from $\varrho_1$, $\varrho_2$ and $\varrho_3$ are
$ \rho_1 = \langle t_1, t_2, t_4, t_5, t_6, t_8 \rangle $,
$ \rho_2 = \langle t_1, t_2, t_4, t_6, t_5, t_9 \rangle $ and
$ \rho_3 = \langle t_1, t_{10} \rangle $, respectively,
because the label of $t_3$ and $t_7$ is $\tau$.
%
To map observable sequences to non-silent symbols, we lift the notion of labeling to strings: $\lambda^* : T^* \to \Sigma^*$.
The application of $\lambda^*$ to observable runs generates strings commonly known as \emph{traces} in process mining.
In the example of \cref{fig:petrinet}, the traces derived from $\rho_1$, $\rho_2$ and $\rho_3$ are
$ \langle{\tasko,\taskp,\taski,\taskeu,\tasks,\taskc}\rangle $,
$ \langle{\tasko,\taskp,\taski,\tasks,\taskeu,\taskr}\rangle $ and
$ \langle{\tasko,\taskx}\rangle $ respectively.
A finite multi-set of traces is an \emph{event log} $L \subset \MultiS{\Sigma^*}$.%
\footnote{With $\MultiS{X}$ we denote the multi-set function $\MultiS{}: X \to \mathbb{N}$ mapping the elements of set $X$ to their multiplicity.}
An example event log for \cref{fig:petrinet} is $L = \{ \langle{\tasko,\taskp,\taski,\taskeu,\tasks,\taskc}\rangle^{10}, \langle{\tasko,\taskp,\taski,\tasks,\taskeu,\taskr}\rangle^5, \langle{\tasko,\taskx}\rangle^1 \}$.

\subsection{Declarative Process Models}
\label{sec:declarative}
%
%
%
\begin{figure}
	\centering
	\includegraphics[width=\textwidth]{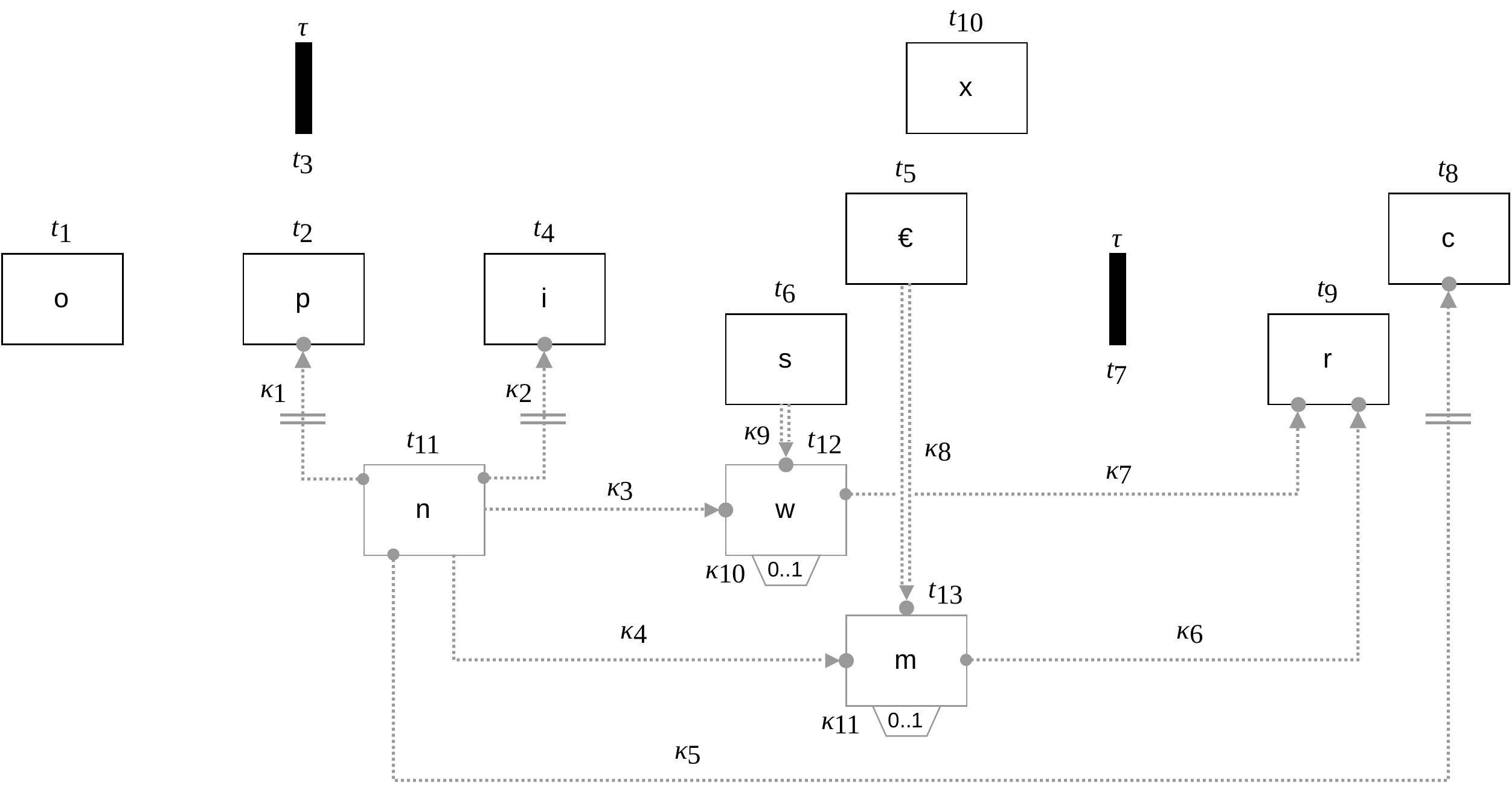}
	\caption[A declarative process model]{A declarative process model. Notice that the model in the picture corresponds to the declarative fragment of the model in \cref{fig:running:example}.}
	\label{fig:running:example:declarative}
\end{figure}

A declarative process model represents the behavior of a process by means of \emph{constraints}, i.e., rules that must not be violated during the execution of process instances.
Such rules are usually exerted over activities. 
Constraints express norms, best practices, and behavioral patterns that restrict the possible behavior.
Declarative processes are thus characterized by the fact that they define process behavior from the outside in: rather than prescribing the workflow of the process upfront, they loosely capture the possible execution scenarios to allow for maximum flexibility at run-time \cite{pesic2008constraint}.

Declarative process modeling languages come endowed with (rule) templates on which constraints are based.
Templates have a graphical representation and their semantics can be formalized using formal logics, making them verifiable and executable.
Each constraint inherits the graphical representation and semantics from its template.  The major benefit of using templates is that analysts do not have to be aware of the underlying logic-based formalization to understand the models. They work with the graphical representation of templates, while the underlying formulae remain hidden.
To date, \gls{declare} is one of the most well-established declarative process modelling languages.
It provides a standard library of templates (a.k.a.\ repertoire~\cite{Polyvyanyy.etal/FAOC2016:ExpressivePowerBehavioralProfiles,DBLP:journals/is/CiccioMMM18}), i.e., behavioral constraints parameterized over activities.
In our investigation, we will consider \gls{declare} but our results can be readily extended to other declarative approaches, such as 
DCR Graphs~\cite{Hildebrandt2011161}
or 
DPIL~\cite{Schonig2015125}. 

Taking inspiration from the formalization proposed in \cite{DBLP:journals/is/CiccioMMM17}, we formally define a declarative process model as follows.
\begin{definition}[Declarative process model]\label{def:dp}
    Let $\Sigma$ be an alphabet, and  $\tau \notin \Sigma$ a special symbol denoting a \emph{silent} action.
	A declarative process model is a tuple $\Dp=(R,T,A,\ell,K)$ where
	\begin{asparadesc}
		\item[$R$] is a repertoire of templates, i.e., predicates $\DclrSty{r}(x_1, \ldots, x_n) \in R$ on variables $x_1, \ldots, x_n$ (we say $n \in \mathbb{N}$ is the \emph{arity} of $\DclrSty{r}$),
		\item[$T$] is a finite non-empty set of transitions,
		\item[$A = \Sigma \cup \{ \tau \}$],
		\item[$\ell$] is a labelling function $\ell : T \to A$, and
		\item[$K \ni \kappa$] is a set of constraints, namely templates of arity $n$ whose variables are assigned by a mapping with labelled transitions $x_i \stackrel{\kappa}{\mapsfrom} t_i$ with $t_i \in T$, $1 \leqslant i \leqslant n$. 
		We shall compactly denote a constraint $\kappa \in K$ as $\DclrSty{r}(t_1, \ldots, t_n)$.
	\end{asparadesc}
\end{definition}
\begin{sloppypar}
\Cref{fig:running:example:declarative} shows a {\Declare} model that encompasses the declarative constraints of the mixed-paradigm model in \cref{fig:running:example}. 
For example, \begin{scriptsize}
	\begin{tikzpicture}[baseline]
	\node[DECLARE.task,anchor=base,label=left:$t_{11}$] (1) {$\Task{n}$};
	\node[DECLARE.task,right=of 1,label=right:$t_4$] (2) {$\Task{i}$};
	
	\path (1) edge [DECLARE.succ,DECLARE.neg,densely dotted,gray] node {} (2);
	\end{tikzpicture}%
\end{scriptsize}
uses the {\NotSuccTmp} template over activities \Task{Receive cancellation} ($t_{11}/\Task{n}$) and \Task{Emit invoice} ($t_{4}/\Task{i}$).
We recall that we use $t_{11}/\Task{n}$ as a compact notation to indicate that $t_{11}$ is labeled with \Task{n}.
Despite the constraints are exerted over transitions, we shall also use transition labels to denote the assigned parameters of a constraint whenever it eases the readability of examples: for instance, we shall denote $\NotSucc{t_{11}}{t_4}$ also as \NotSucc{\Task{Receive cancellation}}{\Task{Emit invoice}} or \NotSucc{\Task{n}}{\Task{i}} with the single-symbol abbreviations.
\end{sloppypar}

Different logic-based approaches have been used to define the semantics of the \gls{declare} templates.
In principle, Pesic~et~al.~\cite{pesic2007declare,Hofstede.etal/2010:YAWLBook} adopted \gls{ltl} \cite{Clarke.etal01:ModelCheckingBook}.
Their interpretation on finite traces with \gls{ltlf} has been later clarified by De~Giacomo~et~al.~\cite{DeGiacomo.etal/AAAI2014:ReasoningLTLFinite,DeGiacomo.Vardi/IJCAI2013:LinearTemporalLogic}.
In \cite{desmedt2015ri}, \gls{declare} constraints are translated into equivalent Petri nets with weighted, reset and inhibitor arcs.
In \cite{prescherdeclarative,desmedt2016modelchecking}, \glspl{rex} are used to define the semantics of \gls{declare} templates.
Since \glspl{rex} and \gls{mso} over finite traces have equivalent expressiveness \cite{DeGiacomo.Vardi/IJCAI2013:LinearTemporalLogic,Buechi/MathematicalLogicQuarterly1960:WeakSecondOrderFSA}, \glspl{rex} have a higher expressive power than \gls{ltlf} and, as such, are a suitable language to include the formulation of \gls{declare}. 
In the remainder of this paper, we will formalize \gls{declare} semantics as \glspl{rex}.
\begin{table*}[tbp]
  \resizebox{1.0\textwidth}{!}{%
      %
	\rowcolors{2}{gray!25}{white}
	\begin{tabular}{ l l p{10cm} }
		\hiderowcolors
		\toprule
		\textbf{Template} & \textbf{Regular Expression \cite{desmedt2016modelchecking,westergaardunconstrainedminer}}                  & \textbf{Description}                                                                                                                                                                                \\ \midrule
		\showrowcolors
		\Exi{\letterx}{n}            & \RegExp{.{*}(\letterx.{*})\{\textrm{n}\}}                                                                                                  & Activity {\letterx} happens at least $n$ times.                                                                                                                                                     \\
		\Abse{\letterx}{n}                             & \RegExp{{[}{\textasciicircum}\letterx{]}{*}(\letterx?{[}\^{ }\letterx{]}{*})\{\textrm{n}\}}                                                & Activity {\letterx} happens at most $n$ times.                                                                                                                                                      \\
		\Exac{\letterx}{n}                            & \RegExp{{[}\^{ }\letterx{]}{*}(\letterx{[}\^{ }\letterx{]}{*})\{\textrm{n}\}}                                                              & Activity {\letterx} happens exactly $n$ times.                                                                                                                                                      \\
		\Ini{\letterx}                               & \RegExp{(\letterx.{*})?}                                                                                                                   & Each instance has to start with activity \letterx.                                                                                                                                                  \\
		\End{\letterx}                                & \RegExp{.{*}\letterx}                                                                                                                      & Each instance has to end with activity \letterx.                                                                                                                                                    \\
		\ResEx{\letterx}{\lettery}                    & \RegExp{{[}\^{ }\letterx{]}{*}((\letterx.{*}\lettery.{*})$|$(\lettery.{*}\letterx.{*}))?}                                                  & If {\letterx} happens at least once then {\lettery} has to happen or happened before \letterx.                                                                                                      \\
		\CoExi{\letterx}{\lettery}                    & \RegExp{{[}\^{ }\letterx\lettery{]}{*}((\letterx.{*}\lettery.{*})$|$(\lettery.{*}\letterx.{*}))?}                                          & If {\letterx} happens then {\lettery} has to happen or happened after after {\letterx}, and vice versa.                                                                                             \\
		\Resp{\letterx}{\lettery}                     & \RegExp{{[}\^{ }\letterx{]}{*}(\letterx.{*}\lettery){*}{[}\^{ }\letterx{]}{*}}                                                             & Whenever activity {\letterx} happens, activity {\lettery} has to happen eventually afterward.                                                                                                       \\
		\Prec{\letterx}{\lettery}                     & \RegExp{{[}\^{ }\lettery{]}{*}(\letterx.{*}\lettery){*}{[}\^{ }\lettery{]}{*}}                                                             & Whenever activity {\lettery} happens, activity {\letterx} has to have happened before
		it.                                                                                                         \\
		\AltRes{\letterx}{\lettery}                   & \RegExp{{[}\^{ }\letterx{]}{*}(\letterx{[}\^{ }\letterx{]}{*}\lettery{[}\^{ }\letterx{]}{*}){*}}                                           & After each activity {\letterx}, at least one activity {\lettery} is executed. A following
		activity {\letterx} can be executed again only after the first occurrence of
		activity \lettery.       \\
		\AltPrec{\letterx}{\lettery}                  & \RegExp{{[}\^{ }\lettery{]}{*}(\letterx{[}\^{ }\lettery{]}{*}\lettery{[}\^{ }\lettery{]}{*}){*}}                                           & Before each activity {\lettery}, at least one activity {\letterx} is executed. A following
		activity {\lettery} can be executed again only after the first next occurrence
		of activity \letterx. \\
		\ChaResp{\letterx}{\lettery}                  & \RegExp{{[}\^{ }\letterx{]}{*}(\letterx\lettery{[}\^{ }\letterx{]}{*}){*}}                                                                 & Every time activity {\letterx} happens, it must be directly followed by activity
		{\lettery} (activity {\lettery} can also follow other activities).                                               \\
		\ChaPrec{\letterx}{\lettery}                  & \RegExp{{[}\^{ }\lettery{]}{*}(\letterx\lettery{[}\^{ }\lettery{]}{*}){*}}                                                                 & Every time activity {\lettery} happens, it must be directly preceded by activity
		{\letterx} (activity {\letterx} can also precede other activities).                                              \\
		\NotCoExi{\letterx}{\lettery}                 & \RegExp{{[}\^{ }\letterx\lettery{]}{*}((\letterx{[}\^{ }\lettery{]}{*})$|$(\lettery{[}\^{ }\letterx{]}{*}))?}                              & Either activity {\letterx} or {\lettery} can happen, but not both.                                                                                                                                  \\
		\NotSucc{\letterx}{\lettery}                  & \RegExp{{[}\^{ }\letterx{]}{*}(\letterx{[}\^{ }\lettery{]}{*}){*}}                                                                         & Activity {\letterx} cannot be followed by activity {\lettery}, and activity {\lettery} cannot
		be preceded by activity \letterx.                                                                   \\
		\NotChaSucc{\letterx}{\lettery}               & \RegExp{{[}\^{ }\letterx{]}{*}(\letterx+{[}\^{ }\letterx\lettery{]}{[}\^{ }\letterx{]}{*}){*}\letterx{*}}                                  & Activities {\letterx} and {\lettery} can never directly follow each other.                                                                                                                          \\
		\Choice{\letterx,\lettery}                     & \RegExp{.{*}{[}\letterx\lettery{]}.{*}}                                                                                                    & Activity {\letterx} or activity {\lettery} has to happen at least once, possibly both.                                                                                                              \\
		\ExChoice{\letterx,\lettery}           & \RegExp{([\^{ }\lettery]*\letterx[\^{ }\lettery]*)$|$.{*}{[}\letterx\lettery{]}.{*}({[}\^{ }\letterx{]}{*}\lettery{[}\^{ }\letterx{]}{*})} & Activity {\letterx} or activity {\lettery} has to happen at least once, but not both.                                                                                                               \\ \bottomrule
	\end{tabular}
 %
%
  }
  \caption{An overview of \gls{declare} constraint templates with their corresponding regular expressions, and verbose descriptions.}
  \label{tab:declare}
\end{table*}

An overview of the semantics tied to the most commonly used constraints in the \gls{declare} language is given in \cref{tab:declare}.
As {\Declare} constraints are expressible as regular expressions, their semantics are fully captured by \glspl{fsa}. This leads us to the following formal definition.

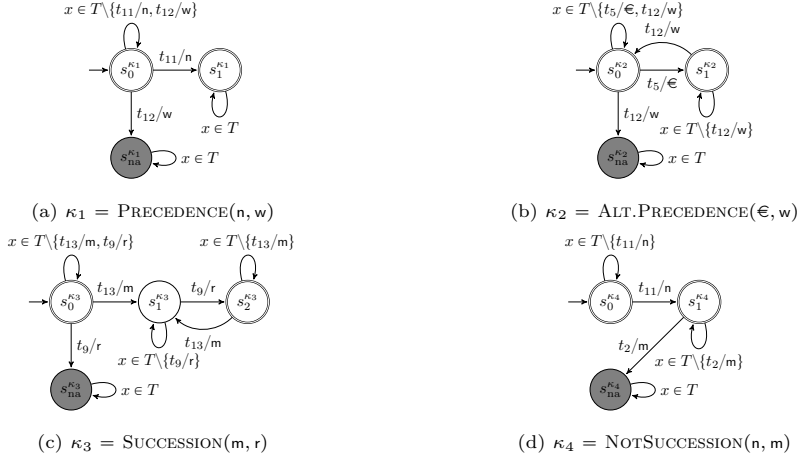
\begin{figure}[tbp]%
	\centering%
	\begin{subfigure}{0.45\textwidth}%
		\centering
		\resizebox{!}{0.125\textheight}{%
			%
\begin{tikzpicture}[->, >=stealth', shorten >=1pt, auto, bend angle=45, initial text = {}]
  \tikzstyle{every state}=[minimum size=1em]

  \node[state, initial, accepting]        (0) {$s^{\kappa_1}_0$};
  \node[state, right=of 0, accepting]     (1) {$s^{\kappa_1}_1$};
  \node[state, below=of 0,fill=gray] (2) {$s^{\kappa_1}_{\textrm{na}}$};

  \path
  (0) edge [loop above] node [] {$x \in T \setminus\! \left\lbrace t_{11}/\Task{n},t_{12}/\Task{w} \right\rbrace$}       (0)
  (0) edge [          ] node [above] {$t_{11}/\Task{n}$}                                                      (1)
  (1) edge [loop below] node         {$x \in T $} (1)
  (0) edge [          ] node [right] {$t_{12}/\Task{w}$}                                                      (2)
  (2) edge [loop right] node [] {$x \in T$}                                                     (2)
  ;  
\end{tikzpicture}
 		}%
		\caption{$\kappa_1 = $ \Prec{\Task{n}}{\Task{w}}}
		\label{fig:constraint:automaton:prec:n:w}
	\end{subfigure}%
	\hfill
	\begin{subfigure}{0.45\textwidth}%
		\centering
		\resizebox{!}{0.125\textheight}{%
			%
\begin{tikzpicture}[->, >=stealth', shorten >=1pt, auto, bend angle=45, initial text = {}]
  \tikzstyle{every state}=[minimum size=1em]

  \node[state, initial, accepting]        (0) {$s^{\kappa_2}_0$};
  \node[state, right=of 0, accepting]     (1) {$s^{\kappa_2}_1$};
  \node[state, below=of 0,fill=gray] (2) {$s^{\kappa_2}_{\textrm{na}}$};

  \path
  (0) edge [loop above] node []      {$x \in T \setminus\! \left\lbrace t_{5}/\Task{\EUR},t_{12}/\Task{w} \right\rbrace$} (0)
  (0) edge [          ] node [below] {$t_{5}/\Task{\EUR}$}                                                            (1)
  (0) edge [          ] node [] {$t_{12}/\Task{w}$}                                                                 (2)
  (2) edge [loop right] node [] {$x \in T$}                                                      (2)
  (1) edge [bend right] node [above] {$t_{12}/\Task{w}$}                                                            (0)
  (1) edge [loop below] node         {$x \in T \setminus\! \left\lbrace t_{12}/\Task{w} \right\rbrace$}        (1)
  ;  
\end{tikzpicture}		
         }%
		\caption{$\kappa_2 = $ \AltPrecShort{\Task{\EUR}}{\Task{w}}}
		\label{fig:constraint:automaton:altprec:eur:w}
	\end{subfigure}%
	\\
	\begin{subfigure}{0.45\textwidth}%
		\centering
		\resizebox{!}{0.125\textheight}{%
			%
\begin{tikzpicture}[->, >=stealth', shorten >=1pt, auto, bend angle=45, initial text = {}]
  \tikzstyle{every state}=[minimum size=1em]

  \node[state, initial, accepting]        (0) {$s^{\kappa_3}_0$};
  \node[state, right=of 0]                (1) {$s^{\kappa_3}_1$};
  \node[state, right=of 1, accepting]     (2) {$s^{\kappa_3}_2$};
  \node[state, below=of 0,fill=gray] (3) {$s^{\kappa_3}_{\textrm{na}}$};

  \path
  (0) edge [loop above] node []      {$x \in T \setminus\! \left\lbrace t_{13}/\Task{m},t_{9}/\Task{r} \right\rbrace$} (0)
  (0) edge [          ] node [above] {$t_{13}/\Task{m}$}                                                            (1)
  (0) edge [          ] node [] {$t_{9}/\Task{r}$}                                                                 (3)
  (3) edge [loop right] node [] {$x \in T$}                                                           (3)
  (1) edge [loop below] node         {$x \in T \setminus\! \left\lbrace t_{9}/\Task{r} \right\rbrace$}        (1)
  (1) edge [          ] node [above] {$t_{9}/\Task{r}$}                                                            (2)
  (2) edge [bend left] node [below] {$t_{13}/\Task{m}$}                                                            (1)
  (2) edge [loop above] node         {$x \in T \setminus\! \left\lbrace t_{13}/\Task{m} \right\rbrace$}        (2)
  ;  
\end{tikzpicture}		
         }%
		\caption{$\kappa_3 = $ \Succ{\Task{m}}{\Task{r}}}
		\label{fig:constraint:automaton:succ:m:r}
	\end{subfigure}%
	\hfill
	\begin{subfigure}{0.45\textwidth}%
		\centering
		\resizebox{!}{0.125\textheight}{%
			%
\begin{tikzpicture}[->, >=stealth', shorten >=1pt, auto, bend angle=45, initial text = {}]
  \tikzstyle{every state}=[minimum size=1em]

  \node[state, initial, accepting]        (0) {$s^{\kappa_4}_0$};
  \node[state, right=of 0, accepting]     (1) {$s^{\kappa_4}_1$};
  \node[state, below=of 0,fill=gray] (2) {$s^{\kappa_4}_{\textrm{na}}$};

  \path
  (0) edge [loop above] node [] {$x \in T \setminus\! \left\lbrace t_{11}/\Task{n} \right\rbrace$}       (0)
  (0) edge [          ] node [above] {$t_{11}/\Task{n}$}                                                      (1)
  (1) edge [loop below] node         {$x \in T \setminus\! \left\lbrace t_{2}/\Task{m} \right\rbrace $} (1)
  (1) edge [          ] node [left ] {$t_{2}/\Task{m}$}                                                      (2)
  (2) edge [loop right] node [] {$x \in T$}                                                     (2)
  ;  
\end{tikzpicture}
 		}%
		\caption{$\kappa_4 = $ \NotSucc{\Task{n}}{\Task{m}}}
		\label{fig:constraint:automaton:notsucc:n:m}
	\end{subfigure}%
	\caption[Constraint automata]{Constraint automata derived from some constraints of the model in \cref{fig:running:example}.}\label{fig:constraint:automata}
\end{figure}

\begin{definition}[Constraint automaton]\label{def:con:aut}
	\label{def:ca}
	Let $\kappa = \DclrSty{r}(t_1, \ldots, t_n) \in K$ be a constraint of a declarative process model $\Dp = (R,T,A,\ell,K)$ as per \cref{def:dp}. The \emph{constraint automaton} of $\kappa$ is a tuple $\Au\!(\kappa) = (T,S,\delta,s_0,S_\textrm{F},A,\ell)$ wherein 
	\begin{iiilist}
	  \item $T$, $S$, $\delta$, $s_0$ and $S_\textrm{F}$ are defined as for standard \glspl{fsa} (\cref{def:fsa}),
	  \item $A$ and $\ell$ are the labels set and labeling function of $\Dp$, respectively,%
	\end{iiilist}
	such that $\Au\!(\kappa)$ accepts all and only those traces $\sigma \in \Sigma^*$ that satisfy $\kappa$.
\end{definition}
Examples of algorithms that produce the automaton of a given constraint can be found in \cite{DBLP:journals/is/CiccioMMM18,westergaardunconstrainedminer}.
\Cref{fig:constraint:automata} illustrates some constraint automata for constraints that are part of the mixed-paradigm model in \cref{fig:running:example}. For the sake of readability, we decorate states with the constraint they refer to as a suffix (e.g., $s^{\kappa_1}_0$ in \cref{fig:constraint:automaton:prec:n:w}).

Inspired by the approach in \cite{DBLP:journals/is/CiccioMMM18}, we classify every state $s^\kappa$ of a constraint automaton based on the truth value of the corresponding constraint in that state:
\begin{compactdesc}
    \item[\bf Permanent violation:] No computation that traverses state $s^\kappa$ can be a prefix of a run -- thus, no final state is reachable; we mark those states with the $_\textrm{na}$ subscript, as can be noticed in \cref{fig:constraint:automaton:prec:n:w,fig:constraint:automaton:altprec:eur:w,fig:constraint:automaton:succ:m:r,fig:constraint:automaton:notsucc:n:m};
    \item[\bf Temporary violation:] The computations that traverse state $s^\kappa$ can be a prefix of a run but $s^\kappa \notin S_\textrm{F}$, i.e., $s^\kappa$ is not a final state; this is the case, for instance, of $s^{\kappa_3}_1$ in \cref{fig:constraint:automaton:succ:m:r};
    \item[\bf Temporary satisfaction:] The computations ending in $s^\kappa$ are runs ($s^\kappa \in S_\textrm{F}$) but can also be prefixes of computations ending in $s^\kappa_\textrm{na}$; this is the case, e.g., of $s^{\kappa_2}_1$ in
    \cref{fig:constraint:automaton:altprec:eur:w};
    \item[\bf Permanent satisfaction:] All computations that traverse state $s^\kappa$ are runs – thus, no other states than final ones are reachable from $s^\kappa$; this is the case, e.g., of $s^{\kappa_1}_1$ in
    \cref{fig:constraint:automaton:prec:n:w}.
\end{compactdesc}
Our framework checks for the truth value 
in the state of every constraint (automaton) upon the replay of a trace, as it allows us to (1) single out the constraints being violated by a trace and (2) modularize the computation of alignments between the trace and the mixed-paradigm model without resorting on a cross-product of all constraint automata, which is known to be a barely tractable operation~\cite{DBLP:journals/is/CiccioMMM17}.
%
%
  
\section{Approach}
\label{sec:approach}
%
%
%
In this section, we describe the approach to align mixed-paradigm models with event logs, together with examples that illustrate the findings.
We begin with the formalization and illustration of mixed-paradigm models and their alignments with the traces of an event log.
Thereupon, based on existing techniques for aligning traces with procedural models, two new techniques are proposed to deal with mixed-paradigm models. First, we show how we can compute alignments while guaranteeing that all constraints are satisfied at the end of the trace. Second, we show how to allow for the violation of constraints if that leads to a better explanation of the trace.
%
%
%
\subsection{Mixed-Paradigm Models}
\label{sec:MPPM}
%
%
%
Based on the above definitions, we can now introduce the notion of mixed-paradigm models.
\begin{figure}[tb]%
	\centering%
	\includegraphics[width=\textwidth]{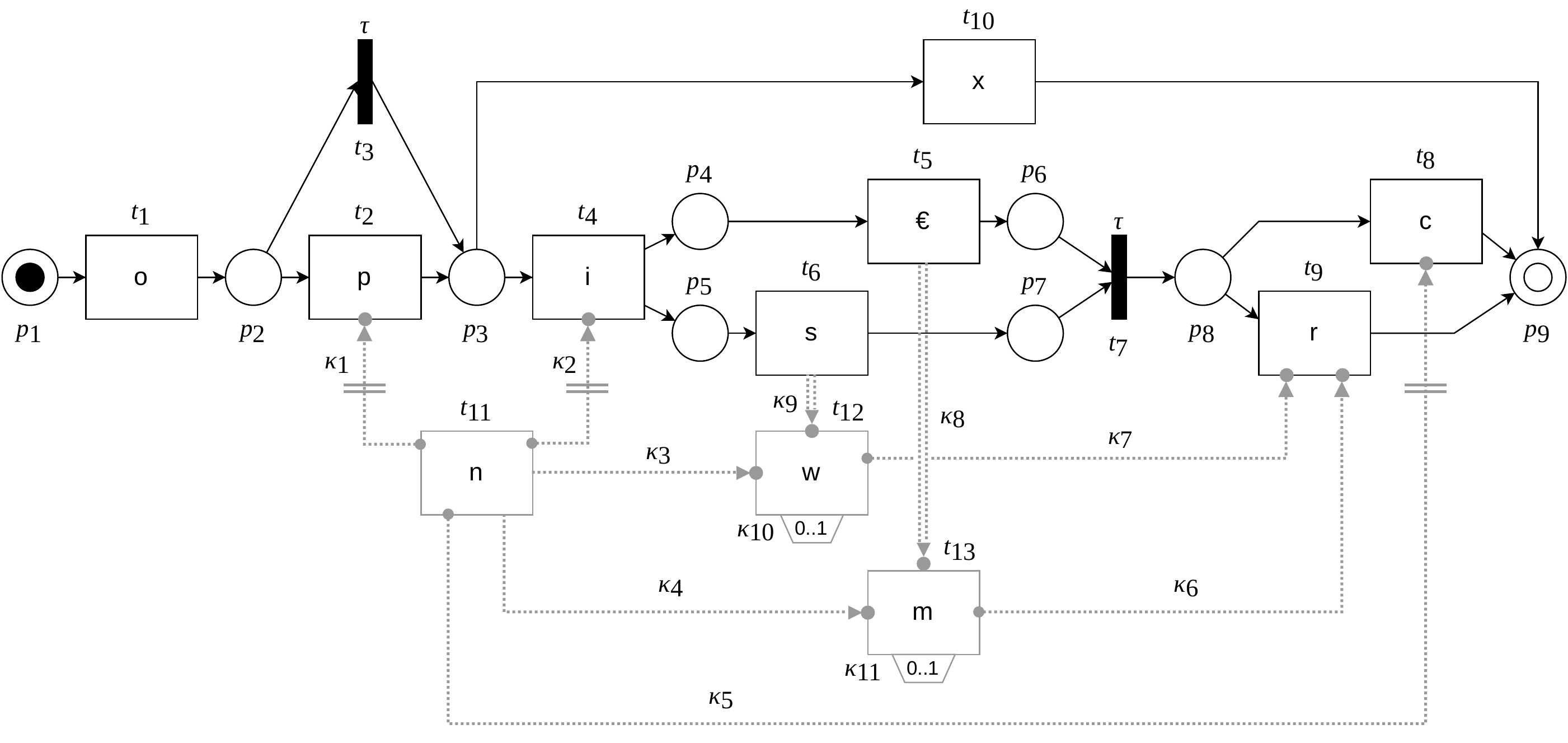}
	\caption{A mixed-paradigm model.}
	\label{fig:mpexample}
\end{figure}
\Cref{fig:mpexample} depicts a mixed-paradigm process model (notice that its structure is the same as that of \cref{fig:running:example} with one-letter abbreviations for the activities). A mixed-paradigm process model consists of
\begin{iiilist}
  \item a Workflow net and
  \item a finite set of \gls{declare} constraints.
\end{iiilist}
Other mixed-paradigm solutions exist as well, such as 
BPMN with \gls{declare} (BPMN-D)~\cite{DeGiacomo201584}. 
Furthermore, conversions between both types of models have been introduced, e.g., from \gls{declare} to Petri nets~\cite{prescherdeclarative,desmedt2015ri}.
Nevertheless, both approaches can be converted to models that are compatible with the proposed conformance checking technique.
Formally, we define a mixed-paradigm model as follows.
\begin{definition}[Mixed-paradigm model]\label{def:mp}
	A mixed-paradigm model is a tuple $\Mp = (\WfN, \Dp)$, where 
	\begin{asparadesc}
		\item[$\WfN$]$= ( P, T, F, A, \ell, \mu_0, M_\textrm{F} )$ is a Workflow net (as per \cref{def:wfn}),
		\item[$\Dp$]$= ( R, T, A, \ell, K )$ is a declarative process model (as per \cref{def:dp}),
	\end{asparadesc}
	such that {\WfN} and {\Dp} share the same transitions set ($T$), labels ($A = \Sigma \cup \tau$) and labeling function ($\ell$).
\end{definition}
We assume that Workflow nets and \gls{declare} models are defined over the same set of activities so that the models' state spaces are synchronized during the execution.
The separation of the mixed-paradigm model in its procedural and declarative parts, as well as the fact that the two fragments of the model share the same transitions and respective labels, can be observed by comparing \cref{fig:mpexample} with \cref{fig:petrinet} and \cref{fig:running:example:declarative}.
Without loss of generality, we take {\Declare} as the repertoire of $\Dp$.

A workflow run (or simply \emph{run}, for short) of a mixed-paradigm model is a sequence of transitions that is a run to the Workflow net (i.e., it leads from its initial marking to the final one).
If the run ends in a configuration that satisfies every constraint, it is a \emph{full run} of the mixed-paradigm model.
We formalise these notions as follows.
\begin{sloppypar}
\begin{definition}[Run of a mixed-paradigm model]\label{def:mp:run}
Let $\Mp = (\WfN, \Dp)$ be a mixed-paradigm model composed of a Workflow net $\WfN = ( P, T, F, A, \ell, \mu_0, M_\textrm{F} )$ and a declarative process model $\Dp = ( R, T, A, \ell, K ) $ as per \cref{def:mp}.
Let $m = |K| \in \mathbb{N}$ be the number of constraints in $\Dp$
and
$\Au\!(\kappa) = (T,S,\delta,s_0,S_\textrm{F},A,\ell)$
be the constraint automaton of $\kappa \in K$ as per \cref{def:ca}.
A \emph{computation} $\Pi$ of $\Mp$ is a finite sequence of steps
$\left\langle \Pi_1,\ldots,\Pi_n\right\rangle$
of length $n \in \mathbb{N}$
of tuples
${ \Pi_i = \left(
    \Fires{\mu_{i-1}}{t_i}{\mu_i} ,
    \quad
    \StepAu{s^{\kappa_1}_{i-1}}{\Au\!(\kappa_1)}{t_i}{s^{\kappa_1}_i},
    \ldots, 
    \StepAu{s^{\kappa_m}_{i-1}}{\Au\!(\kappa_m)}{t_i}{s^{\kappa_m}_i}
    \right)
}$
for $ 1 \leqslant i \leqslant n$,
starting at the initial marking $\mu_0$ of $\WfN$ and the initial states $s^{\kappa_1}_0,\ldots,s^{\kappa_m}_0$ of all constraint automata $\Au\!(\kappa_1),\ldots,\Au\!(\kappa_m)$ of the constraints in $\Dp$.
A computation $\Pi$ of length $n$ having as its last step
${ \Pi_n = \left(
    \Fires{\mu_{n-1}}{t_n}{\mu_n} ,
    \quad
    \StepAu{s^{\kappa_1}_{n-1}}{\Au\!(\kappa_1)}{t_n}{s^{\kappa_1}_n},
    \ldots, 
    \StepAu{s^{\kappa_m}_{n-1}}{\Au\!(\kappa_m)}{t_n}{s^{\kappa_m}_n}
    \right)
}$,
such that
$\mu_n \in M_\textrm{F}$
is the final marking of $\WfN$,
is a \emph{workflow run} (or simply \emph{run}, for short) of \Mp.
Let $s^K_i$ be the tuple $(s^{\kappa_1}_{i},\ldots,s^{\kappa_m}_i)$ of states of the constraint automata $\Au\!(\kappa_1), \ldots, \Au\!(\kappa_m)$ at the $i$-th step,
$S^K_\textrm{F}$ be the union of the sets of accepting states of $\Au\!(\kappa_1), \ldots, \Au\!(\kappa_m)$, and
$S^K_\textrm{0}$ the set consisting of their initial states.
If $\Pi$ is a run and
$s^{\kappa_1}_n, \ldots, s^{\kappa_m}_n$
are final states of $\Au\!(\kappa_1), \ldots, \Au\!(\kappa_m)$, i.e., $s^K_n \in S^K_\textrm{F}$, then $\Pi$ is a \emph{full run} of \Mp.
\end{definition}
\end{sloppypar}
In essence, a step of a mixed-paradigm model is an ensemble of firings in the procedural fragment and of a step in each of the constraint automata of its declarative fragment.
Considering the example model of \cref{fig:mpexample}, by replaying
$\left\langle t_1, t_2, t_{11}, t_{10} \right\rangle$
and
$\left\langle t_1, t_2, t_4, t_{11}, t_6, t_5, t_{13}, t_{12}, t_7, t_9 \right\rangle$
we attain two full runs.
By replaying
$\left\langle t_1, t_2, t_4, t_{11}, t_6, t_5, t_7, t_{12}, t_{13} \right\rangle$
we have a computation: notice that the constraint automata of $\Succ{t_{12}}{t_9}$ and $\Succ{t_{13}}{t_9}$ are not in a final state (both are in a state of temporary violation, waiting for $t_9$ to occur), and marking
$ \{ p_8^1 \} $
is not final.
Finally,
$\left\langle t_1, t_2, t_4, t_{11}, t_6, t_5, t_{13}, t_{12}, t_7, t_8 \right\rangle$
yields a workflow run but not a full run because, although the final marking is reached, the constraint automata of
$\Succ{t_{12}}{t_9}$ and
$\Succ{t_{13}}{t_9}$
are not in a final state (temporary violation)
and the constraint automaton of
$\NotSucc{t_{11}}{t_8}$
is in a state of permanent violation.

Notice that the state space exploration of these models is not trivial, as in a next state the Workflow net might still be capable of reaching a final marking, but might not be able to reach an accepting state for all the constraints' automata any longer.
Although we can use constraints to reduce the state space of the Workflow net, in every state we need to check which transitions can still be fired in order to guarantee that the constraints can be brought to an accepting state in any future state.
Consider for example the firing sequence $\varrho_1=\langle t_1,t_2,t_4,t_{11},t_{12},t_5,t_6,t_7\rangle$ over \cref{fig:mpexample}.
It leads to a marking with a single token in $p_8$. However, $t_8/\taskc$ is not enabled to fire due to the \NotSucc{\Task{n}}{\Task{c}} constraint and the only possible next firing is that of $t_9/\Task{r}$.

An initial approach to calculate the future satisfiability of constraints on-the-fly was presented in \cite{westergaard2013mixing}. However, this approach might prove intractable in case of larger sets of constraints and is not needed for conformance purposes.
By using alignments we avoid a full state space exploration by iteratively polling queues of future states as will be discussed in \cref{subsec:satisfied,subsec:violating}, where we also explore the impact of leaving constraints to be violated (at a cost).

We recall that, as this paper focuses on mixed-paradigm models consisting of (\gls{declare}) constraint automata and Petri nets, other mixed-paradigm approaches can readily apply these insights.
Indeed, many procedural models can be converted into Petri nets (e.g., EPCs, YAWL, and BPMN). 
Secondly, the use of FSAs allows the use of an array of other declarative constraints.
Aside from \gls{declare}, e.g., DCR Graphs can be converted into 
automata~\cite{hildebrandt2011declarative}.
Also, notice that the translation from \gls{declare} constraints to BPMN-D fragments~\cite{DeGiacomo201584} is based on the production of FSAs as an intermediate step.

\subsection{Alignments}
\label{sec:alignments}
%
%
%
Conformance checking has been tackled in many ways \cite{vandenbroucke2013cobefra}, using concepts such as token replay \cite{rozinat2008conformance}, behavioral profiles \cite{DBLP:journals/is/WeidlichPDMW11}, alignments \cite{van2012replaying}, negative events \cite{vandenbroucke2013determining}, or topological entropy \cite{DBLP:journals/tosem/PolyvyanyySWCM20}, among others.
A general framework for introducing probabilistic weights to either recordings in the model and in the event log was proposed in \cite{Rogge-Solti.etal/BPM2016:InLogandModelWeTrust}.
While each of these approaches has its merits, the alignment-based approach  \cite{adriansyah2013memory,DBLP:conf/caise/DongenCCT17} is elaborated and used in the rest of the paper for its overall efficiency.
An alignment is a sequence of moves that synchronizes the computation of a model and a log's trace.
As we distinguish a (workflow) run from a full run of a mixed-paradigm model, according to whether the declarative constraints are satisfied or not at the end, so we make a distinction between a (workflow) alignment and a full alignment in the following.
\begin{definition}[Alignment]\label{def:alignment}
Let $\Mp = (\WfN, \Dp)$ be a mixed-paradigm model as per \cref{def:mp}.
Let $\sigma\in\Sigma^*$ be a sequence $\langle\sigma_1,\ldots,\sigma_n\rangle$ of length $n \in \mathbb{N}$ 
and $\gg$ be the symbol that denotes the skipping of an activity (i.e., no move is performed). 
Let $T^\gg$ denote $T\,\cup\,\{\gg\}$ and $A^\gg$ denote $A\,\cup\{\gg\}$.
Then, a workflow alignment (or \emph{alignment} for short) of $\sigma$ and $\Mp$ is a sequence of \emph{moves} $\gamma = \langle \gamma_1,\ldots,\gamma_m\rangle \in (T^{\gg}\times A^{\gg})^*$ with $\gamma_i=(t_i, a_i)$ for $1 \leqslant i \leqslant m$, with $m \geqslant n \in \mathbb{N}$, between the trace $\sigma$ and the mixed-paradigm model $\Mp$ if and only if:
	\begin{compactenum}
	\item For each $1 \leq i\leq m$ it holds that:
		\begin{compactitem}
		  \item $\gamma_i$ is a \emph{move on model}: $t_i \in T$, $a_i = \gg$ (with a $\tau$-move being the special case where $\ell(t)=\tau$),
		  \item $\gamma_i$ is a \emph{move on log}: $t_i = \gg$, $a_i \in A$, or
		  \item $\gamma_i$ is a \emph{synchronous move}: $t_i \in T$, $a_i \in A$;
		\end{compactitem}
	\item By applying an order-preserving endomorphism that removes moves on model, we obtain
	$\gamma' = \langle \gamma'_1, \ldots, \gamma'_{m'}\rangle$ with $m' \leqslant m$
	such that its sequence of events
	$\langle a'_1, \ldots, a'_{m'} \rangle$ is the trace $\sigma$;
	\item By applying an order-preserving endomorphism that removes moves on log, we obtain
	$\gamma'' = \langle \gamma''_1, \ldots, \gamma''_{m'}\rangle$ with $m'' \leqslant m$
	such that its sequence of transitions
	$\langle t''_1, \ldots, t''_{m''} \rangle$ is a run for $\WfN$, i.e., $\gamma$ induces a workflow run of $\Mp$ as per \cref{def:mp:run}.
	\end{compactenum}
	If $\gamma$ is a workflow alignment that induces a full run of $\Mp$ as per \cref{def:mp:run}, then $\gamma$ is a \emph{full alignment}.
\end{definition}
Given a cost function for alignments, an \emph{optimal} (workflow or full) \emph{alignment} is defined as an alignment that minimizes this function, as we formalize in the following.
\begin{definition}[Optimal alignment]\label{def:alignment:optimal}
Let $\Mp = (\WfN, \Dp)$ be a mixed-paradigm model as per \cref{def:mp}
and
$\sigma\in\Sigma^*$ be a sequence $\langle\sigma_1,\ldots,\sigma_n\rangle$ of length $n \in \mathbb{N}$.
Given a function
$\textrm{cost}: (T^{\gg}\times A^{\gg})^* \to \mathbb{R}^+$,
an alignment $\gamma^{\textrm{opt}} \in (T^{\gg}\times A^{\gg})^*$ of $\sigma$ and $\Mp$ is an \emph{optimal alignment} if and only if, for any other alignment $\gamma \in (T^{\gg}\times A^{\gg})^*$, it holds true that
$\textrm{cost}(\gamma^{\textrm{opt}}) \leqslant \textrm{cost}(\gamma)$.
\end{definition}
Notice that there may be more than one optimal alignment for any given sequence of events. 
The default cost function for alignments is based on a function $c: T^{\gg}\times A^{\gg} \to \mathbb{R}^+$ associating \num{0} to every synchronous move and $\tau$-move, and \num{1} to every move on log or model. The default total cost of the alignment, then, is the sum of the partial costs on the moves.

Consider for example the model in \cref{fig:mpexample} and a trace $\langle\tasko,\taskp,\taskn,\taski,\taskn\rangle$, i.e., a trace showing that an order was received ($\Task{o}$) and a product assembled ($\Task{p}$), then a cancellation was received ($\Task{n}$), the invoice emitted ($\Task{i}$) and another cancellation received ($\Task{n}$). If we assume a standard cost function, 
an optimal alignment for reaching the final marking
$ \mu_\textrm{F}=\{ p_9^1 \} $
from the initial marking
$ \mu_0 = \{ p_1^1 \} $
is
$\gamma^{\textrm{opt}}=\langle(t_1,\tasko),(t_2,\taskp),(t_{11},\taskn),(\gg,\taski),(t_{11},\taskn),(t_{10},\gg)\rangle$.
This alignment is full as it ensures that all constraints are satisfied at the end of the execution of the model.
This alignment shows that the \Task{Emit invoice} activity ($\Task{i}$) cannot happen in the model, even though it is in the event log. This is because of the \NotSucc{\Task{Receive cancellation}}{\Task{Emit invoice}} constraint. Finally, in order to reach the final marking, the model needs to execute the \Task{Abort order} transition ($\Task{x}$) which is not part of the event log. 

Now consider another trace,
$\langle\tasko,\taskp,\taski,\tasks,\taskeu,\taskm,\taskn\rangle$.
In this trace, an order is received ($\Task{o}$), a product assembled ($\Task{p}$), an invoice emitted ($\Task{i}$). The product is then shipped ($\Task{s}$) and payed ($\Task{\EUR}$) and finally, money is returned ($\Task{m})$, a cancellation is received ($\Task{n}$) and the process stops. 
Again using the standard cost function, an optimal alignment would be $\gamma^{\textrm{opt}}=\langle(t_1,\tasko),(t_2,\taskp),(t_4,\taski),(t_6,\tasks),(t_5,\taskeu),(t_7,\gg),(\gg,\taskm),(\gg,\taskn),(t_8,\gg)\rangle$ with three deviations, i.e., the move on log for returning the money, the move on log for the cancellation, and a move on model on \Task{Register completion} ($t_8$). This alignment suggests that the model cannot explain that cancellation took place and money was returned, and that, additionally, the case completed regularly although it was not observed in the log. 

\begin{sloppypar}
Sometimes, however, we could prefer an explanation stating that the cancellation did take place, money was returned and then the item return was registered, but the constraint \Prec{\Task{Receive cancellation}}{\Task{Return money}} is violated at the end of the case. The alignment in that case would be 
$\gamma^{\textrm{opt}}=\langle(t_1,\tasko),(t_2,\taskp),(t_4,\taski),(t_6,\tasks),(t_5,\taskeu),(t_{12},\taskm),(t_{11},\taskn)(t_{13},\gg),(t_7,\gg),(t_9,\gg)\rangle$ 
again with three deviations, i.e., the synchronous move $(t_{12},\taskm)$ which breaks the {\PrecTmp} constraint and the two model moves on \Task{Withdraw product} after cancellation and on \Task{Register item return}, indicating that these tasks have not happened yet, but are required to happen to complete the process. 
\end{sloppypar}

In the next section, we show that existing techniques for finding alignments can be modified to solve both types of problems, i.e., finding alignments that guarantee that all constraints are satisfied and finding alignments in which constraints may be violated at the end of the trace. %
%
%
%
\subsection{Computing workflow alignments}
\label{subsec:wfalignment}

Computing alignments for a trace and a procedural model is a non-trivial task. Many techniques exist in literature and for a full overview, we refer to \cite{DBLP:books/sp/CarmonaDSW18}. All search-based techniques work in a similar fashion. The initial state used by the search algorithm is the initial marking of the model and the beginning of the trace to replay. From that state on, all successor states are computed by executing all possible moves, i.e., moves on model, moves on log or synchronous moves. This leads to a number of successor states which are then queued for further investigation.

Several techniques have been proposed to decide which states to investigate in which order. In \cite{DBLP:conf/bpm/Dongen18a} an overview of parameters is provided which influence this order. However, the search in general relies on a fundamental element: the estimation of the remaining cost. The underlying search algorithm for alignment computation is A$^\star$. This algorithm works best in the presence of a heuristic which underestimates the remaining costs of replaying the (remainder of the) given trace from the current marking. Such an estimator exists for Petri nets in the form of a marking equation that abstracts from the order in which transitions are fired to get an estimate.

\Cref{alg:pseudoalignbas} describes in pseudo-code how our algorithm yields a workflow alignment -- thus, not considering the satisfaction of constraints. States are stored in a priority queue and investigated one by one. In each iteration, new successor states are added to a priority queue and this queue is polled in further iterations. Note that we abstract from the closed set here, yet no state has to be visited more than once and no state has to be represented in the queue more than once (for algorithmic details, we refer to \cite{DBLP:conf/bpm/Dongen18}). Notice also that the algorithm assumes that at least one alignment exists.

\begin{algorithm2e}[tb]
	\caption{Workflow alignment.}
	\label{alg:pseudoalignbas}
\SetKwInOut{Input}{input}
\SetKwInOut{Output}{output}
\SetKwProg{AlignTraceAndModel}{AlignTraceAndModel}{}{}
\AlignTraceAndModel{$(\sigma=\langle \sigma_1,\ldots,\sigma_n\rangle \in \Sigma^*,\Mp,c)$}{
  \Input{A trace $\sigma$; a mixed-paradigm model $\Mp=(\WfN,\Dp)$; a partial cost function $c:(T^{\gg}\times A^{\gg})\to\mathbb{R}^+$}
  \Output{An alignment $\gamma$}
  $i \gets 1$  \tcp*{$i$ keeps track of the index in the trace}
  $\mu \gets \mu_0$ \tcp*{$\mu$ keeps track of the state in the Workflow net \WfN}
  $g \gets 0 $ \tcp*{$g$ keeps track of the cost so far}
  $h \gets \mathrm{estimateRemainingCost}(\sigma, \WfN, i,\mu, c) $ \tcp*{$h$ underestimates the remaining cost}
  $p \gets \mathrm{NULL}$  \tcp*{$p$ stores the predecessor of the current state}
  $Q \gets \{\}$  \tcp*{$Q$ is a queue of states to investigate}
  $q \gets (i,\mu,g,h,p)$ \tcp*{$q$ is the current head of the queue}
  \While{$i \leq n+1\,\vee\, \mu \neq \mu_\textrm{F}$}{ \label{algline:term}
  $T_{\textrm{enabled}}\gets \mathrm{getEnabledTransitionsPN}(\WfN,\mu)$ \label{algline:enablement}\;
    \For{$t\in T_{\textnormal{enabled}}$}{
      $\mu' \gets \mu-\bullet t+t\bullet$ \;
      \If(\tcp*[f]{Compute next state for synchronous move}){$l(t)=\sigma_i$}{ 
        $g' \gets g + c(t,\sigma_i) $\;
        $h' \gets \mathrm{estimateRemainingCost}(\sigma, \WfN, i',\mu', c) $\;
        $Q \gets \mathrm{enqueue}((i+1,\mu',g,h,q))$  \tcp*{Add new state to the queue}
      }
      $g' \gets g + c(t,\gg) $ \tcp*{Compute next state for model move}
      $h' \gets \mathrm{estimateRemainingCost}(\sigma, \WfN, i, \mu', c) $\;
      $Q \gets \mathrm{enqueue}( (i,\mu',g',h',q) )$  \tcp*{Add new state to the queue}
    }
    \If(\tcp*[f]{Compute next state for log move}){$i \leq n$} { 
      $g' \gets g + c(\gg,l(\sigma_i)) $\;
      $h' \gets \mathrm{estimateRemainingCost}(\sigma, \WfN, i+1, \mu, c) $\;
      $Q \gets \mathrm{enqueue}( (i+1, \mu, g', h',q) )$  \tcp*{Add new state to the queue}
    }
    $(i, \mu, g, h, p) \gets \mathrm{pullNextBestState}(Q)$  \tcp*{Pull next state to investigate from the queue}
    $q \gets (i, \mu, g, h, p)$\; 
  }    
  $\gamma \gets \mathrm{extractAlignment}(i,\mu,g,h,p)$\; \tcp*[f]{By recursion from the final state to the initial state with $p=\mathrm{NULL}$, the alignment is constructed}
  \KwRet{$\gamma$}\;
}
 \end{algorithm2e}

\begin{figure}
  \centering
  \includegraphics[width=\textwidth]{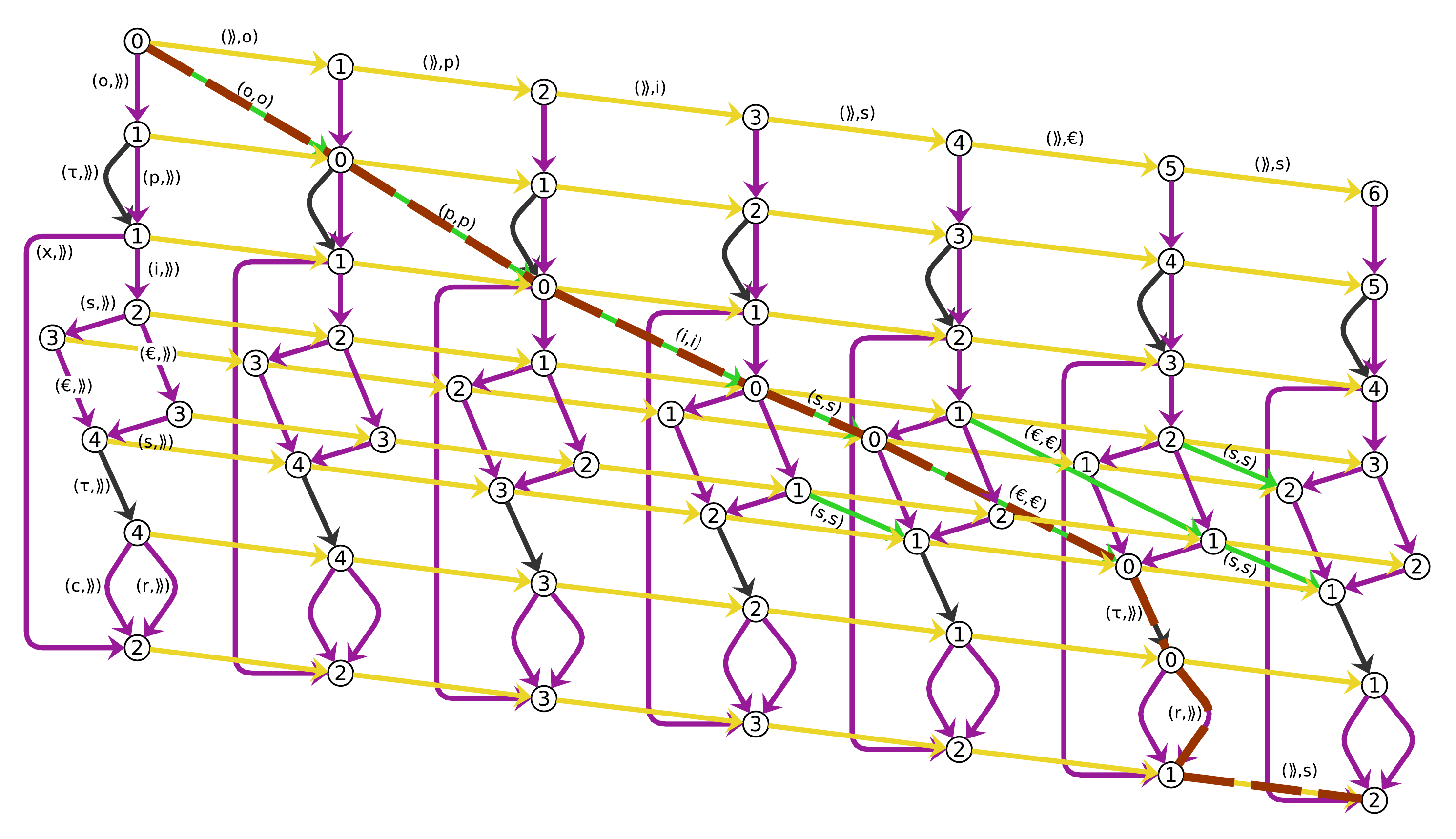}
  \caption[The A$^\star$ search space]{The full search space of the basic A$^\star$ algorithm for the trace $\langle\tasko,\taskp,\taski,\tasks,\taskeu,\tasks\rangle$ and the Workflow net of \cref{fig:petrinet}. Notice that this search space is never fully expanded.}
  \label{fig:searchspaceWF}
\end{figure}

Let us consider the trace
$\langle\tasko,\taskp,\taski,\tasks,\taskeu,\tasks\rangle$
and the Workflow net of \cref{fig:petrinet}. \Cref{fig:searchspaceWF} depicts the entire search space considered by the A$^\star$ algorithm in \cref{alg:pseudoalignbas}. For clarity of presentation, the moves are identified using the transition labels, not the identifiers. The color scheme we use follows the typical scheme of alignment-based conformance checking tools. The moves on model are depicted top-down in purple and the moves on log are depicted in yellow from left to right. The green synchronous moves are also shown and, for each node, the shortest distance from the top-left node is written. $\tau$-moves are drawn in dark gray. Finally, the dashed brown line shows a shortest path in this graph, which corresponds to the following alignment: $\gamma^{opt}=\langle(t_1,\tasko),(t_2,\taskp),(t_4,\taski),(t_6,\tasks),(t_5,\taskeu),(t_7,\gg),(t_8,\gg),(\gg,\tasks)\rangle$ with cost \num{2}. The A$^\star$ algorithm will not expand the full search space, but because of the quality of the estimation function, it will in this case only enqueue the nodes on this shortest path and their direct neighbors. Without this estimation function, however, all nodes with cost less or equal to \num{3} will be queued before this shortest path is returned.

The two techniques presented in this section adopt the existing search-based techniques in two ways. First, we provide a technique that guarantees that all constraints are satisfied at the end of alignment. Second, we provide a technique that allows for the violation of constraints using a cost for doing so, similar to the cost for moves on model and log.

\subsection{Computing full alignments}
\label{subsec:satisfied}

Consider a mixed-paradigm process model which consists of a procedural model and additional declarative constraints on the transitions in this model. As we have seen, the model without any constraints is a straightforward procedural model. 
To calculate alignments for this model, we can simply use the existing A$^\star$-based search techniques.

The declarative constraints added to that model cause its language to shrink, i.e., certain traces are no longer part of the language as they violate one or more constraints (see \cref{sec:MPPM}). To yield a full run, namely the run that ends in a final marking for the workflow net and in the final states of constraint automata (\cref{def:mp:run}), 
we modify the algorithm by only changing Line~\ref{algline:enablement} of \cref{alg:pseudoalignbas} to check for the admitted transitions as well as Line~\ref{algline:term} for the termination condition.
With these modifications in place, we obtain a full alignment (\cref{def:alignment}).

\begin{algorithm2e}[tb]
	\caption{Full alignment.}
	\label{alg:pseudoalignsatisfied}
\SetKwInOut{Input}{input}
\SetKwInOut{Output}{output}
\SetKwProg{AlignTraceAndModel}{AlignTraceAndModel}{}{}
\AlignTraceAndModel{$(\sigma=\langle \sigma_1,\ldots,\sigma_n\rangle \in \Sigma^*,\Mp,c)$}{
  \Input{A trace $\sigma$; a mixed-paradigm model $\Mp=(\WfN,\Dp)$; a partial cost function $c:(T^{\gg}\times A^{\gg})\to\mathbb{R}^+$}
  \Output{An alignment $\gamma$}
  $i \gets 1$  \tcp*{$i$ keeps track of the index in the trace}
  $(\mu,s) \gets (\mu_0,S^K_0)$ \tcp*{$(\mu,s)$ keeps track of the state in the Workflow net and $s$ the state of each automaton}
  $g \gets 0 $ \tcp*{$g$ keeps track of the cost so far}
  $h \gets \mathrm{estimateRemainingCost}(\sigma, \WfN, i,(\mu,s), c) $ \tcp*{$h$ underestimates the remaining cost}
  $p \gets \mathrm{NULL}$  \tcp*{$p$ stores the predecessor of the current state}
  $Q \gets \{\}$  \tcp*{$Q$ is a queue of states to investigate}
  $q \gets (i,(\mu,s),g,h,p)$ \tcp*{$q$ is the current head of the queue}
  \While(\tcp*[f]{Do not stop if there is an unsatisfied constraint}){$i \leq n+1\,\vee\, \mu \nsubseteq M_\textrm{F}\, \vee\, s\not \in S^K_F $}{ \label{algline:stopsatisfied}
  $T_{\textrm{enabled}}\gets \mathrm{getEnabledTransitionsPN}(\WfN,\mu)$\;
    \For{$t\in T_{\textrm{enabled}}$}{
      \If(\tcp*[f]{Transition $t$ cannot lead to any non-accepting state}){$\forall_{\kappa \in K} \delta^\kappa(s^{\kappa},t) \not = s^\kappa_{F}$} {\label{algline:checkEnabled}
         $\mu' \gets \mu-\bullet t+t\bullet$  \tcp*{Compute next marking in the net}
         $s' \gets \{ \delta^\kappa(s^\kappa,t) \mid \kappa \in K\} $ \tcp*{Compute next state for all automata}
         \If(\tcp*[f]{Compute next state for synchronous move}){$l(t)=\sigma_i$}{  
           $g' \gets g + c(t,\sigma_i) $\; 
           $h' \gets \mathrm{estimateRemainingCost}(\sigma, \WfN, i',(\mu',s'), c) $\;
           $Q \gets \mathrm{enqueue}((i+1,(\mu',s'),g,h,q))$ \tcp*{Add new state to the queue}
         } 
         $g' \gets g + c(t,\gg) $ \tcp*{Compute next state for move on model}
         $h' \gets \mathrm{estimateRemainingCost}(\sigma, \WfN, i,(\mu',s'), c) $\;
         $Q \gets \mathrm{enqueue}((i,(\mu',s'),g',h',q))$  \tcp*{Add new state to the queue}
      }
    }
    \If(\tcp*[f]{Compute next state for move on log}){$i \leq n$}{
      $g' \gets g + c(\gg,l(\sigma_i)) $\;
      $h' \gets \mathrm{estimateRemainingCost}(\sigma, \WfN, i+1,(\mu,s), c) $\;
      $Q \gets \mathrm{enqueue}((i+1,(\mu,s),g',h',q))$  \tcp*{Add new state to the queue}
    }
    $(i,(\mu,s),g,h,p) \gets \mathrm{pullNextBestState}(Q)$ \tcp*{Pull next state to investigate from the queue}
    $q \gets (i,\mu,g,h,p)$\;
  }    
  $\gamma \gets \mathrm{extractAlignment}(i,(\mu,s),g,h,p)$\;
  \KwRet{$\gamma$}\;
}
 \end{algorithm2e}

The algorithm is described in \cref{alg:pseudoalignsatisfied}. Notice that it is only slightly different than the base algorithm. For each automaton, we need to keep track of the state in which the automaton is. In Line~\ref{algline:stopsatisfied}, we check if any of the automata is in a non-accepting state. If this is the case, the search has to continue as the stop criterion is that the final marking in the model is reached, the trace is fully explained, and all constraints are satisfied. In Line~\ref{algline:checkEnabled} we check, for a transition $t$ which is enabled in the model, if any of the automata could end up in a state of permanent violation $s^\kappa_{\textrm{na}}$. If so, this transition cannot be executed and the algorithm continues with the next transition.

By ensuring that (a) no state of permanent violation is ever reached and (b) the algorithm only terminates if all constraints are satisfied, the resulting alignment algorithm is correct if and only if the estimate of the remaining cost is still an underestimate. This is trivially true since it is true for the larger language of the procedural model without constraints~\cite{adriansyahreplay}. 

Let us consider the trace
$\langle\tasko,\taskp,\taski,\tasks,\taskeu,\taskm,\taskn\rangle$
again -- we already discussed it at the end of \cref{sec:alignments}. Recall that, in this trace, an order is received ($t_1/\Task{o}$), a product assembled ($t_{2}/\Task{m}$), an invoice emitted ($t_{4}/\Task{i}$). The product is then shipped ($t_{6}/\Task{s}$) and payed ($t_{5}/\Task{\EUR}$). Finally, money is returned ($t_{13}/\Task{m})$ after which a cancellation is received ($t_{11}/\Task{n}$) and the process stops.

\begin{figure}
  \centering
  \includegraphics[height=0.5\textheight]{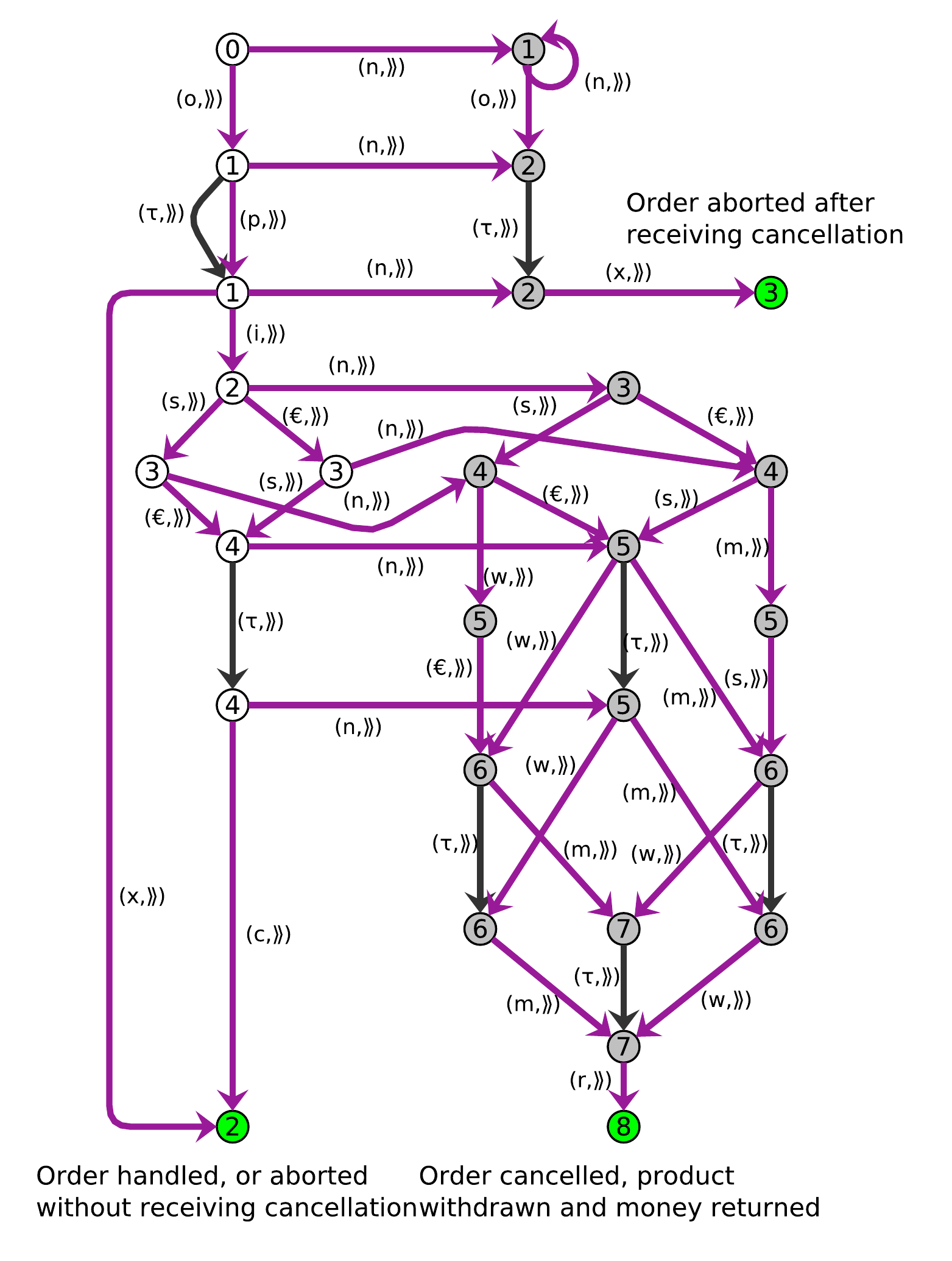}
  \caption[Moves on model guaranteeing satisfied constraints]{The state space for the mixed paradigm model of \cref{fig:running:example}. The green states are termination states. In all gray states, activity \Task{Receive cancellation} can happen multiple times (self-loops are omitted for the sake of readability).}
  \label{fig:modelspaceSatisfied}
\end{figure}

\begin{figure}
  \centering
  \includegraphics[width=\textwidth]{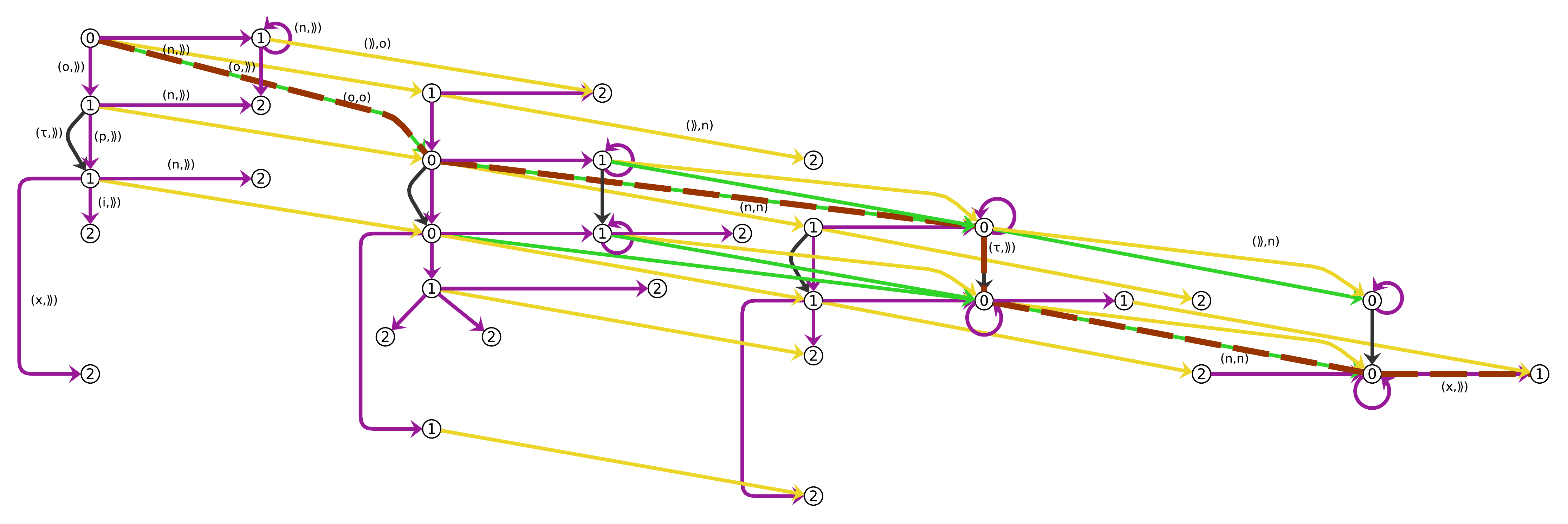}
  \caption[SearchSpace for Mixed Paradigm]{The worst-case expansion of the search space of the basic A$^\star$ algorithm for the trace $\langle\tasko,\taskn,\taskn\rangle$ and the Workflow net of \cref{fig:running:example}. The shortest path corresponding to $\gamma^{opt}=\langle(t_1,\tasko),(t_{11},\taskn),(t_3,\gg),(t_{11},\taskn),(t_{10},\gg)\rangle$ is highlighted with a superimposed dashed line.}
  \label{fig:searchspaceSatisfied}
\end{figure}

\Cref{fig:modelspaceSatisfied} shows part of the search space when computing alignments in the mixed paradigm model of \cref{fig:running:example}. This figure shows the behavior of the model without any events in the log. There are three possible final states. The first final state, on the bottom-left corner of the figure, represents all executions in which activity \Task{Receive cancellation} (\Task{n}) does not happen. Notice that this state cannot be reached by executing \Task{Register item return} (\Task{r}), since that activity requires a cancellation to have occurred due to constraint \Prec{\Task{Receive cancellation}}{\Task{Register item return}}. The second final state, on the top-right corner, represents the case where cancellation is received before the invoice is emitted. In this case, the model is forced to abort the order. The third final state represents the case where cancellation was received after emitting the invoice. This part of the model requires handling the withdrawal of the product and the returning of the money as well as the registration of the item's return. 
\Cref{fig:searchspaceSatisfied} shows the worst case expansion of the search space for \cref{alg:pseudoalignsatisfied} when aligning the model of the running example of \cref{fig:running:example} to the trace $\langle\tasko,\taskn,\taskn\rangle$, i.e., the trace in which, after receiving an order, the order is canceled twice by the customer. The shortest past is highlighted and corresponds to the alignment $\gamma^{opt}=\langle(t_1,o),(t_{11},n),(t_3,\gg),(t_{11},n),(t_{10},\gg)\rangle$ with cost \num{1} for the move on model of \Task{Abort order}.

The full search space for the A$^\star$ algorithm for trace $\langle\tasko,\taskp,\taski,\tasks,\taskeu,\taskm,\taskn\rangle$ contains $27 \times 8 = 216$ states and is structured the same way as \cref{fig:searchspaceWF}, i.e., with synchronous moves and moves on log between identical copies of the model's state space.

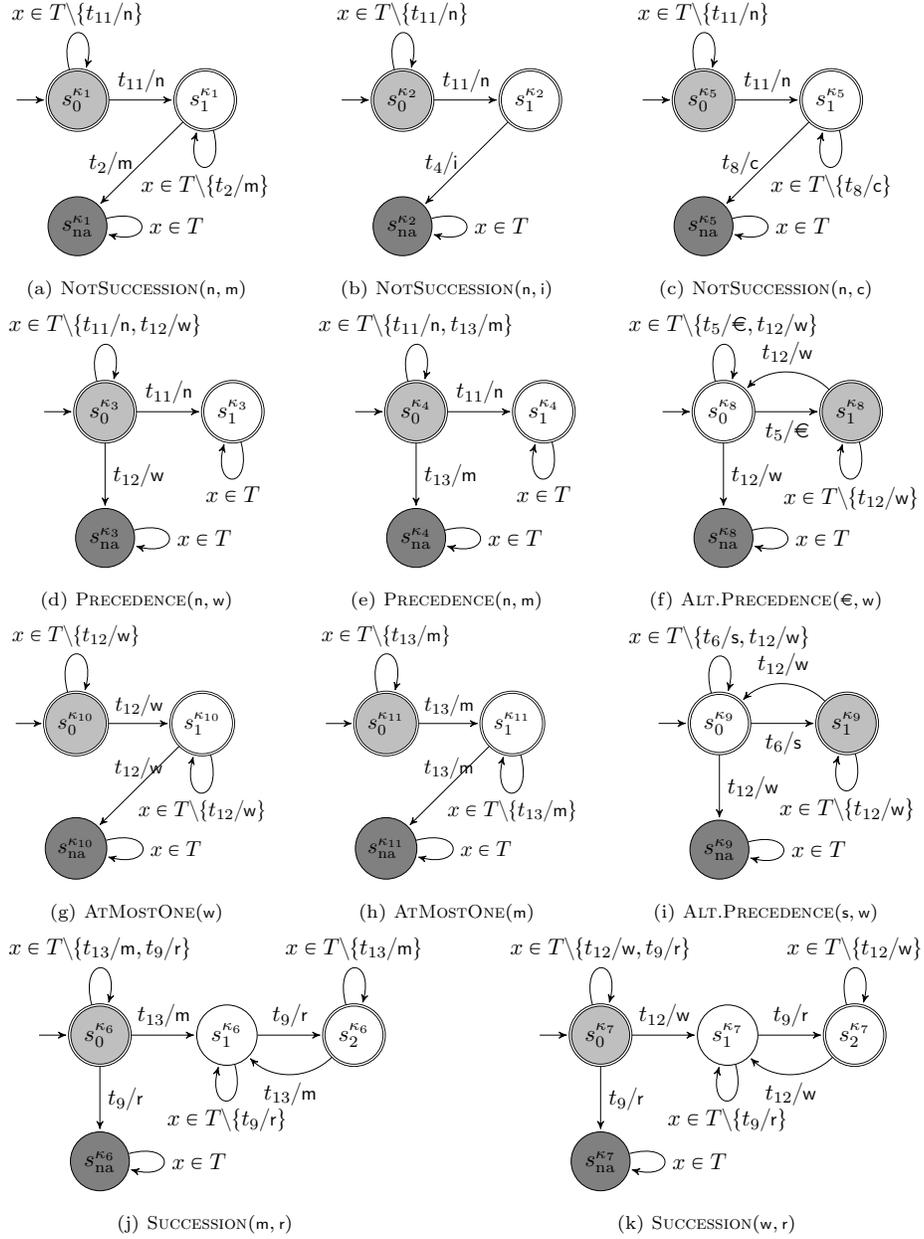
\begin{figure}[p]%
	\centering%
\begin{subfigure}{0.3\textwidth}%
	\centering%
	\resizebox{!}{0.18\textheight}{%
        \begin{tikzpicture}[->, >=stealth', shorten >=1pt, auto, bend angle=45, initial text = {}]
          \tikzstyle{every state}=[minimum size=1em]
        
          \node[state, initial, accepting, fill=lightgray]        (0) {$s^{\kappa_1}_0$};
          \node[state, right=of 0, accepting]     (1) {$s^{\kappa_1}_1$};
          \node[state, below=of 0,fill=gray] (2) {$s^{\kappa_1}_{\textrm{na}}$};
        
          \path
          (0) edge [loop above] node [] {$x \in T \setminus\! \left\lbrace t_{11}/\Task{n} \right\rbrace$}       (0)
          (0) edge [          ] node [above] {$t_{11}/\Task{n}$}                                                      (1)
          (1) edge [loop below] node         {$x \in T \setminus\! \left\lbrace t_{2}/\Task{m} \right\rbrace $} (1)
          (1) edge [          ] node [left ] {$t_{2}/\Task{m}$}                                                      (2)
          (2) edge [loop right] node [] {$x \in T$}                                                     (2)
          ;  
        \end{tikzpicture}
	}%
	\caption{\NotSucc{\Task{n}}{\Task{m}}}
\end{subfigure}%
\hfill
\begin{subfigure}{0.3\textwidth}%
	\centering%
	\resizebox{!}{0.18\textheight}{%
        \begin{tikzpicture}[->, >=stealth', shorten >=1pt, auto, bend angle=45, initial text = {}]
          \tikzstyle{every state}=[minimum size=1em]
        
          \node[state, initial, accepting, fill=lightgray]        (0) {$s^{\kappa_2}_0$};
          \node[state, right=of 0, accepting]     (1) {$s^{\kappa_2}_1$};
          \node[state, below=of 0,fill=gray] (2) {$s^{\kappa_2}_{\textrm{na}}$};
        
          \path
          (0) edge [loop above] node [] {$x \in T \setminus\! \left\lbrace t_{11}/\Task{n} \right\rbrace$}       (0)
          (0) edge [          ] node [above] {$t_{11}/\Task{n}$}                                                      (1)
          (1) edge [          ] node [left ] {$t_{4}/\Task{i}$}                                                      (2)
          (2) edge [loop right] node [] {$x \in T$}                                                     (2)
          ;  
        \end{tikzpicture}
	}%
	\caption{\NotSucc{\Task{n}}{\Task{i}}}
\end{subfigure}%
\hfill
\begin{subfigure}{0.32\textwidth}%
	\centering%
	\resizebox{!}{0.18\textheight}{%
        \begin{tikzpicture}[->, >=stealth', shorten >=1pt, auto, bend angle=45, initial text = {}]
          \tikzstyle{every state}=[minimum size=1em]
        
          \node[state, initial, accepting, fill=lightgray]        (0) {$s^{\kappa_5}_0$};
          \node[state, right=of 0, accepting]     (1) {$s^{\kappa_5}_1$};
          \node[state, below=of 0,fill=gray] (2) {$s^{\kappa_5}_{\textrm{na}}$};
        
          \path
          (0) edge [loop above] node [] {$x \in T \setminus\! \left\lbrace t_{11}/\Task{n} \right\rbrace$}       (0)
          (0) edge [          ] node [above] {$t_{11}/\Task{n}$}                                                      (1)
          (1) edge [loop below] node         {$x \in T \setminus\! \left\lbrace t_{8}/\Task{c} \right\rbrace $} (1)
          (1) edge [          ] node [left ] {$t_{8}/\Task{c}$}                                                      (2)
          (2) edge [loop right] node [] {$x \in T$}                                                     (2)
          ;  
        \end{tikzpicture}
	}%
	\caption{\NotSucc{\Task{n}}{\Task{c}}}
\end{subfigure}%

\begin{subfigure}{0.3\textwidth}%
	\centering%
	\resizebox{!}{0.18\textheight}{%
        \begin{tikzpicture}[->, >=stealth', shorten >=1pt, auto, bend angle=45, initial text = {}]
          \tikzstyle{every state}=[minimum size=1em]
        
          \node[state, initial, accepting, fill=lightgray]        (0) {$s^{\kappa_3}_0$};
          \node[state, right=of 0, accepting]     (1) {$s^{\kappa_3}_1$};
          \node[state, below=of 0,fill=gray] (2) {$s^{\kappa_3}_{\textrm{na}}$};
        
          \path
          (0) edge [loop above] node [] {$x \in T \setminus\! \left\lbrace t_{11}/\Task{n},t_{12}/\Task{w} \right\rbrace$}       (0)
          (0) edge [          ] node [above] {$t_{11}/\Task{n}$}                                                      (1)
          (1) edge [loop below] node         {$x \in T $} (1)
          (0) edge [          ] node [right] {$t_{12}/\Task{w}$}                                                      (2)
          (2) edge [loop right] node [] {$x \in T$}                                                     (2)
          ;  
        \end{tikzpicture}
	}%
	\caption{\Prec{\Task{n}}{\Task{w}}}
\end{subfigure}%
\hfill
\begin{subfigure}{0.3\textwidth}%
	\centering%
	\resizebox{!}{0.18\textheight}{%
        \begin{tikzpicture}[->, >=stealth', shorten >=1pt, auto, bend angle=45, initial text = {}]
          \tikzstyle{every state}=[minimum size=1em]
        
          \node[state, initial, accepting, fill=lightgray]        (0) {$s^{\kappa_4}_0$};
          \node[state, right=of 0, accepting]     (1) {$s^{\kappa_4}_1$};
          \node[state, below=of 0,fill=gray] (2) {$s^{\kappa_4}_{\textrm{na}}$};
        
          \path
          (0) edge [loop above] node [] {$x \in T \setminus\! \left\lbrace t_{11}/\Task{n},t_{13}/\Task{m} \right\rbrace$}       (0)
          (0) edge [          ] node [above] {$t_{11}/\Task{n}$}                                                      (1)
          (1) edge [loop below] node         {$x \in T $} (1)
          (0) edge [          ] node [right] {$t_{13}/\Task{m}$}                                                      (2)
          (2) edge [loop right] node [] {$x \in T$}                                                     (2)
          ;  
        \end{tikzpicture}
	}%
	\caption{\Prec{\Task{n}}{\Task{m}}}
\end{subfigure}%
\hfill
\begin{subfigure}{0.32\textwidth}%
	\centering%
	\resizebox{!}{0.18\textheight}{%
        \begin{tikzpicture}[->, >=stealth', shorten >=1pt, auto, bend angle=45, initial text = {}]
          \tikzstyle{every state}=[minimum size=1em]
        
          \node[state, initial, accepting]        (0) {$s^{\kappa_8}_0$};
          \node[state, right=of 0, accepting, fill=lightgray]     (1) {$s^{\kappa_8}_1$};
          \node[state, below=of 0,fill=gray] (2) {$s^{\kappa_8}_{\textrm{na}}$};
        
          \path
          (0) edge [loop above] node []      {$x \in T \setminus\! \left\lbrace t_{5}/\Task{\EUR},t_{12}/\Task{w} \right\rbrace$} (0)
          (0) edge [          ] node [below] {$t_{5}/\Task{\EUR}$}                                                            (1)
          (0) edge [          ] node [] {$t_{12}/\Task{w}$}                                                                 (2)
          (2) edge [loop right] node [] {$x \in T$}                                                      (2)
          (1) edge [bend right] node [above] {$t_{12}/\Task{w}$}                                                            (0)
          (1) edge [loop below] node         {$x \in T \setminus\! \left\lbrace t_{12}/\Task{w} \right\rbrace$}        (1)
          ;  
        \end{tikzpicture}		
    }%
	\caption{\AltPrecShort{\Task{\EUR}}{\Task{w}}}
\end{subfigure}%

	\begin{subfigure}{0.3\textwidth}%
	\centering%
  		\resizebox{!}{0.18\textheight}{%
        \begin{tikzpicture}[->, >=stealth', shorten >=1pt, auto, bend angle=45, initial text = {}]
          \tikzstyle{every state}=[minimum size=1em]
        
          \node[state, initial, accepting, fill=lightgray]        (0) {$s^{\kappa_{10}}_0$};
          \node[state, right=of 0, accepting]     (1) {$s^{\kappa_{10}}_1$};
          \node[state, below=of 0,fill=gray] (2) {$s^{\kappa_{10}}_{\textrm{na}}$};
        
          \path
          (0) edge [loop above] node [] {$x \in T \setminus\! \left\lbrace t_{12}/\Task{w} \right\rbrace$}       (0)
          (0) edge [          ] node [above] {$t_{12}/\Task{w}$}                                                      (1)
          (1) edge [loop below] node         {$x \in T \setminus\! \left\lbrace t_{12}/\Task{w} \right\rbrace $} (1)
          (1) edge [          ] node [above] {$t_{12}/\Task{w}$}                                                      (2)
          (2) edge [loop right] node [] {$x \in T$}                                                     (2)
          ;  
        \end{tikzpicture}
	}%
	\caption{\Uniq{\Task{w}}}
\end{subfigure}%
\hfill
	\begin{subfigure}{0.3\textwidth}%
	\centering%
  		\resizebox{!}{0.18\textheight}{%
        \begin{tikzpicture}[->, >=stealth', shorten >=1pt, auto, bend angle=45, initial text = {}]
          \tikzstyle{every state}=[minimum size=1em]
        
          \node[state, initial, accepting, fill=lightgray]        (0) {$s^{\kappa_{11}}_0$};
          \node[state, right=of 0, accepting]     (1) {$s^{\kappa_{11}}_1$};
          \node[state, below=of 0,fill=gray] (2) {$s^{\kappa_{11}}_{\textrm{na}}$};
        
          \path
          (0) edge [loop above] node [] {$x \in T \setminus\! \left\lbrace t_{13}/\Task{m} \right\rbrace$}       (0)
          (0) edge [          ] node [above] {$t_{13}/\Task{m}$}                                                      (1)
          (1) edge [loop below] node         {$x \in T \setminus\! \left\lbrace t_{13}/\Task{m} \right\rbrace $} (1)
          (1) edge [          ] node [above] {$t_{13}/\Task{m}$}                                                      (2)
          (2) edge [loop right] node [] {$x \in T$}                                                     (2)
          ;  
        \end{tikzpicture}
	}%
	\caption{\Uniq{\Task{m}}}
\end{subfigure}%
\hfill
\begin{subfigure}{0.32\textwidth}%
	\centering%
	\resizebox{!}{0.18\textheight}{%
        \begin{tikzpicture}[->, >=stealth', shorten >=1pt, auto, bend angle=45, initial text = {}]
          \tikzstyle{every state}=[minimum size=1em]
        
          \node[state, initial, accepting]        (0) {$s^{\kappa_9}_0$};
          \node[state, right=of 0, accepting, fill=lightgray]     (1) {$s^{\kappa_9}_1$};
          \node[state, below=of 0,fill=gray] (2) {$s^{\kappa_9}_{\textrm{na}}$};
        
          \path
          (0) edge [loop above] node []      {$x \in T \setminus\! \left\lbrace t_{6}/\Task{s},t_{12}/\Task{w} \right\rbrace$} (0)
          (0) edge [          ] node [below] {$t_{6}/\Task{s}$}                                                            (1)
          (0) edge [          ] node [] {$t_{12}/\Task{w}$}                                                                 (2)
          (2) edge [loop right] node [] {$x \in T$}                                                      (2)
          (1) edge [bend right] node [above] {$t_{12}/\Task{w}$}                                                            (0)
          (1) edge [loop below] node         {$x \in T \setminus\! \left\lbrace t_{12}/\Task{w} \right\rbrace$}        (1)
          ;  
        \end{tikzpicture}		
    }%
	\caption{\AltPrecShort{\Task{s}}{\Task{w}}}
\end{subfigure}%

\begin{subfigure}{0.45\textwidth}%
	\centering%
	\resizebox{!}{0.18\textheight}{%
        \begin{tikzpicture}[->, >=stealth', shorten >=1pt, auto, bend angle=45, initial text = {}]
          \tikzstyle{every state}=[minimum size=1em]
        
          \node[state, initial, accepting, fill=lightgray]        (0) {$s^{\kappa_6}_0$};
          \node[state, right=of 0]                (1) {$s^{\kappa_6}_1$};
          \node[state, right=of 1, accepting]     (2) {$s^{\kappa_6}_2$};
          \node[state, below=of 0,fill=gray] (3) {$s^{\kappa_6}_{\textrm{na}}$};
        
          \path
          (0) edge [loop above] node []      {$x \in T \setminus\! \left\lbrace t_{13}/\Task{m},t_{9}/\Task{r} \right\rbrace$} (0)
          (0) edge [          ] node [above] {$t_{13}/\Task{m}$}                                                            (1)
          (0) edge [          ] node [] {$t_{9}/\Task{r}$}                                                                 (3)
          (3) edge [loop right] node [] {$x \in T$}                                                           (3)
          (1) edge [loop below] node         {$x \in T \setminus\! \left\lbrace t_{9}/\Task{r} \right\rbrace$}        (1)
          (1) edge [          ] node [above] {$t_{9}/\Task{r}$}                                                            (2)
          (2) edge [bend left] node [below] {$t_{13}/\Task{m}$}                                                            (1)
          (2) edge [loop above] node         {$x \in T \setminus\! \left\lbrace t_{13}/\Task{m} \right\rbrace$}        (2)
          ;  
        \end{tikzpicture}		
    }%
	\caption{\Succ{\Task{m}}{\Task{r}}}
\end{subfigure}%
\hfill
\begin{subfigure}{0.45\textwidth}%
	\centering%
	\resizebox{!}{0.18\textheight}{%
        \begin{tikzpicture}[->, >=stealth', shorten >=1pt, auto, bend angle=45, initial text = {}]
          \tikzstyle{every state}=[minimum size=1em]
        
          \node[state, initial, accepting, fill=lightgray]        (0) {$s^{\kappa_7}_0$};
          \node[state, right=of 0]                (1) {$s^{\kappa_7}_1$};
          \node[state, right=of 1, accepting]     (2) {$s^{\kappa_7}_2$};
          \node[state, below=of 0,fill=gray] (3) {$s^{\kappa_7}_{\textrm{na}}$};
        
          \path
          (0) edge [loop above] node []      {$x \in T \setminus\! \left\lbrace t_{12}/\Task{w},t_{9}/\Task{r} \right\rbrace$} (0)
          (0) edge [          ] node [above] {$t_{12}/\Task{w}$}                                                            (1)
          (0) edge [          ] node [] {$t_{9}/\Task{r}$}                                                                 (3)
          (3) edge [loop right] node [] {$x \in T$}                                                           (3)
          (1) edge [loop below] node         {$x \in T \setminus\! \left\lbrace t_{9}/\Task{r} \right\rbrace$}        (1)
          (1) edge [          ] node [above] {$t_{9}/\Task{r}$}                                                            (2)
          (2) edge [bend left] node [below] {$t_{12}/\Task{w}$}                                                            (1)
          (2) edge [loop above] node         {$x \in T \setminus\! \left\lbrace t_{12}/\Task{w} \right\rbrace$}        (2)
          ;  
        \end{tikzpicture}		
    }%
	\caption{\Succ{\Task{w}}{\Task{r}}}
\end{subfigure}%
 	\caption[Constraint automata during trace replay]{All the automata for the constraints in the model of \cref{fig:running:example}. The light grey states are the states the automata are in when transitions $t_1/\Task{o}$, $t_{2}/\Task{m}$, $t_{4}/\Task{i}$, $t_{6}/\Task{s}$ and $t_{5}/\Task{\EUR}$ have been executed.}\label{fig:running:automata}
\end{figure}

Let us consider all the constraints in the model. There are \num{11} constraints in total: $\kappa_1$ to $\kappa_{11}$. Their respective automata are depicted in \cref{fig:running:automata}. The initial state of the search algorithm puts all these automata in the initial state $s_0$. Then, in Line~\ref{algline:checkEnabled} of \cref{alg:pseudoalignsatisfied}, we make sure that the current transition does not lead to any of the non-accepting states $s_{\textrm{na}}$ in any of the automata. If it does not, then this transition can be executed; otherwise, it cannot be executed.

Assume we have reached a state in the search where we have explained the fitting prefix of the sequence, i.e., we executed transitions $t_1/\Task{o}$, $t_{2}/\Task{m}$, $t_{4}/\Task{i}$, $t_{6}/\Task{s}$ and $t_{5}/\Task{\EUR}$  in the model. The next event to explain is $\Task{Return money}$. In the model, the marking is
$\{ p_6^1, p_7^1 \}$
and, as shown in \cref{fig:running:automata}, the automata are in their initial state, except for the automata of $\kappa_8$ and $\kappa_9$ which are in states $s^{\kappa_8}_1$ and $s^{\kappa_9}_1$ respectively. The corresponding transition $t_{13}$ for task $\Task{m}$ is enabled according to the workflow model. However, if we executed this transition, automaton $\kappa_4$ would transition from state $s^{\kappa_4}_0$ to $s^{\kappa_4}_{\textrm{na}}$ which is a state of permanent violation. In other words, this transition is not allowed at this point in time and the only way to explain the event is by labeling it as a move on log, which does not cause the state of the Workflow net to change, neither does it change the state of any of the automata.

The next event to explain would be the event \Task{Receive cancellation} ($t_{11}/\Task{n}$). The execution of $t_{11}/\Task{n}$ is allowed as it would not violate any of the constraints. The algorithm will therefore explore this option. However, after executing $t_{11}/\Task{n}$, it is no longer possible to execute $t_{8}/\Task{c}$ without reaching a state of permanent violation because of $\kappa_5$. To terminate the run, this forces the execution of $t_{9}/\Task{r}$, which, in turn, requires $t_{13}/\Task{m}$ and $t_{12}/\Task{w}$ to be fired through $\kappa_6$ and $\kappa_7$. This would lead to an alignment
$ \gamma = \langle (t_1,o),(t_2,p),(t_4,i),(t_6,s),(t_5,\Task{\EUR}),(t_7,\gg),(\gg,m),(t_{11},n),(t_{12},\gg),(t_{13},\gg),(t_9,\gg) \rangle $
with four deviations, namely a move on log on $t_{13}/\Task{m}$ and three moves on model on $t_{12}/\Task{w}$, $t_{13}/\Task{m}$ and $t_{9}/\Task{r}$. We already know that there is a better alignment with only three deviations, namely $\gamma^{opt}=\langle(t_1,o),(t_2,p),(t_4,i),(t_6,s),(t_5,\Task{\EUR}),(t_7,\gg),(\gg,m),(\gg,n),(t_8,\gg)\rangle$, so the alignment $\gamma$ will not be selected as the optimal one.
To overcome this issue, we modify the algorithm as follows.

\subsection{Allowing for constraint violation at a cost}\label{subsec:violating}
In this subsection, we show how we modify our algorithm to 
allow constraints to become permanently violated at a cost. To this end, we introduce a cost function that, for each constraint in the model, associates a cost to its permanent violation. When executing a transition in the model, we then no longer need to check \emph{if} any of the automata becomes permanently violated, but we need to add costs for \emph{when} they do. However, this alone is not sufficient as we also need to update the termination condition.

In \cref{alg:pseudoalignsatisfied}, the main loop terminates if (1) the model reaches the final marking, (2) the trace is fully explained and (3) all constraints are satisfied. The latter condition needs to change to ``all constraints are satisfied \textit{or permanently violated}''. We cannot leave an automaton in a temporarily violated state and simply add the costs of violating that constraint in the final step as there may be better alignments which we have not yet explored. 

\begin{algorithm2e}[tbp]
	\caption{Full alignment with costs on permanent violations.}
	\label{alg:pseudoalignviolating}
\SetKwInOut{Input}{input}
\SetKwInOut{Output}{output}
\SetKwProg{AlignTraceAndModel}{AlignTraceAndModel}{}{}
\AlignTraceAndModel{$(\sigma=\langle \sigma_1,\ldots,\sigma_n\rangle\in \Sigma^*,\Mp,c, C)$}{
  \Input{A trace $\sigma$; a mixed-paradigm model $\Mp=(\WfN,\Dp)$; a partial cost function $c:(T^{\gg}\times A^{\gg}) \to \mathbb{R}^+$; a partial cost function $C: K \to \mathbb{R}^+$}
  \Output{An alignment $\gamma$}
  $i \gets 1$  \tcp*{$i$ keeps track of the index in the trace}
  $(\mu,s) \gets (\mu_0,S^K_0)$ \tcp*{$(\mu,s)$ keeps track of the state in the Workflow net and $s$ the state of each automaton}
  $g \gets 0 $ \tcp*{$g$ keeps track of the cost so far}
  $h \gets \mathrm{estimateRemainingCost}(\sigma, \WfN, i,(\mu,s), c) $ \tcp*{$h$ underestimates the remaining cost}
  $p \gets \mathrm{NULL}$  \tcp*{$p$ stores the predecessor of the current state}
  $Q \gets \{\}$  \tcp*{$Q$ is a queue of states to investigate}
  $q \gets (i,(\mu,s),g,h,p)$ \tcp*{$q$ is the current head of the queue}
  \While(\tcp*[f]{\parbox[t]{4cm}{\raggedright Stop if all constraints are permanently violated or satisfied}}){$i \leq n+1\,\vee\, \mu \neq \mu_\textrm{F}\, \vee\, \forall_{\kappa \in K} s^\kappa \in s^\kappa_F \cup \{s^\kappa_{\textrm{na}}\} $}{ \label{algline:stopsatisfiedorviolated}
    \If(\tcp*[f]{Special case if some automaton is temporarily violated}){$i = n+1\,\wedge\, \mu = \mu_\textrm{F}$} { \label{algline:specialcase}
      $s' \gets \{\}$, $g' \gets 0$\;
      \For(\tcp*[f]{Iterate over all Automata}){$\kappa \in K$}{
        \If{$s^\kappa \in S^\kappa_{F},\vee\, s^\kappa=s^\kappa_{\textrm{na}} $}{
          $s' \gets s' \cup \{s^\kappa\} $ \tcp*{keep satisfied and permanently violated states}
        } 
        \Else{
          $s' \gets s' \cup \{s^\kappa_{\textrm{na}}\} $ \tcp*{Change temporarily violated to permanently violated}
          $g' \gets g' + C(\kappa)$ \tcp*{Update the cost of permanently violating this constraint}
        }
      }
      $Q \gets \mathrm{enqueue}((i,(\mu,s'),g',0,q))$  \tcp*{Add final state to the queue (note $h=0$)}
    }
    \Else {
      $T_{\textrm{enabled}}\gets \mathrm{getEnabledTransitionsPN}(\WfN,\mu)$\;
      \For{$t\in T_{\textrm{enabled}}$}{
        $\mu' \gets \mu-\bullet t+t\bullet$  \tcp*{Compute next marking in the net}
        $s' \gets \{ \delta^\kappa(s^\kappa,t) \mid \kappa \in K\} $ \tcp*{Compute next state for all automata}
        \If(\tcp*[f]{Compute next state for synchronous move}){$l(t)=\sigma_i$}{ 
          $g' \gets g + c(t,\sigma_i) + \sum_{\kappa \in K,  s^\kappa \not = s^\kappa_{\textrm{na}} \wedge s'^\kappa = s^\kappa_{\textrm{na}}} C(\kappa) $ \label{algline:costupdate1} \tcp*[f]{\parbox[t]{2.5cm}{\raggedright Add cost for all permanently violated constraints}}
          $h' \gets \mathrm{estimateRemainingCost}(\sigma, \WfN, i',(\mu',s'), c) $\;
          $Q \gets \mathrm{enqueue}((i+1,(\mu',s'),g,h,q))$ \tcp*{Add new state to the queue}
        } 
        $g' \gets g + c(t,\gg) + \sum_{\kappa \in K,  s^\kappa \not = s^\kappa_{\textrm{na}} \wedge s'^\kappa = s^\kappa_{\textrm{na}}} C(\kappa) $ \label{algline:costupdate2} \tcp*[f]{\parbox[t]{3cm}{\raggedright Add cost for all permanently violated constraints}}
        $h' \gets \mathrm{estimateRemainingCost}(\sigma, \WfN, i,(\mu',s'), c) $\;
        $Q \gets \mathrm{enqueue}((i,(\mu',s'),g',h',q))$  \tcp*{Add new state to the queue}
      }
      \If(\tcp*[f]{Compute next state for move on log}){$i \leq n$}{
        $g' \gets g + c(\gg,l(\sigma_i)) $\;
        $h' \gets \mathrm{estimateRemainingCost}(\sigma, \WfN, i+1,(\mu,s), c) $\;
        $Q \gets \mathrm{enqueue}((i+1,(\mu,s),g',h',q))$  \tcp*{Add new state to the queue}
      }
    }
    $(i,(\mu,s),g,h,p) \gets \mathrm{pullNextBestState}(Q)$ \tcp*{Pull next state to investigate from the queue}
    $q \gets (i,\mu,g,h,p)$\;
  }    
  $\gamma \gets \mathrm{extractAlignment}(i,(\mu,s),g,h,p)$\;
  \KwRet{$\gamma$}\;
} \end{algorithm2e}

\Cref{alg:pseudoalignviolating} shows the final algorithm that allows for the violation of constraints while making alignments. An additional parameter $C$ includes the cost for permanently violating a constraint at the end of the alignment. The part after Line~\ref{algline:specialcase} updates any automaton that is in a state of temporary violation to a permanent violation, updates the cost, and re-queues the state. Again, it is trivial to see that this algorithm produces optimal alignments. The underestimation function is correct for the model without constraints, and when we allow constraints to be violated, the language of the model does not change when adding constraints. Furthermore, the additional costs for violating constraints does not break the monotonicity of the heuristic required for the A$^\star$ algorithm to be correct~\cite{adriansyahreplay}.

It is important to notice that this algorithm introduces two new move types, namely moves on model and synchronous moves that violate a constraint -- as opposed to the existing moves on model and synchronous moves that do not violate constraints. Because the constraints can now be permanently violated, the size of the search space for A$^\star$ also becomes larger. In the worst case (if all constraint automata are defined on disjoint sets of transitions), the size of the search space corresponds to the product of the sizes of the constraint automata and of the state space of the Workflow net. However, in practice, this rarely leads to problems, as we will see in \cref{sec:evaluation}.

Let us again consider the example of before ($\langle\tasko,\taskp,\taski,\tasks,\taskeu,\taskm,\taskn\rangle$). Again, assume we have reached a state in the search where we have explained the fitting prefix of the sequence (i.e., we executed transitions $t_1/\Task{o}$, $t_{2}/\Task{m}$, $t_{4}/\Task{i}$, $t_{6}/\Task{s}$ and $t_{5}/\Task{\EUR}$ in the model). The next event to explain is $\Task{Return money}$. Recall that the marking is 
$\{ p_6^1, p_7^1 \}$
and the automata are in their initial state, except for the automata of $\kappa_8$ and $\kappa_9$ in states $s^{\kappa_8}_1$ and $s^{\kappa_9}_1$, respectively. 
Transition $t_{13}/\Task{m}$ is enabled according to the workflow model. If we executed it, automaton $\kappa_4$ would transition from state $s^{\kappa_4}_0$ to $s^{\kappa_4}_{\textrm{na}}$, which is a permanently violated state. For the sake of argument, let us assume all violation cause an additional cost of \num{1} to be added to the total, i.e., violating a constraint incur a partial cost of \num{1}, just like a move on model or a move on log. In Line~\ref{algline:costupdate1}, the cost of permanently violating constraint $\kappa_4$ is added to the cost of that synchronous move. The new state is put in the queue for further investigation. 
\\
The next event to explain would be the event \Task{Receive cancellation}. Its execution is allowed as it would not violate any of the constraints that are not already permanently violated. The algorithm will therefore explore this option. After executing $t_{11}/\Task{n}$, it is no longer possible to execute $t_{8}/\Task{c}$ without reaching a state of permanent violation, due to $\kappa_5$. To terminate the run, this forces the execution of $t_{9}/\Task{r}$, which, in turn, through $\kappa_7$, requires $t_{12}/\Task{w}$ to happen first. This would lead to an alignment $\gamma^{opt}=\langle(t_1,o),(t_2,p),(t_4,i),(t_6,s),(t_5,\Task{\EUR}),(t_7,\gg),(t_{12},m),(t_{11},n),(t_{13},\gg),(t_9,\gg)\rangle$ with three deviations, namely a synchronous move violating a constraint on $t_{13}/\Task{m}$ and two moves on model on $t_{12}/\Task{w}$ and $t_{9}/\Task{r}$. 
\\
We already know that there is an alignment with three deviations, but this alignment is equally optimal. If the cost of violating a constraint is selected to be lower than the cost of a move on model or log, this alignment would become the only optimal alignment.

Both algorithms presented in this section rely on the original estimation function for alignments, which ignores the declarative constraints. While this is provably correct from an algorithmic point of view, it implies that the part of the search space that needs to be visited by the algorithm increases. In the worst case, the model contains no places and only declarative constraints. Such a model would yield the largest possible search space and cause the performance of the A$^\star$ algorithm to degrade to the level of Dijkstra's shortest path algorithm~\cite{dijkstra}. 

\subsection{Calculating Fitness}
Fitness is a process quality measure that assesses the extent to which a process model explains the executions recorded in the event log~\cite{Aalst.etal/WIRev2012:ConformanceCheckingAndMetrics,DBLP:books/sp/CarmonaDSW18}. In particular, it considers the ability of the model to replay the traces. The measure is expressed as the fraction of the log behavior that is admitted by the model.
As our approach is built upon alignments for conformance checking, we apply the usual calculation for fitness proposed in~\cite{adriansyah2011conformance}. Notice, indeed, that the adaptations in the concepts of runs for mixed-paradigm models, as well as in the cost assignments to cater for possible violations of the declarative part of the mixed-paradigm, are compatible with the standard definition of alignment. We calculate fitness, in particular, as follows.

Given a trace $\sigma = \langle \sigma_1, \ldots, \sigma_n \rangle$ of length $n \in \mathbb{N}$ and a mixed-paradigm model $\Mp$, let $\gamma^{\textrm{opt}}_{\sigma,\Mp}$ be the optimal alignment of $\sigma$ and $\Mp$ as per \cref{def:alignment:optimal}. Let $\gamma^{\textrm{worst}}_{\sigma,\Mp}$ be the worst-case alignment, calculated as the cost of the optimal alignment of the empty trace $\langle\rangle$ and $\Mp$, plus the cost of a sequence of moves on log for every $\sigma_i \in \sigma$ ($1 \leqslant i \leqslant n$). The \emph{trace fitness} $\textrm{fitness}(\sigma,\Mp)$ is calculated as follows:
\begin{equation}
	\textrm{fitness}(\sigma,\Mp) = 1 -
	\frac{ \textrm{cost}\left( \gamma^{\textrm{opt}}_{\sigma,\Mp} \right) }
		{ \textrm{cost}\left( \gamma^{\textrm{worst}}_{\sigma,\Mp} \right) }.
\end{equation}
\noindent
For example, consider again the example trace seen above ($\langle\tasko,\taskp,\taski,\tasks,\taskeu,\taskm,\taskn\rangle$)
and the process model of \cref{fig:mpexample} and a unitary cost for moves on model, moves on log, and constraint violations.
As shown in \cref{subsec:satisfied,subsec:violating}, the optimal alignment is associated to a cost of \num{3}.
The cost of the worst-case alignment is the sum of \num{3} (the full alignment with an empty trace requires three moves on model, corresponding to the firing sequence
$\langle t_1, t_2, t_{10} \rangle$)
and \num{7} (corresponding to a move on log for every event in the trace).
The trace fitness is thus
$ 1 - \frac{3}{10} = \num{0.7} $.

In the next section, we show how the technique performs on a number of real-life event logs on which we discover mixed-paradigm models using an available mixed-paradigm discovery technique.
\section{Evaluation}
\label{sec:evaluation}
%
%
%
In this section, we evaluate our mixed-paradigm conformance checking approach on three real-life event logs. 
We implemented and integrated our algorithm in the ProM environment.\footnote{\url{http://www.promtools.org}}
The tool can be found in the ProM nightly build under the name \emph{MixedParadigm}.

In the following, we introduce the event logs and discovery algorithm used.
Next, we provide a qualitative analysis of one of the logs to show the capabilities of the mixed-paradigm conformance checking approach.
Finally, we mine different mixed-paradigm models mined from each log and evaluate them by making use of our alignment-based algorithm.

\subsection{Event Logs}
\begin{table}[tb]
	\caption{Overview of the event logs used for evaluation.}
	\label{tab:bpic:overview}%
	\centering
	\scriptsize
	\begin{tabular}{r S S S}
		\toprule
		\textbf{Event log} & \textbf{No.\ activities} & \textbf{No.\ traces} & \textbf{Avg.\ trace length}\\ \midrule
		BPI 2012    & 24 	& 13,087 	& 20.04 \\ \grayrow
		BPI 2013 - incidents   & 13 	& 7,554		& 8.68	\\
		SEPSIS		& 16	& 1,050	& 14,49	\\ \grayrow
		\bottomrule
	\end{tabular}%
\end{table}%
The process mining literature has provided and discussed a wide range of exemplary, real-life event logs.
In this paper, we present the results of our experiments on three such logs to test the conformance checking algorithm for its performance. Two of those are part of the Business Process Intelligence Challenge (BPIC) collection. In addition, we consider an extra event log (Sepsis). We chose those logs as the parameters of the discovery algorithm used in subsequent experiments yield a variety of different models, thereby serving as a good illustration of the applicability of the replaying algorithm.
The selection offers difference blends of the main characteristics of event logs, i.e., size, number of distinct activities, and trace length.
\Cref{tab:bpic:overview} provides an overview of the main characteristics of the logs used in this study:
\begin{compactdesc}
\item[BPIC 2012:]
The BPIC 2012 log%
\footnote{\url{https://doi.org/10.4121/uuid:3926db30-f712-4394-aebc-75976070e91f}}
(henceforth, \textbf{BPI12} for short)
contains the application process for personal loans and credit lines at a Dutch financial institution.
It consists of three subprocesses, each pertaining to a different part of the organization, i.e., the (A)pplications, the (O)ffers, and the (W)orkflow system.
Customers submit an application for a loan through an online system. These applications are then fed into a workflow system, where call-agents can pick up work items to either assess the application or to contact the customer with offers. Eventually, every application should either be declined, cancelled or approved and activated. 
%
\item[BPIC 2013:]
The BPIC 2013 log
contains traces from an incident management system at Volvo IT Belgium.
It contains a smaller range of activities, with more repetitive structures compared with the BPI 2012 log.
Three different logs were made available. For the evaluation, we use the one that focuses on the separately raised incidents.%
\footnote{\url{https://doi.org/10.4121/uuid:500573e6-accc-4b0c-9576-aa5468b10cee}}
We will indicate this log as \textbf{BPI13} for short in the remainder of the paper.
\item[Sepsis:]
The Sepsis log%
\footnote{\url{https://doi.org/10.4121/uuid:915d2bfb-7e84-49ad-a286-dc35f063a460}} 
pertains to the treatment of sepsis in a hospital setting and contains various activities stemming from an ERP system supporting the medical as well as administrative tasks.
\end{compactdesc}
We performed similar experiments on other event logs as well. However, the variety of models generated was lower.
The interested reader can find the results at \url{https://github.com/JohannesDeSmedt/mixed-paradigm-conformance-checking}.


\subsubsection{Methodology}
To obtain mixed-paradigm process models, we used the FusionMINERful discovery algorithm \cite{de2015fusion,desmedt2016modelchecking}.
FusionMINERful is capable of retrieving models with different blends of procedural and declarative model constructs, determined by the user-controlled parameter activity entropy ranging from \SI{0}{\percent} to \SI{100}{\percent}.
Using this parameter, it becomes possible to indicate the estimated level of non-procedural behavior, which results in more activities being mined for constraint-based relationships, rather than them being incorporated in the procedural part of the model.
To obtain the Petri-net part of the model, we use the block-structured approach of Inductive Miner~\cite{leemans2013discovering}. We use MINERful~\cite{DiCiccio2015} to connect activities in the declarative model, and to link them with the procedural model.
MINERful employs the concepts of support and confidence to mine constraints from an event log. Support is the proportion of traces that comply with the constraint. 
Confidence is the proportion of satisfying traces in which the activation of the constraint occurred.
The higher the support, the lower the number of constraints, but the higher trace fitness is to be expected, as more traces will allow for the behavior of the constraints.
Next to a varying level of entropy, a varying level of support and confidence will impact the amount and overall applicability of constraints in the mixed-paradigm model.
Not every distinct combination of these parameters yields a different result.
In particular, we report on the results of applying FusionMINERful setting support and confidence levels between \num{0.5} and \num{1.0}, and entropy level between \num{0.0} and \num{1.0}.

For each event log, we display an overview of the different levels of entropy, support, and confidence which churned out a unique model with FusionMINERful, as well as how many transitions and places that are uniquely present in the Workflow net (i.e., that do not undergo any constraint), and the number of constraints in the declarative model.
The evaluation criteria are the final trace fitness level, the calculation time, and the average number of constraints violated per trace (if violations are allowed).

We run the ProM environment on an Intel Xeon E3-1230v5 CPU at 3.4GHz over 8 threads, allowed to use 24GB of memory.
Our implemented algorithm adhered to a cost for non-synchronous moves of \num{1}, and a cost of violating a constraint of \num{1}.
Replaying is performed by both allowing and disallowing constraint violations: we report the results on trace fitness separately, according to the adopted strategy.

\subsection{Qualitative Evaluation}
\label{sec:eval:quali}
As discussed, mixed-paradigm models excel at capturing different layers of flexibility in event logs.
The procedural part of the model represents the more fixed workflows, while the activities that are 
subject to declarative constraints allow for more intricate behavior.
An interesting aspect of adding constraints to procedural models, indeed, 
is the introduction of an immediate insight into this more intricate behavior. 

\begin{figure}[tb]%
	\centering
	\includegraphics[width=\linewidth]{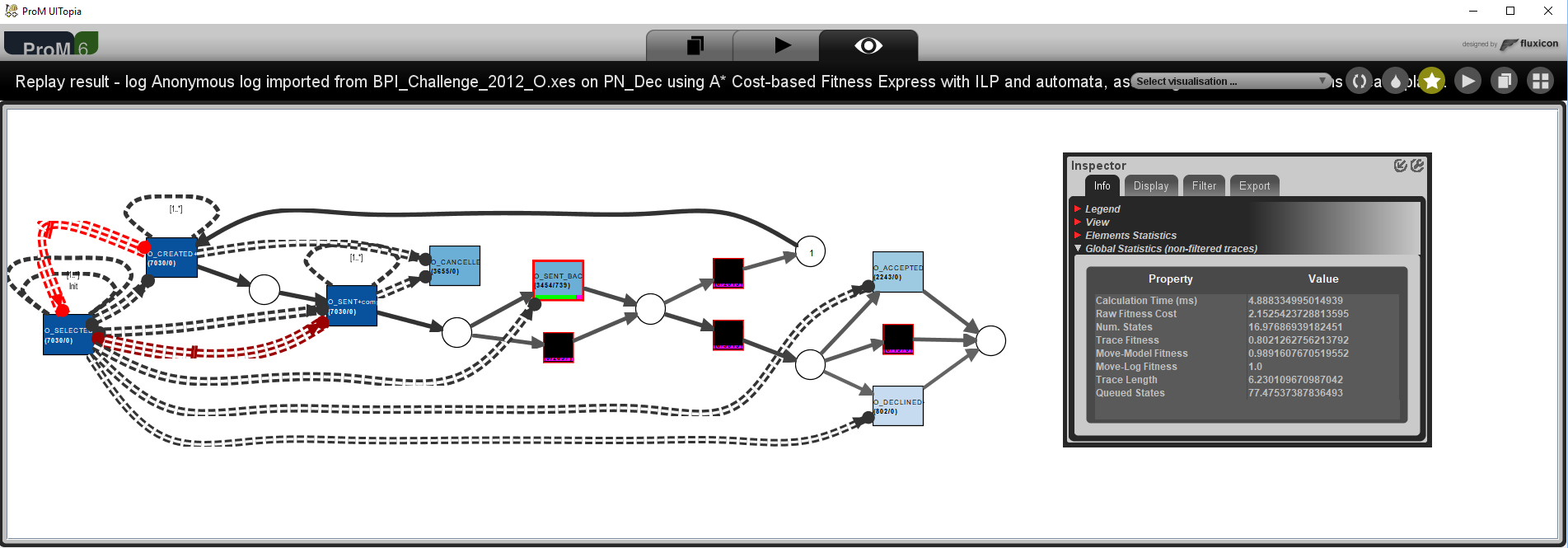}
	\caption{Screenshot of the ProM implementation showing the result of the replay on the offer subprocess in the BPI12 log.}
	\label{fig:screenshot}
\end{figure}
The incorporation of the constraint checking in the replay provides a visual representation of which constraints are violated during the execution.
This is illustrated in \cref{fig:screenshot}. Violated constraints are drawn in red. Additionally, transitions are decorated on the bottom side with lines that denote the occurrence of moves on model (purple).
In order to get a reasonable visual representation in terms of model size (as measured in the number of model constructs), the offer-making subprocess of the BPI12 log is mined for a mixed-paradigm model with a conservative entropy level (\SI{40}{\percent}) and high support and confidence (\SI{100}{\percent}) to illustrate the constraint-checking capabilities of the algorithm.
It can be seen that the {\NotChaSuccTmp} constraints are violated during execution.
This comes as no surprise, given that there is little possibility to circumvent these constraints during the replay without introducing many moves on model.
Besides, all other constraints end in an accepting state. 
Nevertheless, fitting the \Task{O\_SENT\_BACK} activity into the alignments requires many moves on log, which lower the overall trace fitness score to \SI{80.21}{\percent}.

Besides the visual representation of the model, our software shows the detailed information on the fitness scores on the right hand side, as it can be seen in \cref{fig:screenshot}. 
A detailed overview of the alignment steps is also available, highlighting the moves that introduce constraint violations, moves on model, and moves on log.

\subsection{Quantitative Evaluation}
\label{sec:eval:quant}
\begin{figure}[tb]
	\centering
	\begin{subfigure}{0.6\textwidth}
		\includegraphics[width=\linewidth]{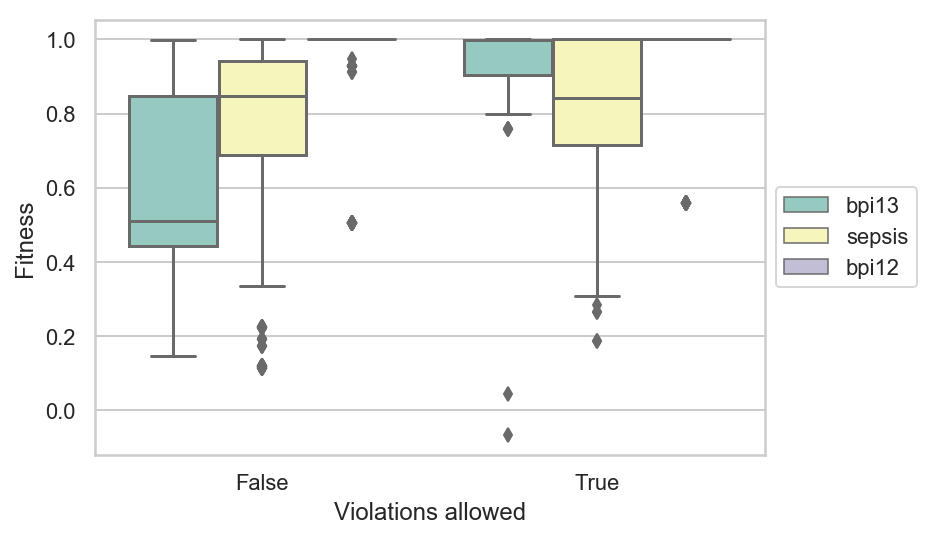}
		\caption{Trace fitness}
	\end{subfigure}
	\begin{subfigure}{0.6\textwidth}
		\includegraphics[width=\linewidth]{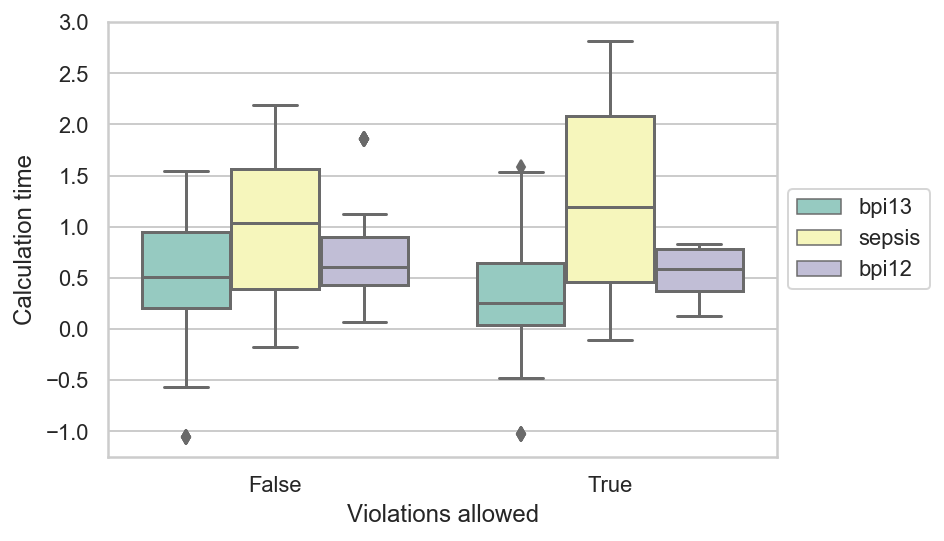}
		\caption{Computation time (in log(s))}
	\end{subfigure}
	\caption{Overview of the results of the mixed-paradigm conformance checking algorithm for all event logs.}
	\label{fig:main_results}
\end{figure}
\begin{table}[tb]
	\caption{Overview of the results for a 100\% procedural model mined with Inductive Miner at a 20\% noise level threshold.}
	\label{tab:proc_baseline}
	\scriptsize
	\centering
	\begin{tabular}{r S S S S}
		\toprule
		\textbf{Event log} & \textbf{Trace fitness} & \textbf{Comp. time} & \textbf{No.\ transitions} & \textbf{No.\ places} \\\midrule
		BPI12 & 0.948 & 5.916 & 40    & 19 \\ \grayrow
		BPI13 & 0.819 & 1.629 & 25    & 14 \\
		Sepsis & 1.000 & 2.682 & 23    & 9 \\
		\bottomrule
	\end{tabular}%
\end{table}
The main goals of trace-based conformance checking is to compute a fitness score for mixed-paradigm models over which event logs are replayed, and to gain further information regarding the events that could not be executed (or constraints that could not be satisfied) in accordance with the underlying model.
In this section, we illustrate how we retrieve this information from numerous models mined and replayed on the event logs.

\begin{figure}[p]
\centering
\begin{subfigure}{\textwidth}\centering
		\resizebox{\linewidth}{!}{
		\includegraphics[width=\linewidth]{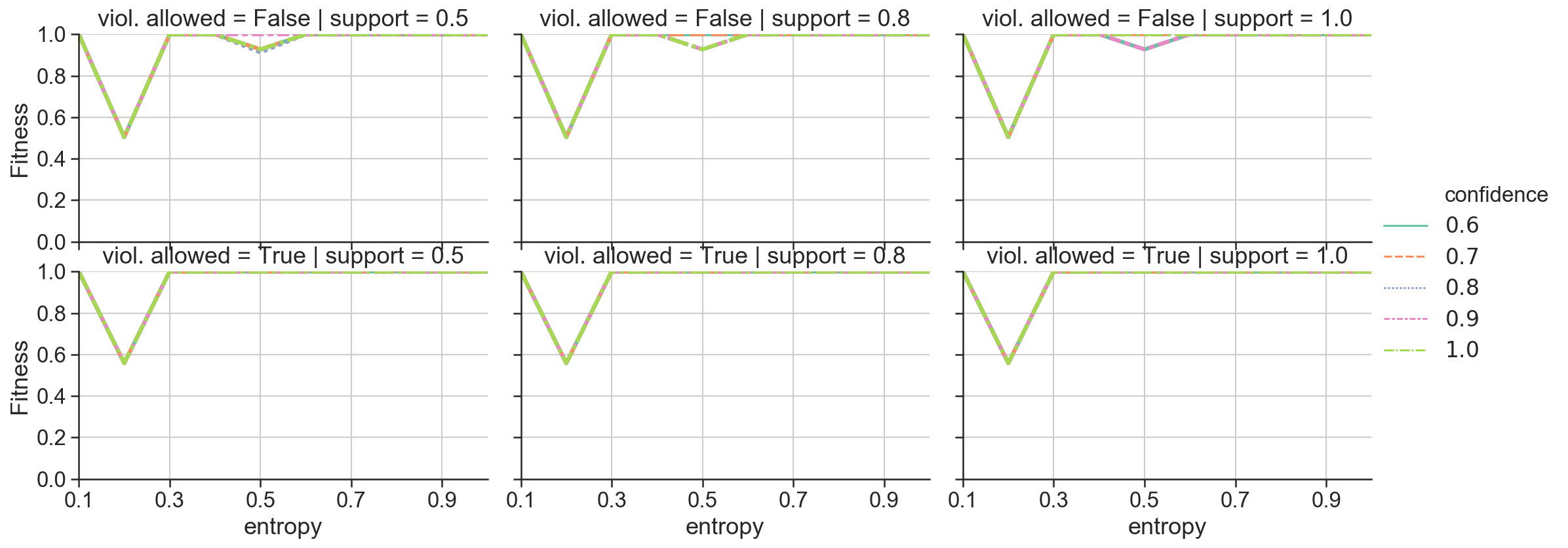}
		}
    \caption{Fitness}
\end{subfigure}\\%
\begin{subfigure}{\textwidth}\centering
		\resizebox{\linewidth}{!}{
		\includegraphics[width=\linewidth]{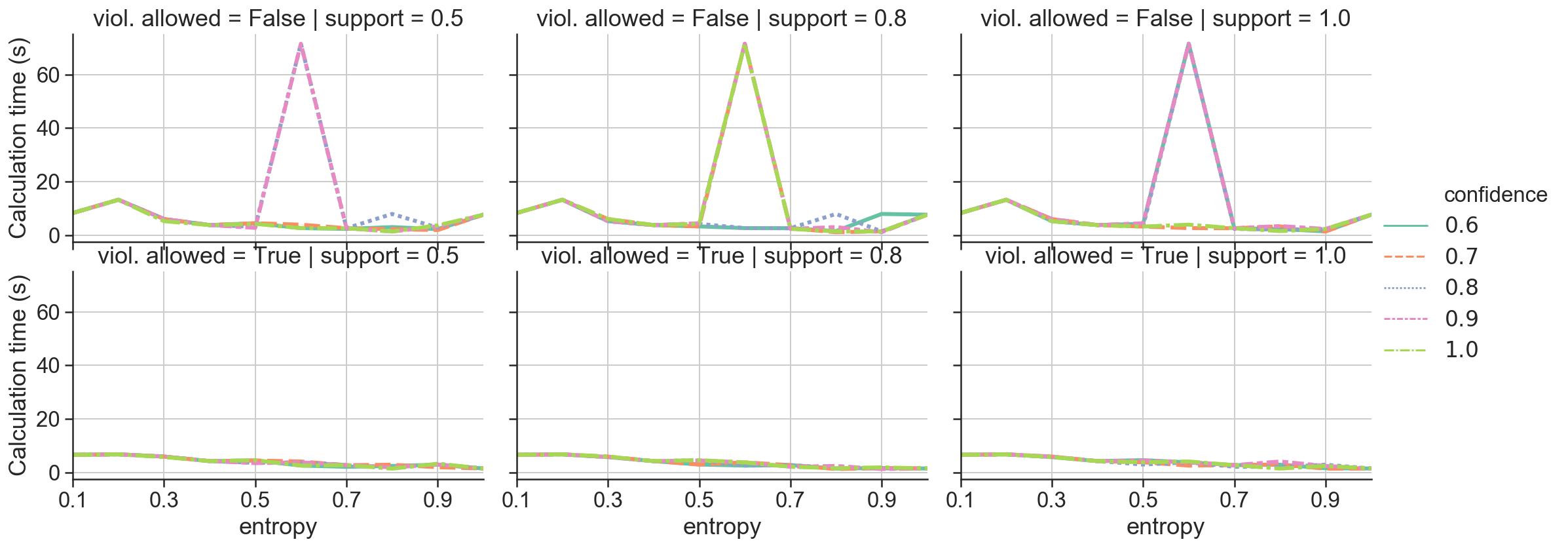}
		}
    \caption{Execution time}
\end{subfigure}\\%
\begin{subfigure}{\textwidth}\centering
		\resizebox{\linewidth}{!}{
		\includegraphics[width=\linewidth]{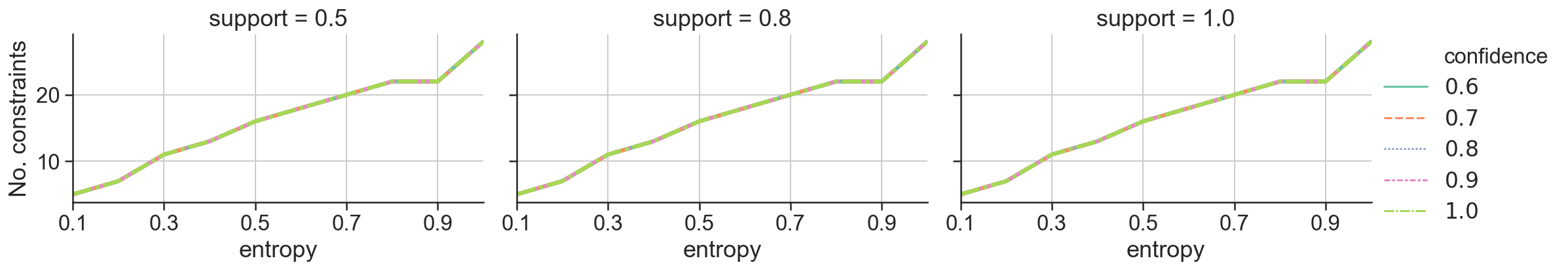}
		}
    \caption{Constraints}
\end{subfigure}\\%
\begin{subfigure}{\textwidth}\centering
		\resizebox{\linewidth}{!}{
		\includegraphics[width=\linewidth]{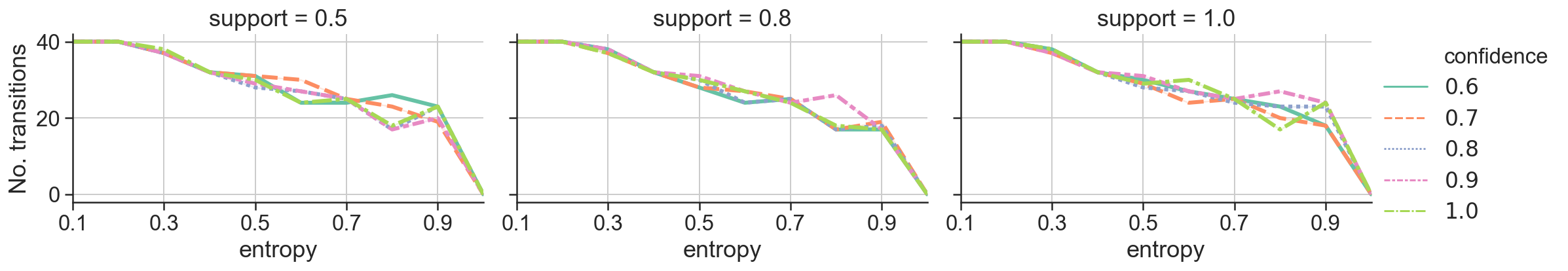}
		}
    \caption{Transitions}
\end{subfigure}\\%
\begin{subfigure}{\textwidth}\centering
		\resizebox{\linewidth}{!}{
		\includegraphics[width=\linewidth]{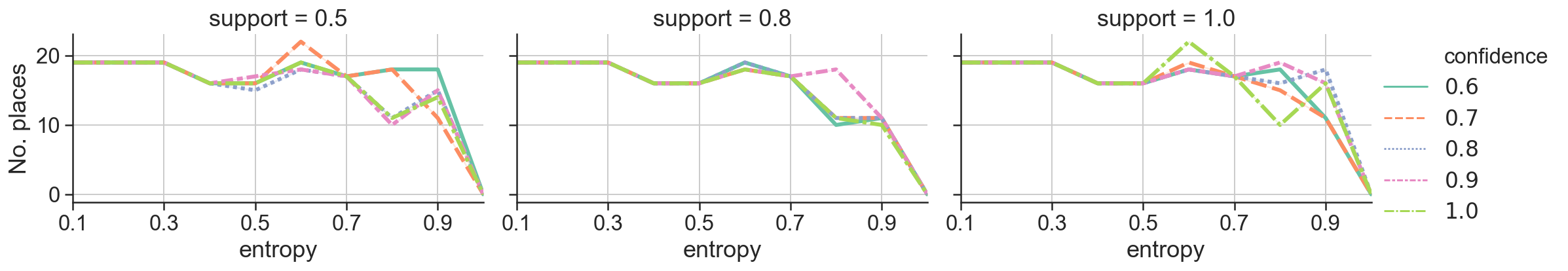}
		}
    \caption{Places}
\end{subfigure} 	\caption{Overview of the results for the BPI12 log.}
	\label{fig:bpi12_res}
\end{figure}

\begin{figure}[tbp]
	\centering
\begin{subfigure}{\textwidth}
		\resizebox{1.0\linewidth}{!}{
		\includegraphics[width=\linewidth]{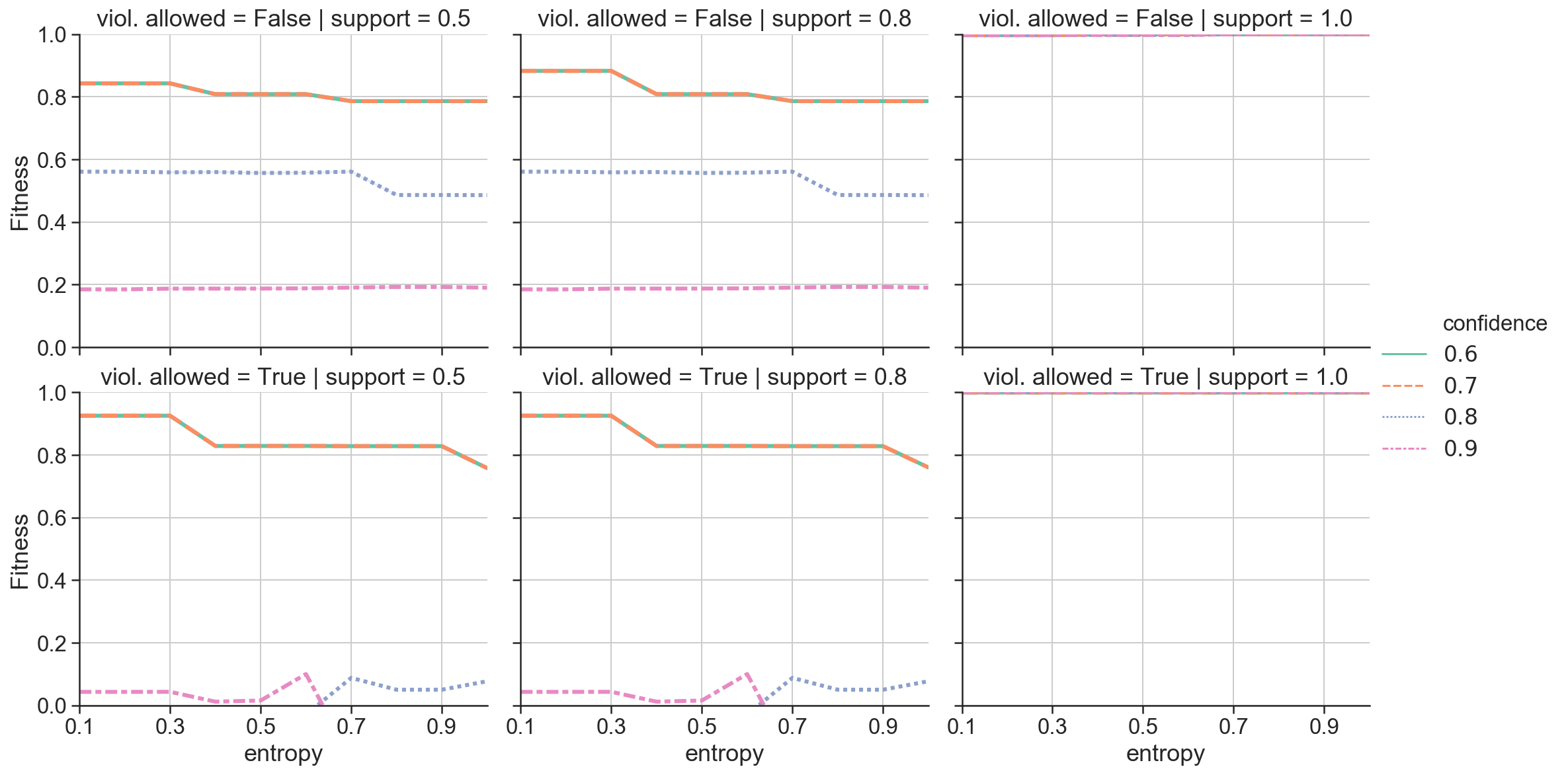}
		}
    \caption{Fitness}
\end{subfigure}
\begin{subfigure}{\textwidth}
		\resizebox{1.0\linewidth}{!}{
		\includegraphics[width=\linewidth]{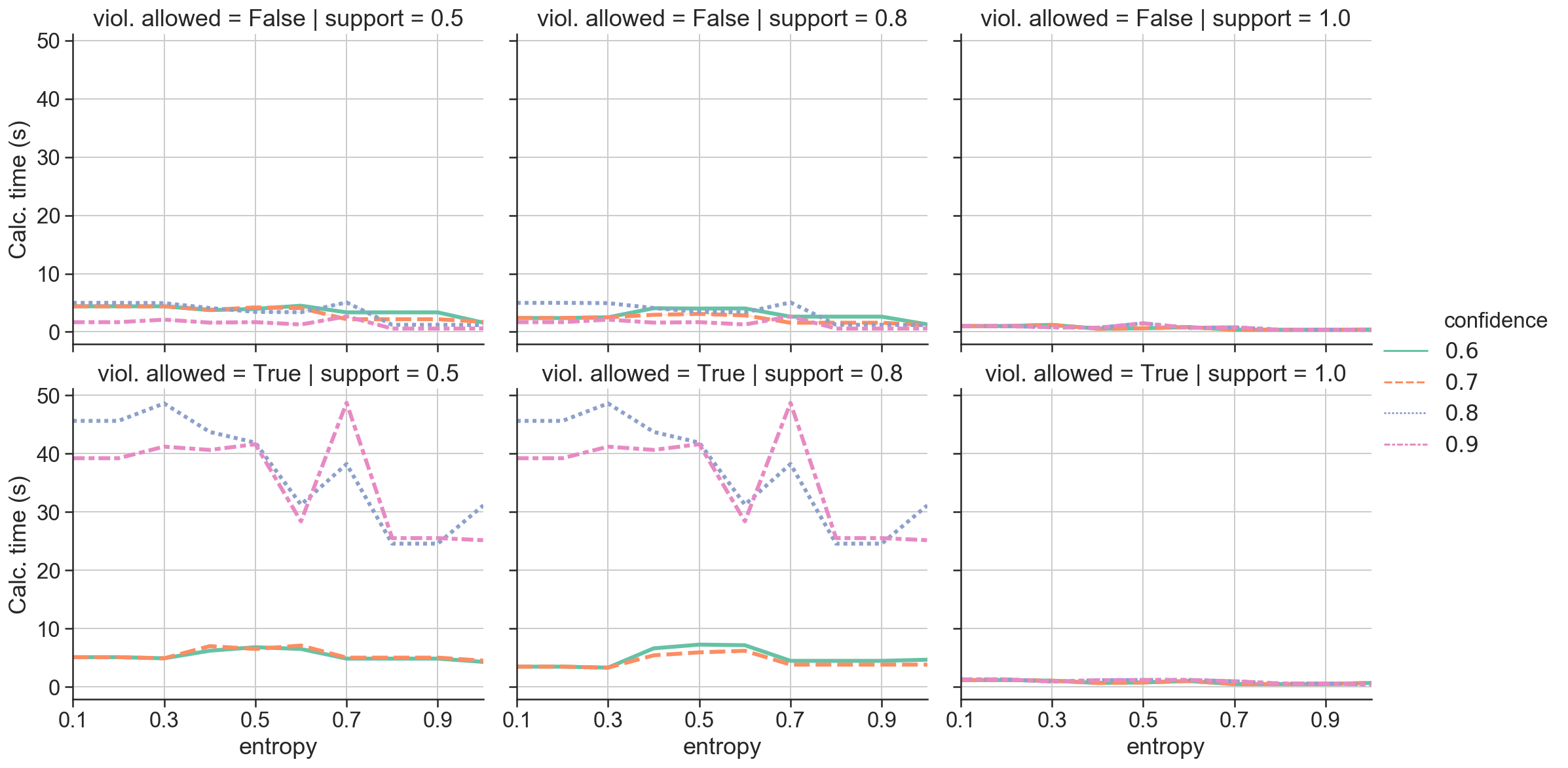}
		}
    \caption{Execution time}
\end{subfigure} 	\caption{Overview of the results for the BPI13 log (i).}
	\label{fig:bpi13_res1}
\end{figure}

\begin{figure}[tbp]
	\centering
\begin{subfigure}{\textwidth}
	\resizebox{1.0\linewidth}{!}{
		\includegraphics[width=\linewidth]{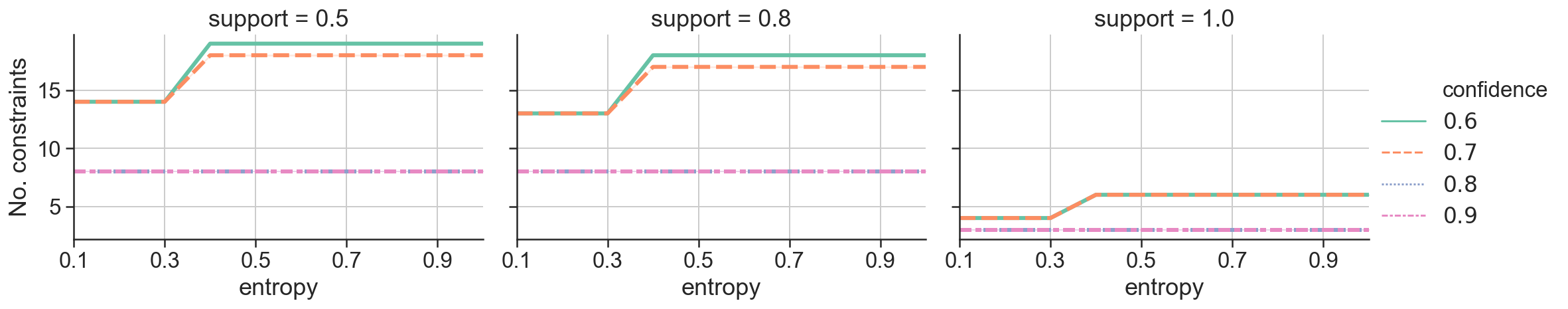}
	}
	\caption{Constraints}
\end{subfigure}
\begin{subfigure}{\textwidth}
	\resizebox{1.0\linewidth}{!}{
		\includegraphics[width=\linewidth]{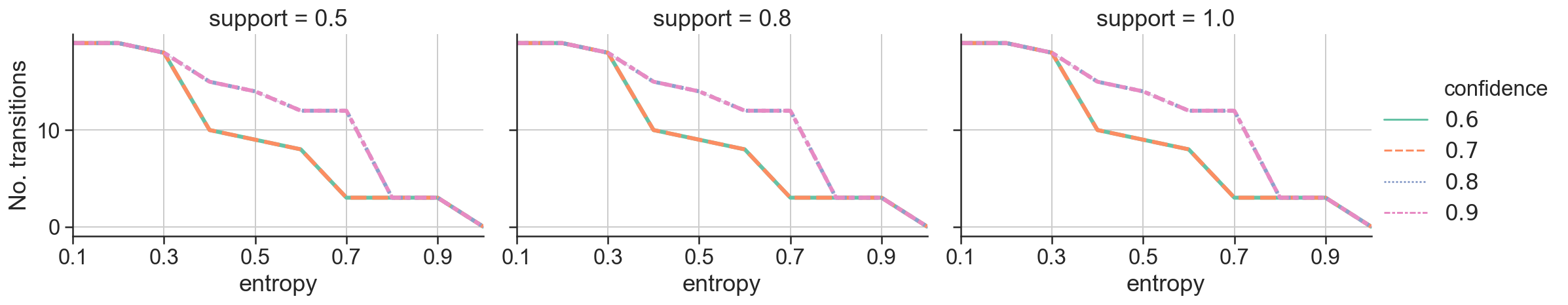}
	}
	\caption{Transitions}
\end{subfigure}
\begin{subfigure}{\textwidth}
	\resizebox{1.0\linewidth}{!}{
		\includegraphics[width=\linewidth]{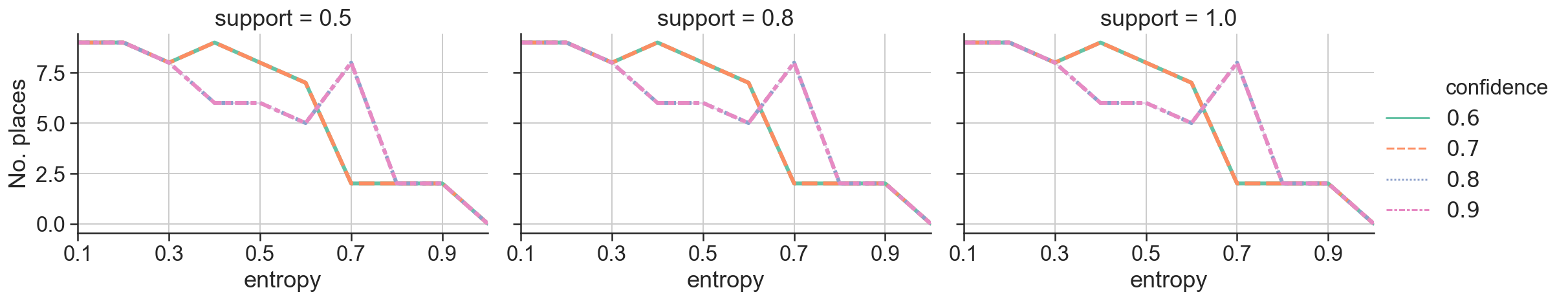}
	}
	\caption{Places}
\end{subfigure}
\begin{subfigure}{\textwidth}
	\resizebox{1.0\linewidth}{!}{
		\includegraphics[width=\linewidth]{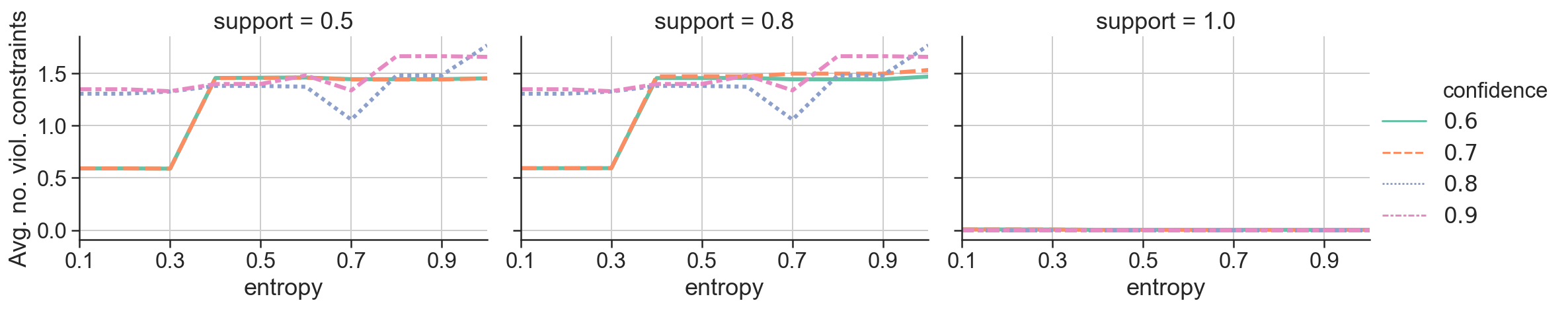}
	}
	\label{fig:bpi13_viol}
	\caption{Constraint violations}
\end{subfigure} 	\caption{Overview of the results for the BPI13 log (ii).}
	\label{fig:bpi13_res2}
\end{figure}

\begin{figure}[tbp]
	\centering
\begin{subfigure}{\textwidth}
		\resizebox{1.0\linewidth}{!}{
		\includegraphics[width=\linewidth]{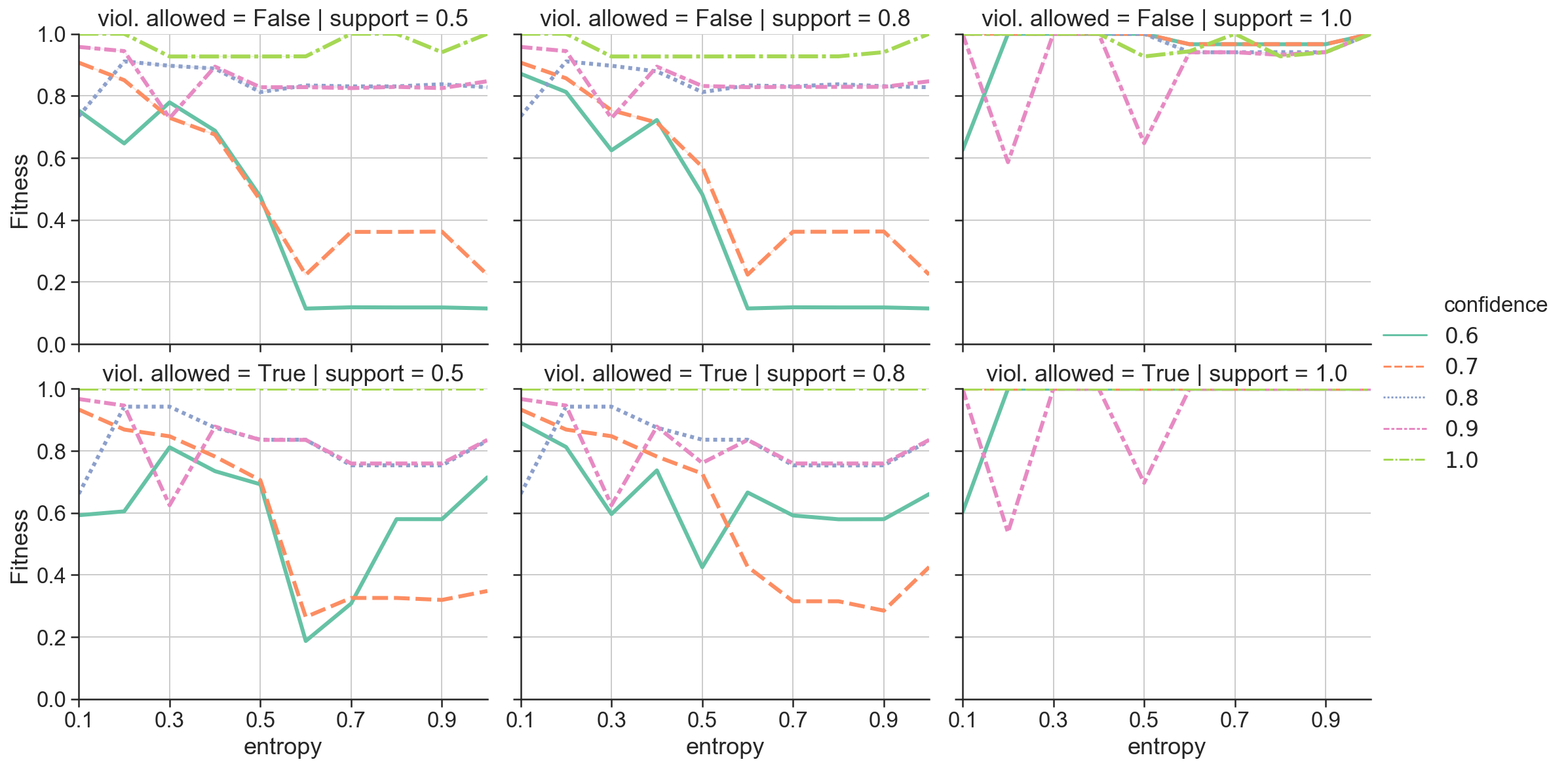}
		}
    \caption{Fitness}
\end{subfigure}
\begin{subfigure}{\textwidth}
		\resizebox{1.0\linewidth}{!}{
		\includegraphics[width=\linewidth]{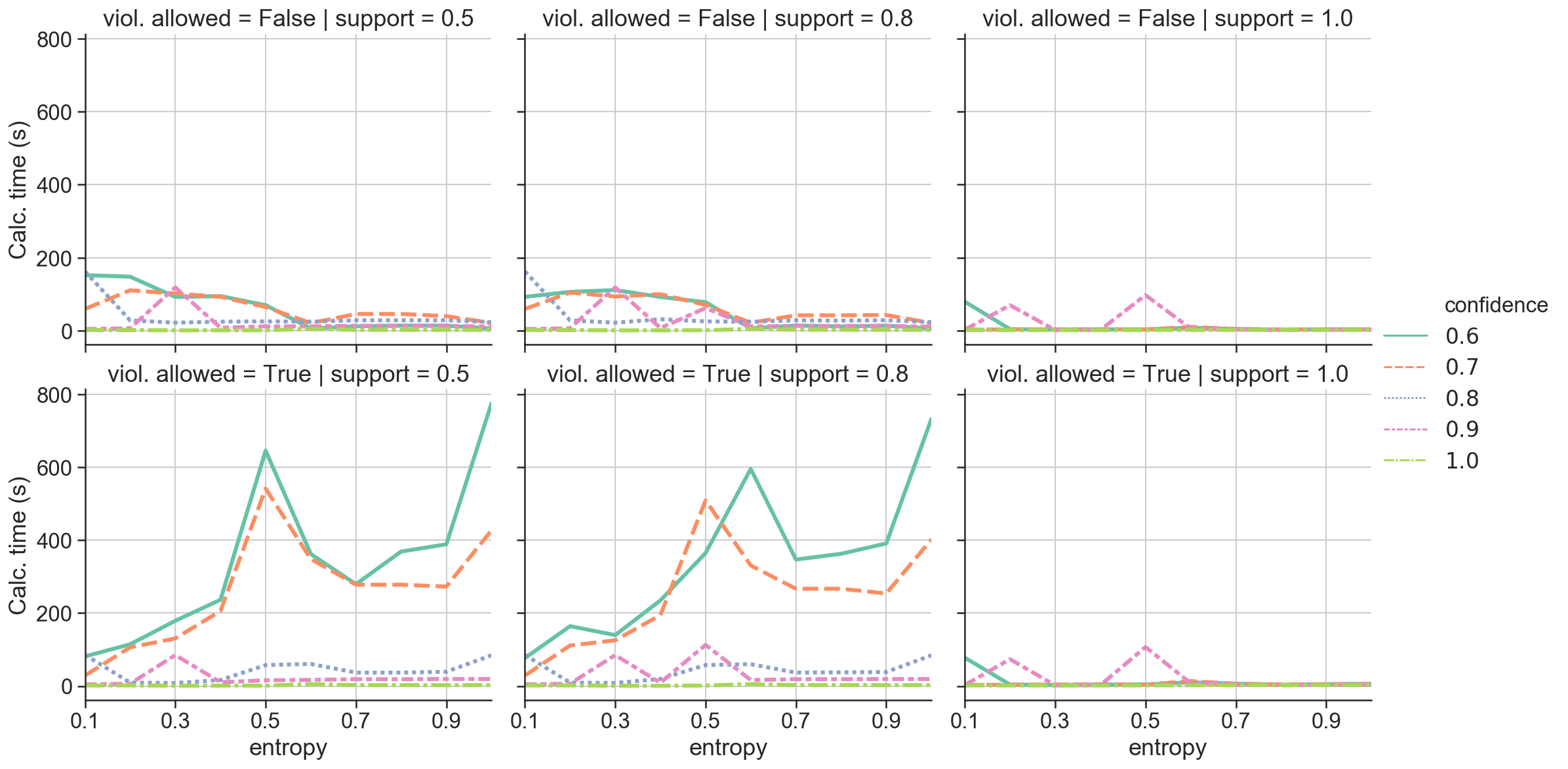}
		}
    \caption{Execution time}
\end{subfigure} 	\caption{Overview of the results for the Sepsis log (i).}
	\label{fig:sepsis_res1}
\end{figure}

\begin{figure}[tbp]
	\centering
\begin{subfigure}{\textwidth}
	\resizebox{1.0\linewidth}{!}{
		\includegraphics[width=\linewidth]{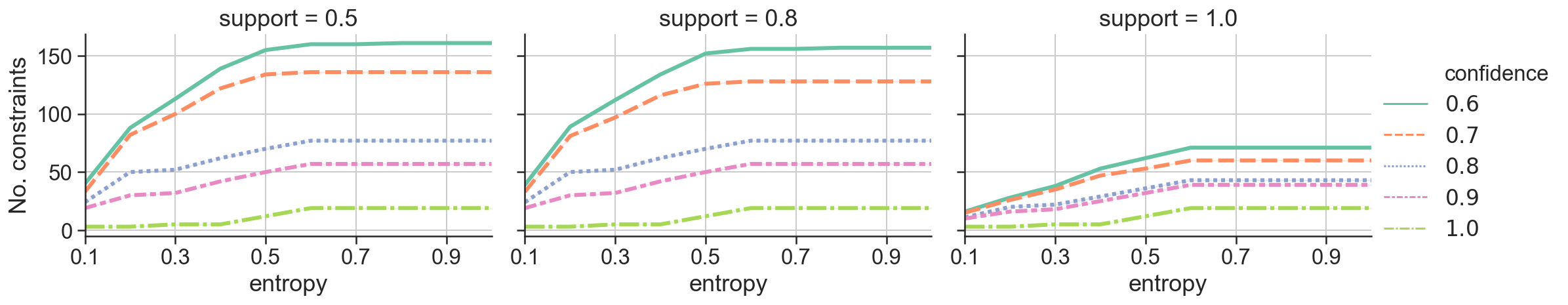}
	}
	\caption{Constraints}
\end{subfigure}
\begin{subfigure}{\textwidth}
	\resizebox{1.0\linewidth}{!}{
		\includegraphics[width=\linewidth]{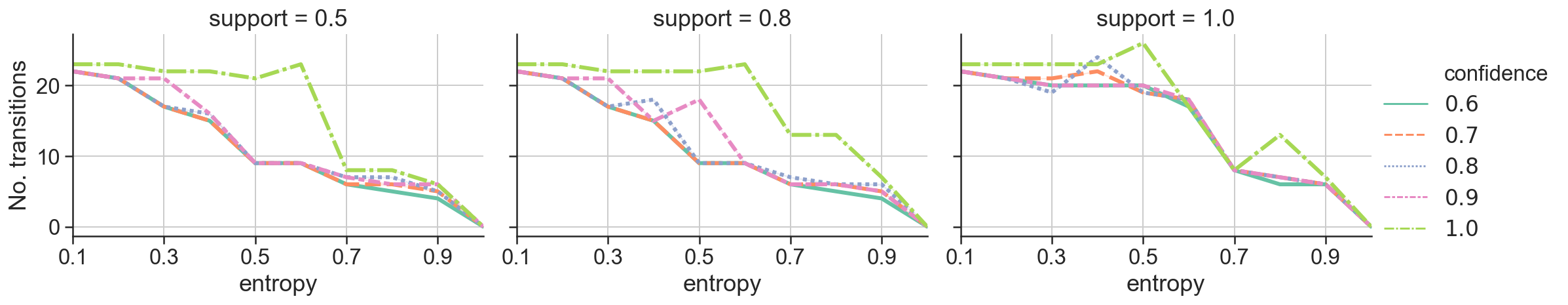}
	}
	\caption{Transitions}
\end{subfigure}
\begin{subfigure}{\textwidth}
	\resizebox{1.0\linewidth}{!}{
		\includegraphics[width=\linewidth]{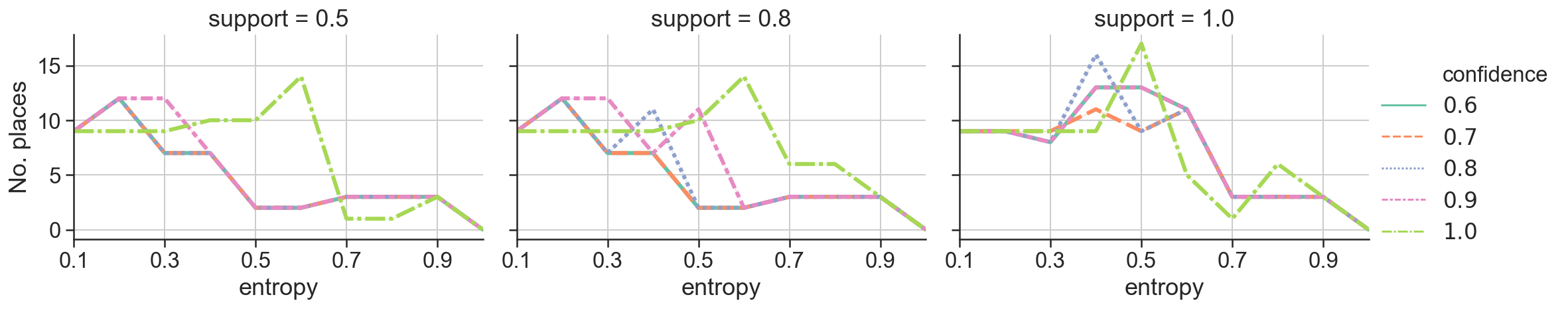}
	}
	\caption{Places}
\end{subfigure}
\begin{subfigure}{\textwidth}
	\resizebox{1.0\linewidth}{!}{
		\includegraphics[width=\linewidth]{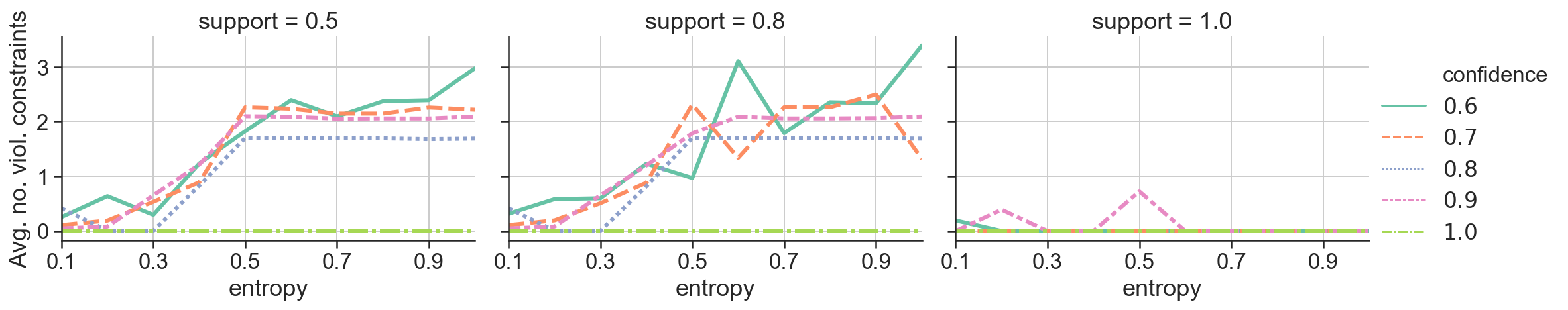}
	}
	\caption{Constraint violations (if allowed)}
\end{subfigure} 	\caption{Overview of the results for the Sepsis log (ii).}
	\label{fig:sepsis_res2}
\end{figure}

\Cref{fig:main_results} depicts an overview of the results for computation time and fitness.
A baseline result for a completely procedural model mined with Inductive Miner is provided in \cref{tab:proc_baseline}.
Overall, the mixed-paradigm conformance checking algorithm is capable of obtaining very high levels of fitness with higher levels reported when constraint violations are allowed -- especially for the BPI13 log.
This might indicate that there are particular constraints which are not compatible with the procedural model at lower levels of support.
Relaxing the need for an execution where constraints can be violated can quickly increase fitness.
It also has a positive impact on run-time for BPI13, with the BPI12 and Sepsis logs reporting similar timings with a larger spread for the latter.
As shown in \cref{fig:bpi12_res,fig:bpi13_res1,fig:bpi13_res2,fig:sepsis_res1,fig:sepsis_res2}, the highest number of constraints are mined for the Sepsis log. Running our implemented algorithm on this log also results in the highest computation time despite the lower number of traces. This is in line with the intuition established in \cref{subsec:violating}, as constraints extending the search space entail worse performance. 

A more detailed overview of fitness, execution time, and the number of modeling constructs in relation to the entropy and support/confidence parameters can be found in \cref{fig:bpi12_res,fig:bpi13_res1,fig:bpi13_res2,fig:sepsis_res1,fig:sepsis_res2}.
Note that the number of transitions and places are reported for the procedural part of the model, i.e., at an entropy of \SI{100}{\percent}, all activities are present in the declarative part of the model.

The high levels of fitness of BPI12 can be found throughout the full entropy range.
Lower levels are reported for an entropy level of \SI{20}{\percent}, suggesting that a particular combination of constraints is present that seems to be incompatible with the Workflow net's state space or rather with each other, as allowing for violations does not restore fitness.
Besides, there is a steady increase in the number of constraints and a similar decrease in the number of transitions over the entropy range.
This means that most activities are quickly eligible to be added to the declarative part of the model given their behavior.
There is little influence of the support and confidence parameters and none of the constraints are violated (and are thus not reported separately).
Compared to the fully procedural model, however, the mixed-paradigm model is capable of achieving higher fitness (\SI{100}{\percent} compared to \SI{95}{\percent}) through adding parts of the model to the declarative part to avoid replay on any presumably overfitting procedural parts.
Nevertheless, reducing the noise threshold of Inductive Miner could lead to perfect fitness, at the price of a less precise model. 
At an entropy level of \SI{10}{\percent}, only a few constraints are added to the procedural model and \SI{100}{\percent} fitness is achieved.
Given the constraints cut into the state space of the procedural model, the mixed-paradigm model by definition gives a more precise result with a few more constructs.
At higher entropy levels (e.g., \SIrange{60}{70}{\percent}) we obtain the best trade-off between higher fitness and fewer model constructs of either paradigm.
The number of constraints does not influence the computation time directly, while there is a single peak for the search  at an entropy level of \SI{60}{\percent} in case constraints cannot be violated.
This indicates that, indeed, some constraint combinations can result in a higher execution time if they result in similar costs in the procedural state space leading to longer queues with equal cost, or are defined over a particularly disjoint set of transitions as explained in \cref{subsec:satisfied}.
This makes convergence towards a unique optimal solution harder.
In general, it is not straightforwardly possible to pinpoint which constraints or combinations are causing difficulties in the state space search as this relates to both the type of constraint(s), the combination of constraint(s) (types), as well as to which transitions they constrain and their connections in the Workflow net.
Besides, the impact of {\Declare} constraints on the size of the state space is non-monotonic \cite{DBLP:journals/is/CiccioMMM17}.
Given that the size of transitions remains relatively stable around this level of entropy, it appears that a particularly disruptive constraint (type) is introduced, which is often part of the negative constraints set (e.g., \NotSuccTmp, \ExChoiceTmp, and the like).

The picture is different for the BPI13 log.
Fitness levels vary slightly over the entropy range but mostly differ according to the level of support and confidence.
Higher levels of fitness are achieved by either allowing violations, or setting support at a threshold of \SI{100}{\percent} to extract fewer yet non-conflicting constraints during the discovery phase~\cite{DBLP:journals/is/CiccioMMM17}.
Both strategies result in models that improve over the baseline fitness of the purely procedural model.
No constraints were mined for \SI{100}{\percent} confidence.
The lower the confidence, the higher the number of constraints though the number of constraints violated on average per trace does not increase (\cref{fig:bpi13_viol}).
The models with the higher confidence levels (\SIrange{80}{90}{\percent}), however, result in low fitness even when violating constraints is allowed (except for \SI{100}{\percent} support).
Hence, it shows that only a small portion of constraints can cause problems for achieving high fitness, whereas allowing them to be violated restores fitness to some extent.
Again, we conclude that there are both incompatible constraints, due to the support threshold set at less than \SI{100}{\percent}, and conflicts in the state space with the procedural model.

\begin{figure}[ptb]
	\centering
	\begin{subfigure}{\textwidth}
		\centering
		\includegraphics[height=0.3\textheight]{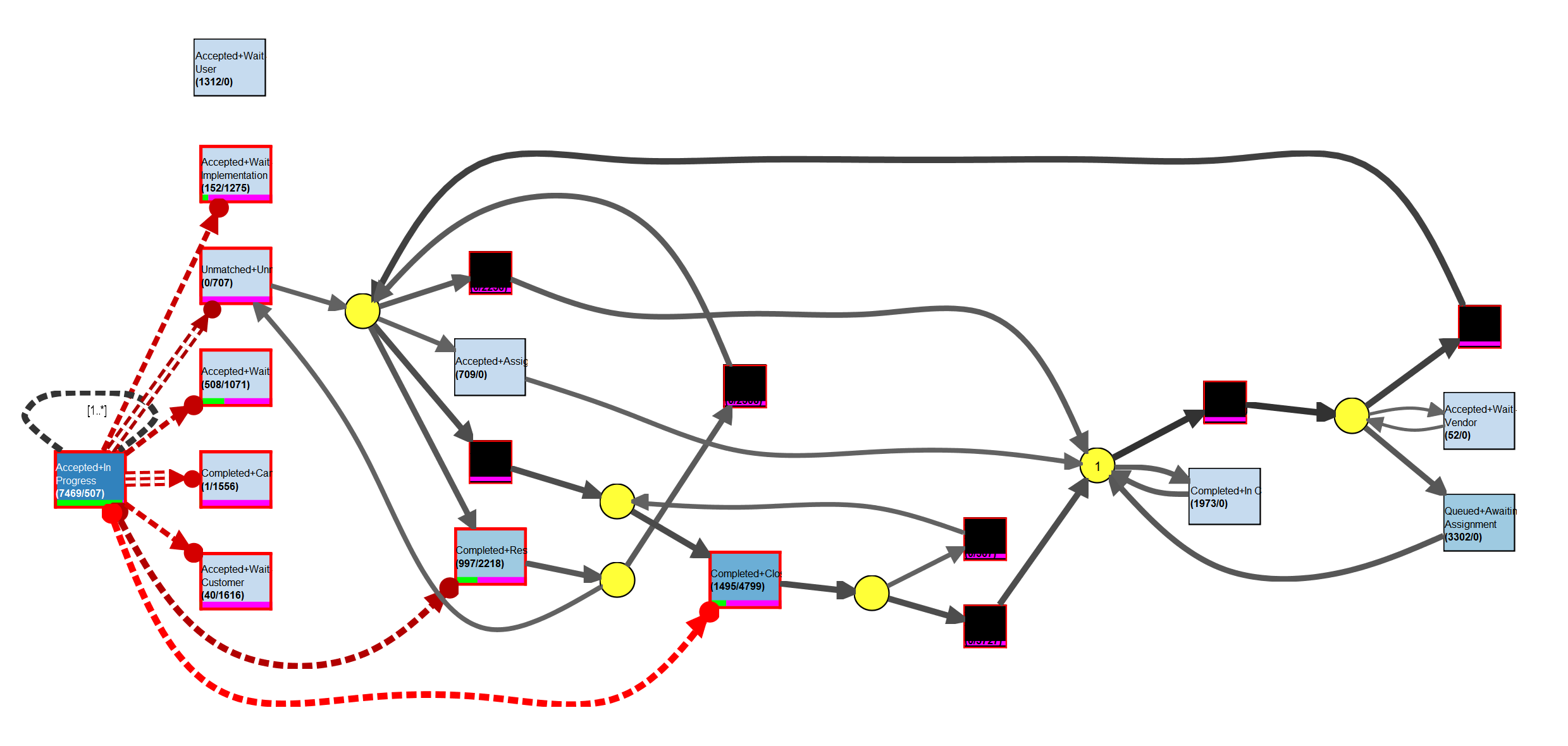}
		\caption{Entropy \SI{50}{\percent}, support \SI{80}{\percent}, confidence \SI{90}{\percent} (\SI{\sim 0}{\percent} fitness).}
	\end{subfigure}
	\begin{subfigure}{\textwidth}
		\centering
		\includegraphics[height=0.3\textheight]{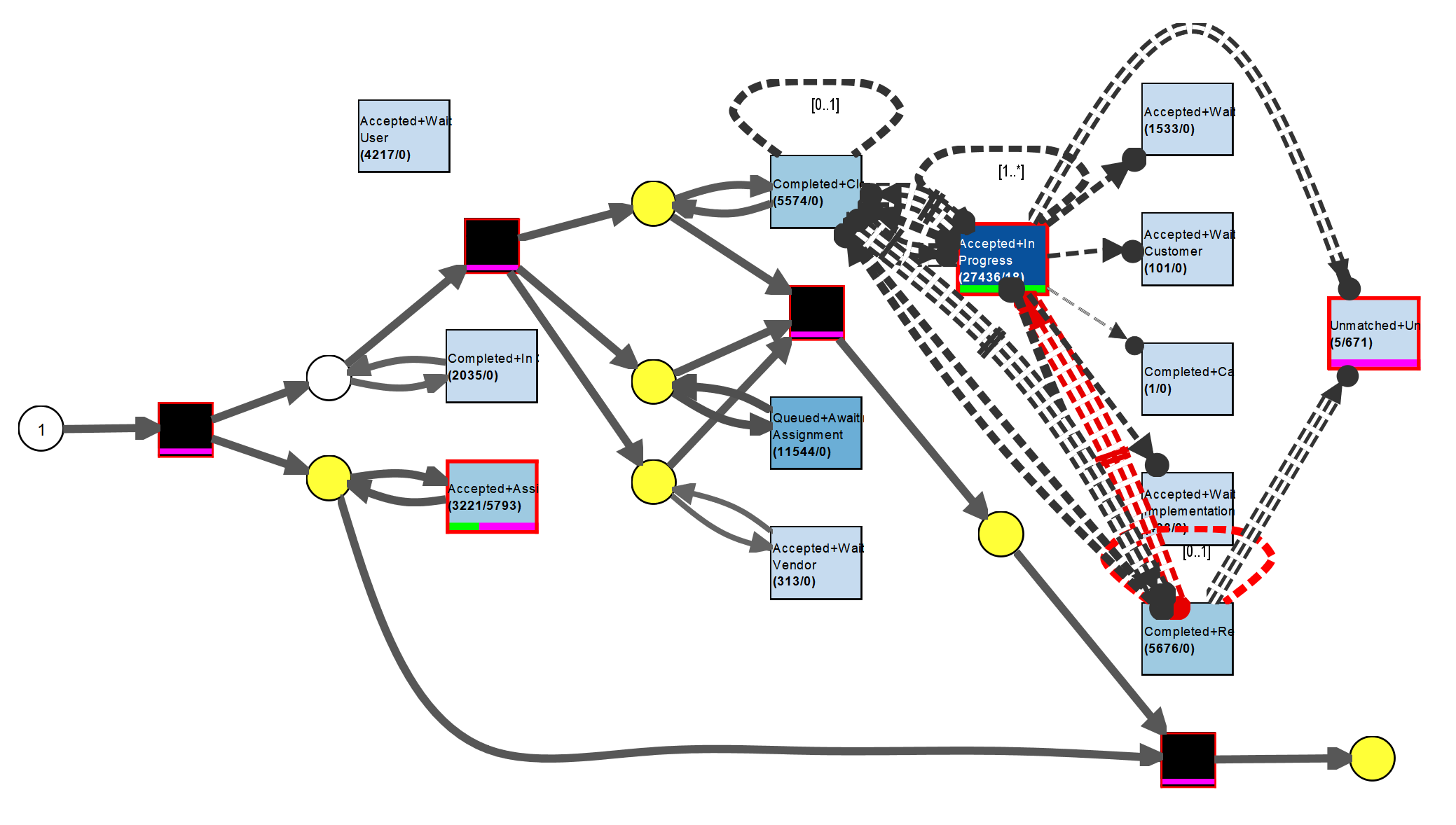}
		\caption{Entropy \SI{50}{\percent}, support \SI{80}{\percent}, confidence \SI{60}{\percent} (\SI{81}{\percent} fitness).}
	\end{subfigure}
	\caption{Output of the mixed-paradigm conformance checker with two parameter settings (violations allowed) on the BPI13 log.}
	\label{fig:bpi13_results}
\end{figure}
\Cref{fig:bpi13_results} shows two models with different fitness levels due to the confidence of constraints mined, leading to different parts of the procedural model being cut away and replaced by constraints.
In the case of the higher confidence level, more of the procedural model is cut away and substituted by incompatible constraints, which are violated, and more invisible transitions.
This shows how the mixed-paradigm conformance checker can give insights into what blends of constructs can work well for a particular event log.

The execution time in this case is higher for exploration when violations are allowed for confidence levels at 80 and 90\%, although this is not related to the number of constraints.
It coincides with the lower fitness levels, meaning a solution is hard to find and may not exist due to a high level of conflicting behavior between the declarative and procedural model, and/or the constraints themselves.

Finally, the Sepsis event log shows the most interesting spread of results over the entire entropy spectrum.
Here, it can be seen that a lower level of confidence increases the number of constraints generated, which has a negative impact on fitness. 
Indeed, the average number of violated constraints per trace generated with support below 100\% rises steadily when more constraints are generated.
At 100\% support, naturally there are close to no violated constraints which is to be expected as these constraints hold for all traces, but given the $<$100\% fitness there still seem to be constraints incompatible with the procedural model.
The addition of constraints mined with high confidence and support do not affect the fitness, but overall the procedural baseline of 100\% is only obtained in a fully declarative model or when violations are allowed.
Hence, the mixed nature of the model cannot offer any particular benefit over a model of either paradigm in this case.
The execution time in this case is again worse for when violations are allowed, meaning there are multiple parts of the procedural state space that result in a similar cost in combination with a large body of constraints.
This makes finding an optimal solution with the least amount of violated constraints hard as is illustrated by the low fitness values.
It is especially noticeable for lower levels of support, which return more constraints but not necessarily result in more constraints being violated per trace on average.
The execution time for completely declarative models is very low, despite potentially resulting in the largest search space (\cref{subsec:violating}) which makes the approach also competitive against the previous work on alignments for declarative models presented in \cite{de2014alignment}.

Overall, the mixed-paradigm conformance checker is capable of offering some interesting results to analyze the trade-offs which exist when blending constructs of either model paradigm.
Most notably, it helps to find the right level of mixture, besides offering a fully procedural or declarative result if needed.
For every event log we have analyzed, a different interplay between the model constructs took place: either the declarative constraints replaced procedural fragments to increase fitness, or the declarative constructs were rather added (with low confidence/support) to achieve lower fitness due to contradicting behavior -- let it be be between the declarative constraints and the procedural model or among constraints themselves.
Finally, in the case of BPI12 and BPI13, the mixed-paradigm model offers a lower number of constructs (i.e., constraints, transitions and places) at certain parameter settings compared to a fully procedural model (consisting of transitions and places), while offering more detailed behavior (higher precision) compared to the \SI{100}{\percent} fitness which is by definition achieved at \SI{100}{\percent} support by declarative models consisting of only a few constraints.
The execution time is influenced by particular combinations of constraints which coincides with low levels of fitness, but is overall low and relatively stable over different combinations of procedural and declarative constructs in a mixed-paradigm model.

\subsection{Implications and Limitations}
The empirical evaluation has shown 
that mixed-paradigm models can 
improve replay performance over the procedural and declarative paradigms, as it can deal with fewer constructs once the right mixture of either paradigm is found.
It is thus possible to obtain alignments for mixed-paradigm models with reasonable computational expense.
The addition of constraints does not drastically increase execution time and often poses no extra requirements on the state space analysis.
Even fully declarative models can be analyzed in comparable time to procedural models with only a few constraints.
Exceptions to this occur mostly when particular combinations of constraints, not necessarily confined to a fixed range of the entropy spectrum, are present although high levels of confidence and support tend to limit this issue.
Therefore, it is best to first check the compatibility of the state spaces of either paradigm~\cite{desmedt2016modelchecking}.

The conformance checking approach, however, is also limited in certain aspects.
The $A^*$-algorithm, while heavily optimized for conformance checking, is still not efficient in every case.
While not apparent from the experimental evaluation, the computational effort can increase in case the activities over which constraints are defined have less overlap with the Workflow net when violations are allowed, or in case a high number of constraints reflect contradictory behavior leading to long state space explorations.
This will especially be the case for process models with a high number of activities.
However, for a mixed-paradigm model with intertwined state spaces to be useful, it is expected to have a reasonable amount of overlapping activities and state spaces that are compatible to a certain extent.
Besides, the current efforts focus on computing fitness and tracking the number of constructs of either paradigm to capture overall readability/simplicity of the models. However, they are only two aspects of conformance checking.
Most notably, it would be interesting to cover precision and generalization to quantify to what extent the mixed-paradigm models, when being more or less fitting, are still performing well in not being overly precise or too general in terms of the allowed behavior.
This is closely related to the number of constraints and often interacts with fitness.
This can be covered in future iterations by incorporating alignment-based approaches such as~\cite{DBLP:conf/bpm/AdriansyahMCDA12}.

Finally, the evaluation has only focused on one particular combination of procedural and declarative languages. 
It would be interesting to further uncover how different languages mixed together can result in the best mixed-paradigm representation, e.g., whether {\Declare} with Workflow nets are better suitable compared to {\Declare} with BPMN, or to DCR Graphs with the same procedural languages.
In this respect, the alignment-based approach can help support evaluating the fitness dimension.
\section{Conclusion}
\label{sec:conclusion}
%
%
%
In this paper, we have presented the first approach enabling conformance checking for mixed-paradigm models. More specifically, our approach handles the intertwined state space of mixed-paradigm models by the help of alignment-based concepts for Petri nets extended with automata for each \gls{declare} constraint. We investigated both options of prohibiting state space exploration that induces constraint violation and of penalizing such violations with costs.
In an experimental evaluation with three real-life event logs, we illustrate how alignments are efficiently calculated over a wide range of mixed-paradigm models with various levels of procedural and declarative constructs.
Our contribution has strong implications for research and practice. Our approach fills the research gap of mixed-paradigm conformance checking and represents an important step for the further development of mixed-paradigm process mining. On the practical side, our approach and its implementation in ProM helps 
process analysts to find the most appropriate mix of either paradigm depending on the characteristics of the event log.

For future work, we intend to investigate other conformance checking metrics besides fitness such as precision and generalization.
Aside from the trade-off that can now be made between the number of constructs generated and trace fitness, it will be interesting to see how much more the reduction in state space by the declarative constructs can improve the precision of procedural models and whether this impacts generalization drastically.
Besides, we will apply the insights from the alignment results to improve mixed-paradigm process mining algorithms.
 %
%
%
%
\section*{\refname}

\end{document}